\documentclass[twocolumn]{aastex63}
\usepackage{amsmath,amssymb,graphicx,longtable,booktabs,color,CJK,natbib}
\usepackage[toc,page]{appendix}
\graphicspath{{./figures/}}

\newcommand{\rfit}{$r_{\rm{fit}}$~}

\submitjournal{The Astrophysical Journal Supplement}

\shorttitle{NGVS XXXIV. Ultra-Compact Dwarfs in the Virgo Cluster}
\shortauthors{Liu et al.}

\begin{document}
\begin{CJK*}{UTF8}{gbsn}

\title{The Next Generation Virgo Cluster Survey. XXXIV. Ultra-Compact Dwarf (UCD) Galaxies in the Virgo Cluster.}

\correspondingauthor{Chengze Liu}
\email{czliu@sjtu.edu.cn}

\author[0000-0002-4718-3428]{Chengze Liu}
\affiliation{Department of Astronomy, School of Physics and Astronomy, and Shanghai Key Laboratory for Particle Physics and Cosmology, Shanghai Jiao Tong University, Shanghai 200240, China}

\author[0000-0003-1184-8114]{Patrick C\^ot\'e}
\affiliation{Herzberg Astronomy and Astrophysics Research Centre, National Research Council of Canada, 5071 W. Saanich Road, Victoria, BC, V9E 2E7, Canada}

\author[0000-0002-2073-2781]{Eric W. Peng}
\affiliation{Department of Astronomy, Peking University, Beijing 100871, China}
\affiliation{Kavli Institute for Astronomy and Astrophysics, Peking University, Beijing 100871, China}

\author[0000-0002-0363-4266]{Joel Roediger}
\affiliation{Herzberg Astronomy and Astrophysics Research Centre, National Research Council of Canada, 5071 W. Saanich Road, Victoria, BC, V9E 2E7, Canada}

\author[0000-0003-1632-2541]{Hongxin Zhang}
\affiliation{School of Astronomy and Space Science, University of Science and Technology of China, Hefei 230026, China}
\affiliation{CAS Key Laboratory for Research in Galaxies and Cosmology, Department of Astronomy, University of Science and Technology of China, Hefei, Anhui 230026, China}

\author[0000-0002-8224-1128]{Laura Ferrarese}
\affiliation{Herzberg Astronomy and Astrophysics Research Centre, National Research Council of Canada, 5071 W. Saanich Road, Victoria, BC, V9E 2E7, Canada}

\author[0000-0003-4945-0056]{Ruben S\'anchez-Janssen}
\affiliation{STFC UK Astronomy Technology Centre, Royal Observatory, Blackford Hill, Edinburgh, EH9 3HJ, UK}

\author[0000-0001-8867-4234]{Puragra Guhathakurta}
\affiliation{UCO/Lick Observatory, Department of Astronomy and Astrophysics, University of California Santa Cruz, 1156 High Street, Santa Cruz, CA 95064, USA}

\author[0000-0003-3997-4606]{Xiaohu Yang}
\affiliation{Department of Astronomy, School of Physics and Astronomy, and Shanghai Key Laboratory for Particle Physics and Cosmology, Shanghai Jiao Tong University, Shanghai 200240, China}

\author[0000-0002-4534-3125]{Yipeng Jing}
\affiliation{Department of Astronomy, School of Physics and Astronomy, and Shanghai Key Laboratory for Particle Physics and Cosmology, Shanghai Jiao Tong University, Shanghai 200240, China}

\author[0000-0002-5897-7813]{Karla Alamo-Mart\'inez}
\affiliation{Instituto de Astrofsica, Pontificia Universidad Cat\'olica de Chile, Av. Vicu\~na Mackenna 4860, 7820436 Macul, Santiago, Chile}

\author[0000-0002-5213-3548]{John P. Blakeslee}
\affiliation{Herzberg Astronomy and Astrophysics Research Centre, National Research Council of Canada, 5071 W. Saanich Road, Victoria, BC, V9E 2E7, Canada}

\author[0000-0002-9795-6433]{Alessandro Boselli}
\affiliation{Aix-Marseille Univ, CNRS, CNES, LAM, Marseille, France}

\author{Jean-Charles Cuilandre}
\affiliation{AIM Paris Saclay, CNRS/INSU, CEA/Irfu, Universit\'e Paris Diderot, Orme des Merisiers, F-91191 Gif-sur-Yvette Cedex, France}

\author[0000-0003-3343-6284]{Pierre-Alain Duc}
\affiliation{Universit\'e de Strasbourg, CNRS, Observatoire astronomique de Strasbourg, UMR 7550, F-67000 Strasbourg, France}

\author[0000-0001-9427-3373]{Patrick Durrell}
\affiliation{Department of Physics and Astronomy, Youngstown State University,  Youngstown, OH 44555, USA}

\author{Stephen Gwyn}
\affiliation{Herzberg Astronomy and Astrophysics Research Centre, National Research Council of Canada, 5071 W. Saanich Road, Victoria, BC, V9E 2E7, Canada}

\author[0000-0002-5389-3944]{Andres Jord\'an}
\affiliation{Facultad de Ingenier\'ia y Ciencias, Universidad Adolfo Ib\'a\~nez, Av.\ Diagonal las Torres 2640, Pe\~nalol\'en, Santiago, Chile}
\affiliation{Millennium Institute for Astrophysics, Chile}

\author[0000-0001-6333-599X]{Youkyung Ko}
\affiliation{Department of Astronomy, Peking University, Beijing 100871, China}

\author[0000-0002-7214-8296]{Ariane Lan\c{c}on}
\affiliation{Universit\'e de Strasbourg, CNRS, Observatoire astronomique de Strasbourg, UMR 7550, F-67000 Strasbourg, France}

\author[0000-0002-5049-4390]{Sungsoon Lim}
\affiliation{Herzberg Astronomy and Astrophysics Research Centre, National Research Council of Canada, 5071 W. Saanich Road, Victoria, BC, V9E 2E7, Canada}
\affiliation{University of Tampa, 401 West Kennedy Boulevard, Tampa, FL 33606, USA}

\author[0000-0001-5569-6584]{Alessia Longobardi}
\affiliation{Aix-Marseille Univ, CNRS, CNES, LAM, Marseille, France}

\author[0000-0002-2849-559X]{Simona Mei}
\affiliation{Universit\'{e} de Paris, F-75013, Paris, France, LERMA, Observatoire de Paris, PSL Research University, CNRS, Sorbonne Universit\'e, F-75014 Paris, France}
\affiliation{Jet Propulsion Laboratory and Cahill Center for Astronomy \& Astrophysics, California Institute of Technology, 4800 Oak Grove Drive, Pasadena, California 91011, USA}

\author[0000-0002-7089-8616]{J. Christopher Mihos}
\affiliation{Department of Astronomy, Case Western Reserve University, 10900 Euclid Ave, Cleveland, OH 44106, USA}

\author{Roberto Mu{\~n}oz}
\affiliation{Instituto de Astrofsica, Pontificia Universidad Cat\'olica de Chile, Av. Vicu\~na Mackenna 4860, 7820436 Macul, Santiago, Chile}

\author[0000-0002-1218-3276]{Mathieu Powalka}
\affiliation{Universit\'e de Strasbourg, CNRS, Observatoire astronomique de Strasbourg, UMR 7550, F-67000 Strasbourg, France}

\author[0000-0003-0350-7061]{Thomas Puzia}
\affiliation{Instituto de Astrofsica, Pontificia Universidad Cat\'olica de Chile, Av. Vicu\~na Mackenna 4860, 7820436 Macul, Santiago, Chile}

\author[0000-0002-1685-4284]{Chelsea Spengler}
\affiliation{Herzberg Astronomy and Astrophysics Research Centre, National Research Council of Canada, 5071 W. Saanich Road, Victoria, BC, V9E 2E7, Canada}

\author[0000-0001-6443-5570]{Elisa Toloba}
\affiliation{Department of Physics, University of the Pacific, 3601 Pacific Avenue, Stockton, CA 95211, USA}

\begin{abstract}
We present a study of ultra compact dwarf (UCD) galaxies in the Virgo cluster based mainly on imaging from  the Next Generation Virgo Cluster Survey (NGVS). Using $\sim$100 deg$^{2}$ of $u^*giz$ imaging, we have identified more than 600 candidate UCDs, from the core of Virgo out to its virial radius. Candidates have been selected through a combination of magnitudes, ellipticities, colors, surface brightnesses, half-light radii and, when available, radial velocities. Candidates were also visually validated from deep NGVS images. Subsamples of varying completeness and purity have been defined to explore the properties of UCDs and compare to those of globular clusters and the nuclei of dwarf galaxies with the aim of delineating the nature and origins of UCDs. From a surface density map, we find the UCDs to be mostly concentrated within Virgo's main subclusters, around its brightest galaxies. We identify several subsamples of UCDs --- i.e., the brightest, largest, and those with the most pronounced and/or asymmetric envelopes --- that could hold clues to the origin of UCDs and possible evolutionary links with dwarf nuclei. We find some evidence for such a connection from the existence of diffuse envelopes around some UCDs, and comparisons of radial distributions of UCDs and nucleated galaxies within the cluster.
\end{abstract}



\section{Introduction}
\label{sec:intro}

Roughly two decades ago, investigators reported the discovery of a potentially new class of stellar system in the Fornax cluster \citep{1999A+AS_134_75Hilker, 2000PASA_17_227Drinkwater, 2001ApJ_560_201Phillipps}. These systems appeared to bridge the gap between normal globular clusters (GCs) and early-type galaxies (including the subset of compact elliptical galaxies), and so were named as ultra-compact dwarf galaxies (UCDs). Since then, such objects have been identified around field galaxies \citep[e.g.,][]{2011MNRAS_414_739Norris, 2014AJ_148_32Jennings} as well as in galaxy groups and clusters: i.e., Virgo \citep{2005ApJ_627_203Hacsegan, 2006AJ_131_312Jones}, Abell 1689 \citep{2005A+A_430_25Mieske}, Centaurus \citep{2007A+A_472_111Mieske}, Hydra \citep{2007ApJ_668_35Wehner}, Abell S0740 \citep{2008AJ_136_2295Blakeslee}, Coma \citep{2010ApJ_722_1707Madrid}, the NGC 1023 group \citep{0603524}, the Dorado group \citep{2007MNRAS_378_1036Evstigneeva}, the NGC 5044 group \citep{2017A+A_599_8Faifer}, the NGC 3613 group \citep{2020MNRAS__87DeBortoli} and the NGC 1132 fossil group \citep{2011ApJ_737_13Madrid}. While UCDs have luminosities comparable to faint dwarf elliptical (dE) galaxies, their sizes ($\sim$10 to 100 pc) are smaller than ``normal" dEs and yet larger than typical GCs. Due to their compact sizes and high stellar densities, they pose significant challenges for standard models of dwarf galaxy formation \cite[see, e.g.,][]{2013ApJL_775_6Strader}.

UCD formation models, which remain mostly qualitative in nature, generally invoke one of two basic scenarios. The first posits that UCDs may simply be the most massive members of the GC population, associated with the high-luminosity tail of the GC luminosity function \citep[e.g.,][]{2002A+A_383_823Mieske} or possibly arising through mergers of massive star clusters \citep[e.g.,][]{2002MNRAS_330_642Fellhauer}. The second asserts that UCDs are the surviving nuclear star clusters of nucleated dwarf elliptical galaxies (dE,Ns) whose surrounding low surface brightness envelopes were removed via tidal stripping \citep[e.g.,][]{2001ApJL_552_105Bekki}. Of course, it is entirely possible that UCDs are not a monolithic population: i.e., that they are manifested through both scenarios \citep[][]{2005ApJ_627_203Hacsegan, 0605447, 2006AJ_131_2442Mieske, 2011A+A_525_86DaRocha}.

In recent years, evidence has mounted in favor of a tidal stripping origin for at least some of these objects. Arguably the strongest evidence comes from studies of the internal kinematics of UCDs: analyses of their integrated light show that UCDs can have high dynamical-to-stellar mass ratios \citep{2014MNRAS_444_2993Forbes, 2015MNRAS_449_1716Janz}, while adaptive optics (AO) assisted integral-field unit (IFU) spectroscopy has enabled the discovery of supermassive black holes (SMBHs) in several systems \citep{2014Natur_513_398Seth, 2017ApJ_839_72Ahn, 2018ApJ_858_102Ahn, 2018MNRAS_477_4856Afanasiev}. Concurrently, a kinematic study of the UCD population around M87 has shown that they follow radially-biased orbits \citep{2015ApJ_802_30Zhang}. Meanwhile, photometric studies have revealed the presence of UCDs with asymmetric/tidal features \citep[e.g.,][]{2015ApJL_812_10Jennings, 2015ApJL_809_21Mihos, 2016A+A_586_102Voggel, 2018ApJ_853_54Schweizer}, UCDs with diffuse envelopes, which populate an apparent sequence in strength from dE,N to UCD \citep[e.g.,][]{2003Natur_423_519Drinkwater, 2005ApJ_627_203Hacsegan, 2014MNRAS_439_3808Penny, 2015ApJ_812_34Liu}, and clustering of GCs around UCDs \citep{2016A+A_586_102Voggel}. With regards to stellar contents, investigators have found color-magnitude and mass-metallicity relations \citep[e.g.,][]{2006ApJS_165_57Cote, 2011AJ_142_199Brodie, 2018ApJ_858_37Zhang}, the absence of color gradients \citep{2015ApJ_812_34Liu}, and similarities in stellar populations to nuclei \cite[e.g.,][]{2010ApJ_724_64Paudel, 2016MNRAS_456_617Janz}. Additionally, N-body simulations and semi-analytic models have demonstrated the viability of tidal stripping (within a cosmological framework) to transform dE,Ns to UCDs \citep[e.g.,][Mayes et al. 2020, in prep.]{2003MNRAS_344_399Bekki, 2013MNRAS_433_1997Pfeffer, 2014MNRAS_444_3670Pfeffer, 2016MNRAS_458_2492Pfeffer}. From this it seems clear that at least some portion of the population (e.g., massive UCDs) represent the stripped remnants of nucleated dwarf galaxies.

A prerequisite for the development and testing of any quantitative UCD formation model is reliable data on the physical properties of these objects, drawn from surveys with well-understood selection functions. Such data has proved elusive, though, and existing UCD samples are usually built from heterogeneous programs. Although they have been across a wide range of environments, most UCDs are located in groups and clusters, or associated with massive galaxies \citep[e.g.,][]{2015ApJ_812_34Liu}. As the richest concentration of galaxies near the Milky Way (MW), the Virgo cluster is an ideal environment for a comprehensive UCD survey. A handful of systems were first discovered in Virgo by \citet{2005ApJ_627_203Hacsegan} through a combination of Keck spectroscopy and HST imaging from the ACS Virgo Cluster Survey \citep{2004ApJS_153_223Cote}. Additional UCDs were later found in both imaging and/or spectroscopic programs \citep[e.g.,][]{2006AJ_131_312Jones, 2008MNRAS_385_83Chilingarian, 2011AJ_142_199Brodie, 2013ApJL_775_6Strader, 2015ApJ_812_34Liu, 2015ApJL_812_2Liu, 2015ApJL_808_32Sandoval, 2015ApJ_802_30Zhang, 2017ApJ_835_212Ko}. These studies have tended to focus on UCDs associated with M87 or a few other of the brightest galaxies in Virgo. Currently, the largest UCD sample in this cluster contains $\sim$150 objects, spread over the M87, M49 and M60 regions \citep{2015ApJ_812_34Liu}.

Given its enormous extent on the sky, a wide-field imaging survey is essential for building a homogeneous and complete sample of Virgo UCDs. The Next Generation Virgo cluster Survey \citep[NGVS;][]{2012ApJS_200_4Ferrarese} is a deep, multi-band ($u^*griz$) imaging campaign of the Virgo cluster carried out with the MegaCam instrument on the Canada France Hawaii Telescope (CFHT). The survey covers an area of 104 deg$^2$ and is typified by excellent image quality, with a median FWHM of $0.54\arcsec$ in the $i$-band (see their Figure~8). In principle, we can use these NGVS images to measure half-light radii for all compact Virgo objects brighter than $g\sim21.5$ mag and larger than $r_h\sim10$ pc \citep[see][]{2015ApJ_812_34Liu}, potentially producing the largest and the most complete sample of UCDs in any environment. The analysis presented here builds on previous NGVS papers that have focused on the photometric and kinematic properties of UCDs \citep[e.g.,][]{2015ApJ_812_34Liu, 2015ApJL_812_2Liu, 2015ApJ_802_30Zhang}. It also complements other papers in the NGVS series dealing with other stellar systems in Virgo, including globular clusters \citep[i.e.,][]{2014ApJ_794_103Durrell, 2016ApJS_227_12Powalka, 2018ApJ_864_36Longobardi}, galaxies \citep{2015ApJ_804_70Guerou, 2016ApJ_820_69Sanchez-Janssen, 2016ApJ_824_10Ferrarese, 2017ApJ_836_120Roediger, 2020ApJ_890_128Ferrarese}, and their nuclei \citep[i.e.,][]{, 2017ApJ_849_55Spengler, 2019ApJ_878_18Sanchez-Janssen}. 

This paper is structured as follows. In \S\ref{sec:data} we provide an overview of the NGVS and the data products used in our analysis, while \S\ref{sec:ucd_selection} describes the methodology we have used to identify UCD candidates. In \S\ref{sec:results} we present our results, including a new catalog of UCD candidates, and draw attention to a number of particularly interesting sub-samples therein. We discuss these findings in \S\ref{sec:discussion}, and in \S\ref{sec:summary}, summarize our conclusions and outline directions for future work. Throughout this study, we adopt a common distance to all UCDs \citep[16.5 Mpc][]{2007ApJ_655_144Mei, 2009ApJ_694_556Blakeslee}, corresponding to a distance modulus of $(m$-$M) = 31.09$ and physical scale of $80$ pc arcsec$^{-1}$.

\section{Data}
\label{sec:data}

\subsection{Overview}

The primary source of data used in this study is the NGVS. The survey footprint covers the two main sub-clusters of Virgo (A and B, centered on M87 and M49, respectively) out to their virial radii (i.e., $R_{\rm 200}$ = 5\fdg38 for Virgo A and 3\fdg33 for Virgo B). As described in \citet{2014ApJS_210_4Munoz} and \citet{2015ApJ_812_34Liu}, the NGVS is an ideal resource for the study of compact stellar systems, e.g., GCs, UCDs and dwarf nuclei. The NGVS imaging consists of short and long exposures, where the former can be used to find UCDs brighter than $g \sim 18.5$ mag. Such objects are interesting given that they define the extreme of UCD formation (and, in some cases, even host supermassive black holes; \citealt{2014Natur_513_398Seth}, \citealt{2018ApJ_858_102Ahn}).

\begin{figure}
\epsscale{1.15}
\plotone{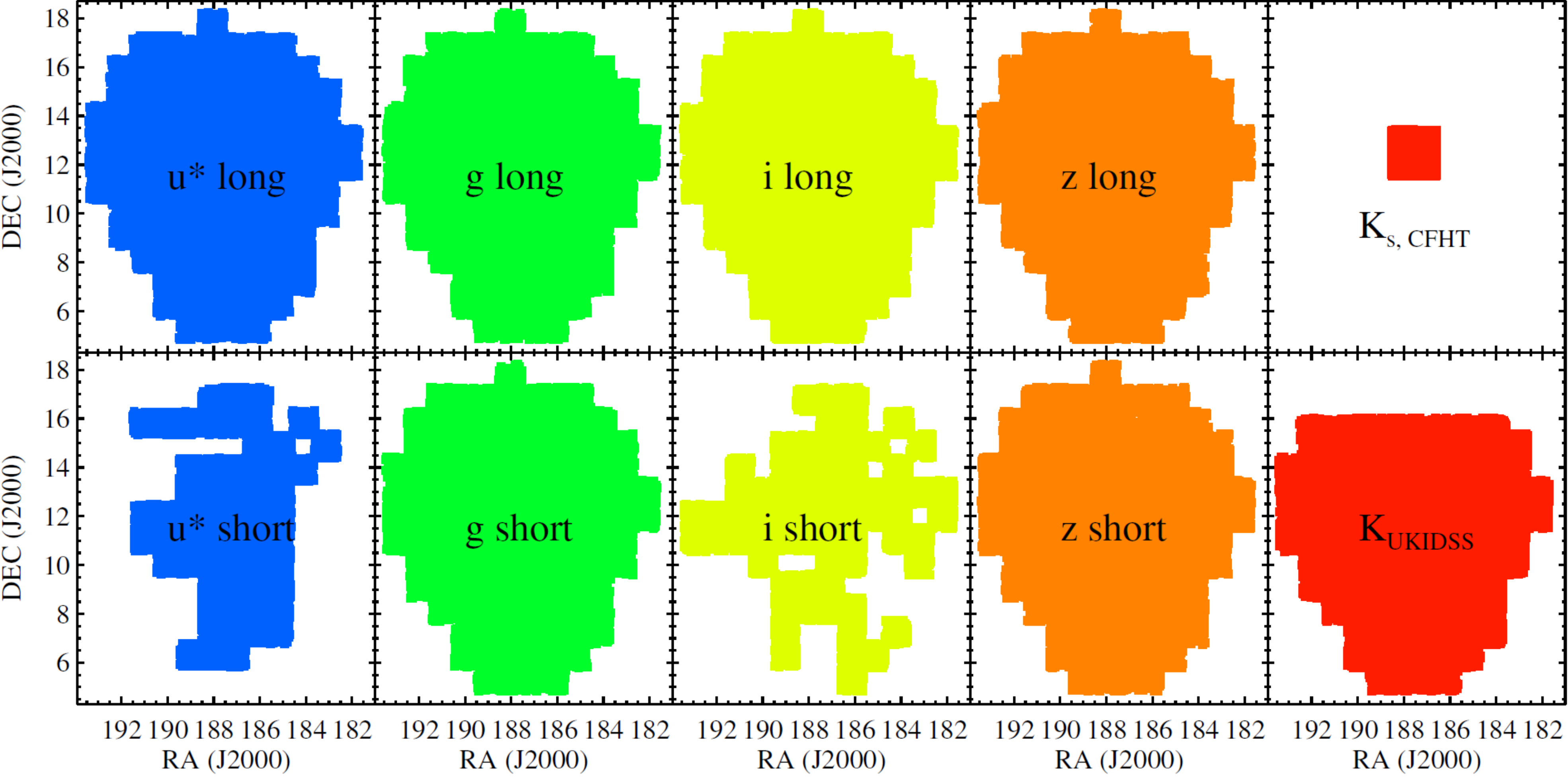}
\caption{The areal coverage of our optical and near-infrared imaging, organized by exposure length ({\it top}: long, {\it bottom}: short). The NGVS achieved 100\% completeness for the long exposures in the $u^*giz$ bands and in the $gz$ bands for the short exposures; short exposures in the $u^*i$ bands were only partially completed. The NGVS-IR ($K_s$-band, \citealt{2014ApJS_210_4Munoz}) imaging only covers the center of sub-cluster A, while the UKIDSS ($K$-band, \citealt{2007MNRAS_379_1599Lawrence}) data cover most of the NGVS footprint.}
\label{fig:map_bands}
\end{figure}

\begin{figure}
\epsscale{1.15}
\plotone{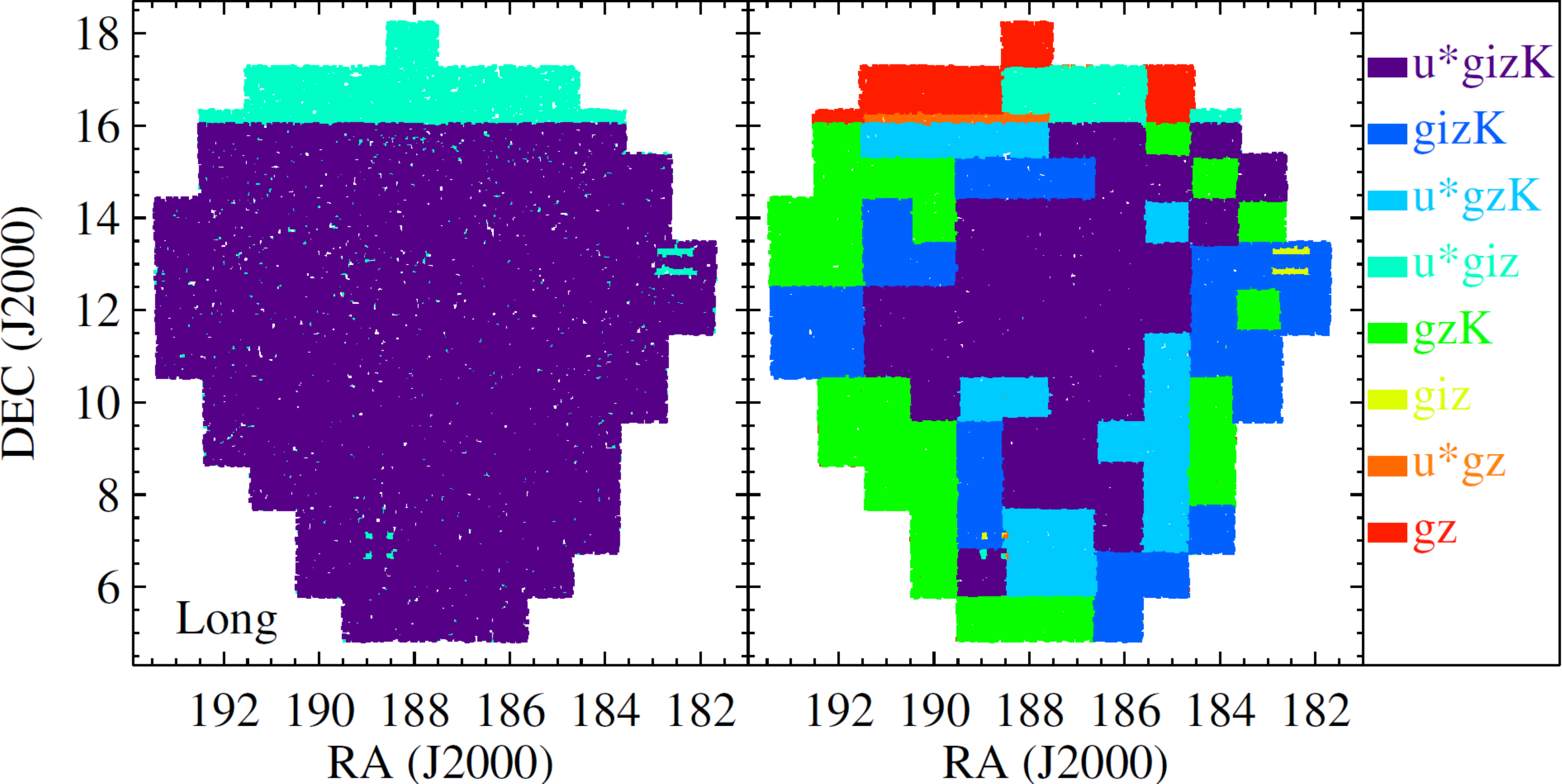}
\caption{Summary of the spatial dependence of the multi-wavelength photometry used to select UCDs, divided by exposure length ({\it left}: long exposure catalog, {\it right}: short exposure catalog).}
\label{fig:map_selection}
\end{figure}

Figure \ref{fig:map_bands} shows the final observing status of the NGVS, organized by exposure length. We have excluded the $r$-band since those observations had to be partially sacrificed due to CFHT's dome shutter failure in 2012A. For the long exposures, the survey is fully complete in the remaining bands, while only partial areal coverage was achieved for the short exposures in the $u^*$ ($\sim50$\% completeness) and $i$ ($\sim57$\% completeness) bands.

Near-infrared imaging has proved to be a powerful tool for UCD selection \citep{2014ApJS_210_4Munoz, 2015ApJ_812_34Liu}. As shown in the upper right corner of Figure \ref{fig:map_bands}, we have deep $K_s$-band images in the central 4 deg$^2$ of sub-cluster~A \citep[NGVS-IR;][]{2014ApJS_210_4Munoz}, which we have previously used to select a high-purity UCD sample around M87 \citep{2015ApJ_812_34Liu}. Alternatively, as shown in the bottom-right panel of Figure \ref{fig:map_bands}, $K$-band imaging from UKIDSS \citep{2007MNRAS_379_1599Lawrence} covers a large fraction of the NGVS footprint. About $\sim70\%$ of the bright objects [$g_0<21.5$ mag, where $g_0$ is the aperture-corrected magnitude measured within a 16-pixel diameter ($\sim 3$\arcsec) and corrected for Galactic extinction] in the NGVS have counterparts in the UKIDSS $K$-band. Thus, although UKIDSS is much shallower than the NGVS-IR ($5\sigma$ limiting magnitude $\sim$18.4 and $\sim$24.4 mag, respectively), it is nonetheless useful for separating UCDs from background galaxies among the bright objects.

We summarize the combinations of imaging data at hand, separated by exposure length, in Figure \ref{fig:map_selection}. For the case of the long exposures (left panel), the footprint is simply divided into two areas depending on the availability of UKIDSS $K$-band imaging. The short exposure map (right panel) is much more complicated owing to the incompleteness in the associated $u^*$ and $i$ band imaging.

Full details on the reduction of NGVS images can be found in \citet{2012ApJS_200_4Ferrarese}. To generate a homogeneous catalog of compact objects, we run SExtractor \citep{1996A+AS_117_393Bertin} in double-image mode. We detect objects in the $g$-band and then measure a set of parameters, including aperture magnitudes, in the $u^*giz$ bands. In this study, we measure the luminosity and color of all detected objects with aperture magnitudes. To minimize systematics, we apply aperture corrections that account for PSF variations within, and between, fields. Specifically, we use corrected 3.0\arcsec-diameter aperture magnitudes to represent total magnitudes and corrected 1.5\arcsec-diameter aperture magnitudes to estimate colors.

For the catalog generation and magnitude correction, we follow the method of \citet{2015ApJ_812_34Liu}, with one exception. \citet{2015ApJ_812_34Liu} subtracted models of the diffuse light from nearby massive galaxies (M87, M49, and M60), whereas this is avoided in the current analysis to have a homogeneous catalog. We generate independent catalogs based on the short and long exposure images and then merge them into one afterwards. We adopt measurements from the short exposure catalog for those objects which are saturated in the long exposures; otherwise measurements are taken from the long exposure catalog.

\begin{figure*}
\epsscale{0.55}
\plotone{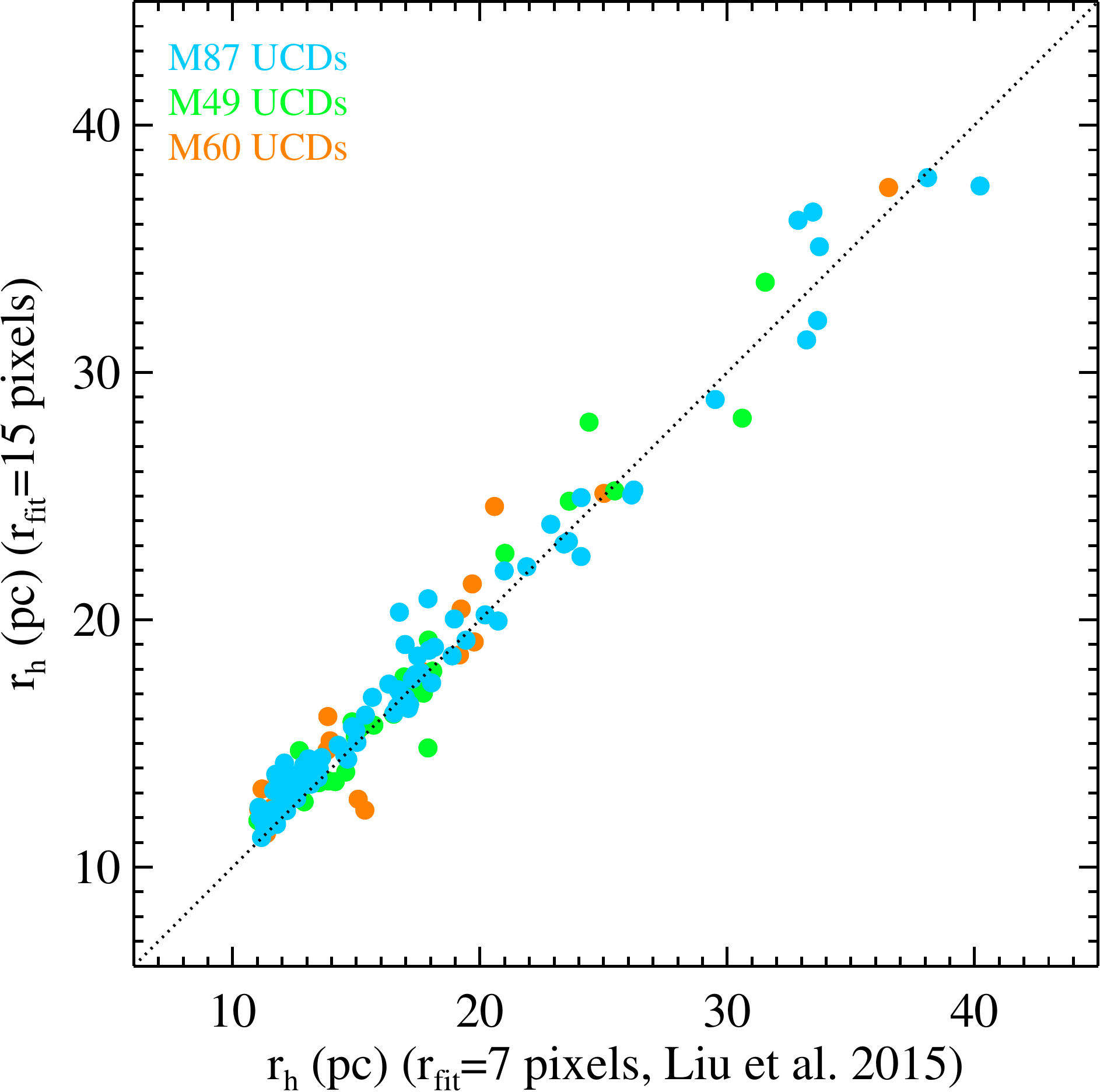}
\epsscale{0.55}
\plotone{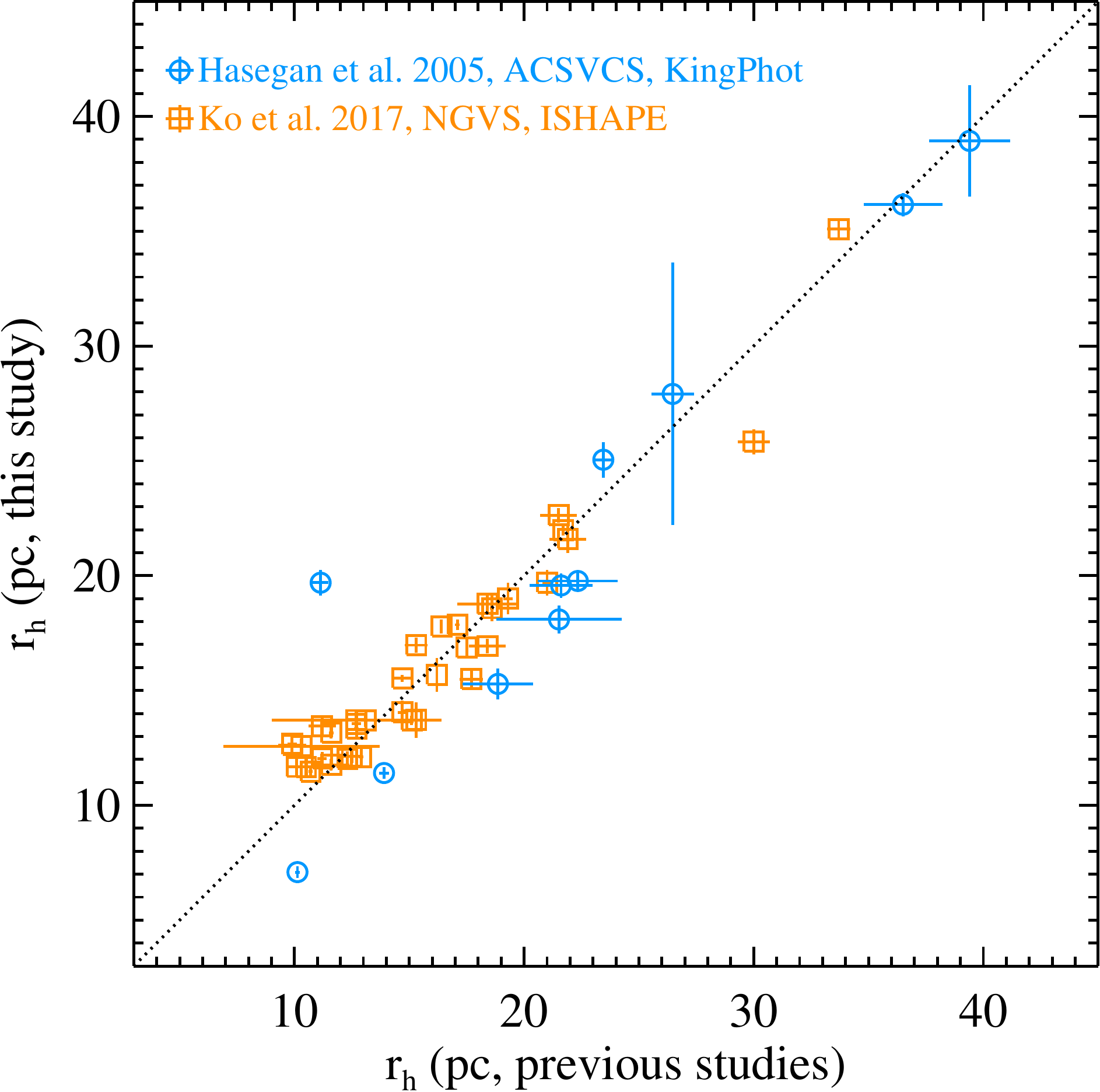}
\caption{\textit{Left panel:} Comparison of $r_h$ measurements for UCDs based on different fitting radii, \rfit (7 vs 15 pixels), with data taken from \cite{2015ApJ_812_34Liu}. \textit{Right panel:} Comparison of $r_h$ measured using different data (blue circles, ACSVCS data, \citealt{2005ApJ_627_203Hacsegan}) or different methods (orange squares, ISHAPE software, \citealt{2017ApJ_835_212Ko}).}
\label{fig:rh_compare}
\end{figure*}

In addition to the imaging that forms the basis of this study, there are many past spectroscopic programs targetting the Virgo cluster that we can draw upon. \citet{2012ApJS_200_4Ferrarese} summarized the relevant programs for Virgo compact stellar systems (i.e., GCs, UCDs and dE,Ns) as of 2012; these include radial velocity measurements from various MMT/Hectospec, Magellan/IMACS, VLT/VIMOS and AAT/AAOmega programs (see the paper for details). Since then, a number of NGVS-motivated spectroscopic programs have been undertaken \citep[see, e.g.,][]{2015ApJ_802_30Zhang, 2018ApJ_858_37Zhang, 2016ApJ_822_51Toloba, 2017ApJ_849_55Spengler, 2018ApJ_864_36Longobardi}. We have collected radial velocities from these and other previous works \citep{1985AJ_90_1681Binggeli, 2001ApJ_559_812Hanes, 2003ApJ_591_850Cote, 2008A+A_485_657Brinchmann, 2011ApJS_197_33Strader, 2012ApJ_760_87Strader, 2013MNRAS_428_389Pota, 2014MNRAS_439_2420Blom, 2014MNRAS_443_1151Norris, 2015ApJ_806_133Li, 2015MNRAS_450_1962Pota, 2017AJ_153_114Forbes, 2017ApJ_835_212Ko, 2018ApJL_856_31Toloba}, as well as from the NASA/IPAC Extragalactic Database (NED)\footnote{https://ned.ipac.caltech.edu/}, the SIMBAD Astronomical Database\footnote{http://simbad.u-strasbg.fr/simbad/} \citep{2000A+AS_143_9Wenger} and the Sloan Digital Sky Survey \citep[SDSS,][]{2018ApJS_235_42Abolfathi}. In all, we have a total of 31,346 velocity measurements for objects in the NGVS footprint brighter than $g_0=21.5$ mag. This database includes foreground stars, GCs, UCDs, galaxies in Virgo, and background galaxies. In what follows, we make use of this large velocity catalog to eliminate contaminants from our photometric UCD selection, as well as to recover UCDs that miss our cuts.

\subsection{Size Measurements}

Size is the defining parameter of UCDs\footnote{Unless stated otherwise, we use the term ``size" to refer to an object's half-light radius, $r_h$, exclusively.}. As shown by \citet{2015ApJ_812_34Liu}, the excellent image quality of the NGVS allows us to measure reliable sizes for compact objects in Virgo (mainly GCs, UCDs, and galactic nuclei) larger than $\sim$10 pc (see their Section 2.3). We measure half-light radii using the KINGPHOT package \citep{2005ApJ_634_1002Jordan},  focusing on the $g$ and $i$ bands because of the former's depth and the latter's exquisite seeing. Comparisons of the two sets of $r_h$ measurements show that they are consistent with each other (see Figure 3 in \citealt{2015ApJ_812_34Liu}). \cite{2005ApJ_634_1002Jordan} show that KINGPHOT $r_h$ measurements are biased to larger values when $r_h \gtrsim r_{\rm{fit}}/2$ (where $r_{\rm{fit}}$ is fitting radius within which we adopt KINGPHOT), which is $\sim$50 pc for an \rfit = 7 pixels (used in \citealt{2015ApJ_812_34Liu}) at the distance of Virgo. This choice of \rfit  is reasonable since previous works shows that most UCDs are smaller than 40 pc \citep{2011AJ_142_199Brodie, 2011ApJ_737_86Chiboucas,  2011ApJS_197_33Strader, 2012MNRAS_422_885Penny}. However, in the interests of determining whether there are larger UCDs in Virgo, we run KINGPHOT with $r_{\rm fit}=15$ pixels in this study. Thus, our KINGPHOT measurements would be biased for objects with $r_h \gtrsim$ 110 pc.

The left panel of Figure \ref{fig:rh_compare} compares our UCD $r_h$ measurements from \cite{2015ApJ_812_34Liu} for the two values of \rfit above. We note that the larger \rfit yields slightly larger sizes when $r_h \lesssim$ 16 pc. Otherwise, the two sets of $r_h$ measurements are statistically equivalent, so we therefore adopt the KINGPHOT measurements made with \rfit = 15 pixels. The right panel of this figure shows a comparison between the $r_h$ measurements from this work and those from previous studies. The blue circles are taken from \citet{2005ApJ_627_203Hacsegan}, who measured $r_h$ using KINGPHOT and HST data (ACSVCS, \citealt{2004ApJS_153_223Cote}). The orange squares denote the $r_h$ measurements from \citet{2017ApJ_835_212Ko}, who used ISHAPE software and NGVS data. From this figure, we can see that our $r_h$ measurements are consistent with the measurements from previous studies, even though they used different methodologies and/or data sets.

\section{UCD Selection}
\label{sec:ucd_selection}

\begin{figure*}
\epsscale{0.539}
\plotone{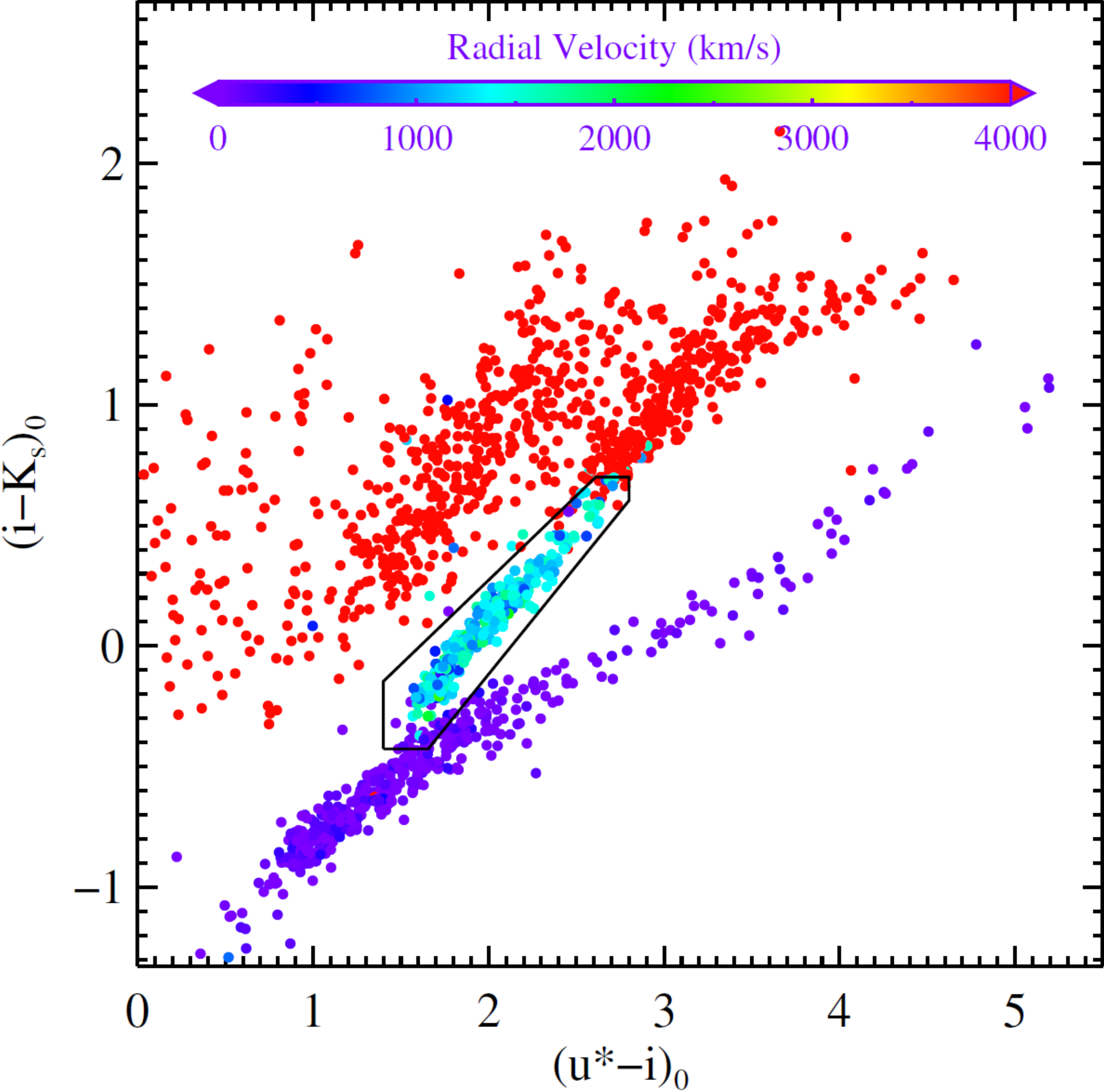}
\epsscale{0.539}
\plotone{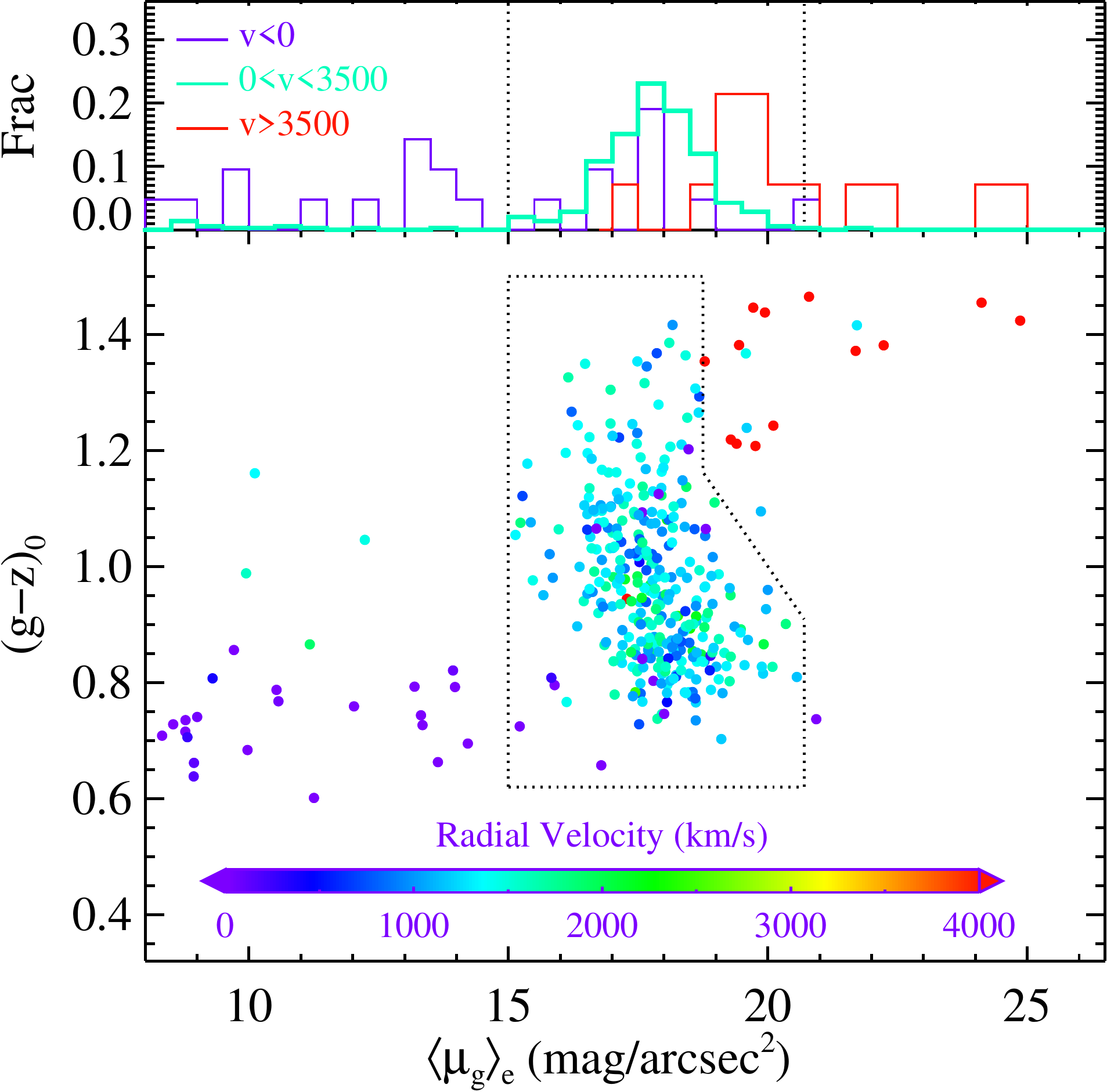}
\caption{{\it Left~panel:} The $u^*iK_s$ color-color diagram. The solid polygon represents our GC+UCD selection box in this plane. The points are colored by their associated radial velocity measurements, reflected in the color bar at top. {\it Right~panel:} The distribution of mean surface brightness, $\langle{\mu_g}\rangle_e$, (in broad bins of radial velocity; top) and $(g-z)_0$ color as a function of $\langle{\mu_g}\rangle_e$ for the objects lying within our selection region on the left. The dotted polygon shows the next layer of our selection, based on surface brightness, and the points are colored by their associated radial velocities.}
\label{fig:uiK_based}
\end{figure*}

We select UCD candidates within the multi-dimensional parameter space of magnitude, ellipticity ($e \equiv 1 - b/a$), color, surface brightness, and half-light radius, which we now describe and justify\footnote{It is worth bearing in mind that UCD selection criteria are often driven as much by observational details, such as limiting magnitudes and angular resolution, as by considerations of formation physics.}. We begin by adopting simple magnitude cuts of $14.0 < g_0 < 21.5$ mag, where the lower bound corresponds to the saturation limit of our $g$-band short exposures and the upper bound to the limit of accurate measurements of half-light radii (see \S2 in \citealt{2015ApJ_812_34Liu} for details). We also apply a simple cut on ellipticity, adhering to the empirical result that most spectroscopically-confirmed UCDs in Virgo (i.e., $v_r < 3500$ km/s) have $e < 0.3$ \citep[see, e.g.,][]{2015ApJ_802_30Zhang}.

As we have previously shown \citep{2014ApJS_210_4Munoz, 2015ApJ_812_34Liu, 2016ApJS_227_12Powalka, 2016ApJL_829_5Powalka}, the combination of $u^*giK_s$ photometry proves to be a highly effective tool for discriminating extragalactic GCs and UCDs from background galaxies and foreground stars. Unfortunately, Figure \ref{fig:map_bands} shows that we only have $K_s$-band imaging for the central 4 deg$^2$ of sub-cluster A from the NGVS-IR (see \citealt{2014ApJS_210_4Munoz} for details). Moreover, in the $u^*$ and $i$ bands, only partial coverage is available in the short exposure category. In an attempt to balance homogeneity and accuracy, we base the color portion of our UCD selection on our $u^*$, $g$, $i$, $z$ and UKIDSS $K$-band photometry. The UKIDSS data, although shallower than that from NGVS-IR, is sufficient for our purpose, i.e. to select UCD candidates with $g_0<21.5$ mag. We will describe our $u^*giK_s$- and $u^*gizK$-based selections and compare their results in the following two subsections.

\subsection{$u^*giK_s$-Based Selection}
\label{sec:ugik_selection}

In the left panel of Figure \ref{fig:uiK_based}, we show the cuts employed as part of our $u^*giK_{s}$-based selection. There, we plot in the $(u^*-i)$ vs. $(i-K_{s})$ plane, the $\sim$2000 objects from our spectroscopic catalog that satisfy our magnitude and ellipticity cuts. The points have been colored by their measured radial velocities and can be divided into three main groups: background galaxies (red dots; $v \gtrsim 3500$ km/s), Virgo members (green and cyan dots; $0 \lesssim v \lesssim 3500$ km/s), and foreground stars (blue and purple dots; $v \lesssim 0$ km/s)\footnote{Note that this very basic redshift classification is not strictly correct; some Virgo members do indeed have negative radial velocities, such as objects belonging to the M86 group \citep[see, e.g.,][]{2018A+A_614_56Boselli, 2012ApJ_757_184Park} and many stars have positive radial velocities \citep{2019A+A_622_205Katz}.}. It is clear that Virgo members can be readily distinguished from background galaxies and foreground stars in the $u^*iK_{s}$ color-color diagram. To isolate Virgo members, we therefore adopt the following color cuts:
  \begin{equation}
    \left\{
    \begin{aligned}
      &  1.400 \le (u^*-i)_0 \le 2.800; \\
      & -0.427 \le (i-K_{s})_0 \le 0.700; \\
      & (i-K_{s})_0 \le -1.127 + 0.700 \times (u^*-i)_0; \\
      & (i-K_{s})_0 \ge -1.917 + 0.900 \times (u^*-i)_0.
   \end{aligned}
   \right.
   \label{eq:uiK}
  \end{equation}
which are indicated by the irregular polygon in the figure.

The right-hand panel of Figure \ref{fig:uiK_based} shows the $(g-z)_0$ color as a function of mean effective surface brightness, $\langle{\mu_g}\rangle_e$, which is the average surface brightness measured within the half-light radius. The points show those objects that passed our color cuts in the $u^*iK_{s}$ diagram and, again, are colored according to their radial velocities. It is clear from the distribution that we can use surface brightness to further improve the purity of our Virgo sample by removing background galaxies and some foreground stars. The dotted polygon shows the exact cuts in surface brightness that we adopt, which are described by the following functions:
  \begin{equation}
    \left\{
    \begin{aligned}
      & 0.620 \le (g-z)_0 \le 1.500; \\
      & \langle{\mu_g}\rangle_e \ge 15.000 {\rm ~mag/arcsec}^2;  \\
      & \langle{\mu_g}\rangle_e \le 18.750 {\rm ~mag/arcsec}^2, \\ 
      & {\rm ~~~~~~~~~~~~when~} 1.163 \le (g-z)_0 \le 1.500; \\
      & \langle{\mu_g}\rangle_e \le 27.692 - 7.692 \times (g-z)_0, \\
      & {\rm ~~~~~~~~~~~~when~} 0.909 \le (g-z)_0 \le 1.163; \\
      & \langle{\mu_g}\rangle_e \le 20.700 {\rm ~mag/arcsec}^2, \\
      & {\rm ~~~~~~~~~~~~when~} 0.620 \le (g-z)_0 \le 0.909. 
   \end{aligned}
   \right.
   \label{eq:mug}
  \end{equation}

The combination of the cuts applied to this point leaves us with a broad sample of Virgo members that includes compact elliptical galaxies, galactic nuclei (in low-mass galaxies), UCDs, and GCs. To isolate the UCDs within this sample, we apply one final set of cuts based on our measured half-light radii, which are:
  \begin{equation}
    \left\{
    \begin{aligned}
      & 11 < \langle{r_h}\rangle < 100 \rm ~pc; \\
      & \frac{|r_{h,g}-r_{h,i}|}{\langle{r_h}\rangle} \le 0.5; \\
      & \frac{r_{h,g,\rm error}}{r_{h,g}} \le 15\%; \\
      & \frac{r_{h,i,\rm error}}{r_{h,i}} \le 15\%.
   \end{aligned}
   \right.
   \label{eq:rh}
  \end{equation}
where $\langle{r_h}\rangle$ represents the weighted mean of the half-light radii measured in the $g$ and $i$ bands. These cuts mimic ones often used in previous studies to separate UCDs and GCs \citep[e.g.,][]{2011AJ_142_199Brodie, 2011ApJS_197_33Strader, 2012MNRAS_422_885Penny}. Though arbitrary, this is not an unreasonable choice since the typical size of either MW or extragalactic GCs is $\sim$3 pc and most GCs are smaller than $\sim 10$ pc \citep[e.g.,][]{1991ApJ_375_594vandenBergh, 2005ApJ_634_1002Jordan}. Furthermore, \citet{2015ApJ_812_34Liu} have shown that $r_h$ measurements based on NGVS imaging are reliable for bright objects ($g_0<21.5$ mag) larger than $r_h \sim 10$ pc (see \S2.3 of their paper). The lower limit on $\langle{r_h}\rangle$ used in this study roughly matches the limiting resolution of NGVS imaging.

\begin{figure*}
\epsscale{0.55}
\plotone{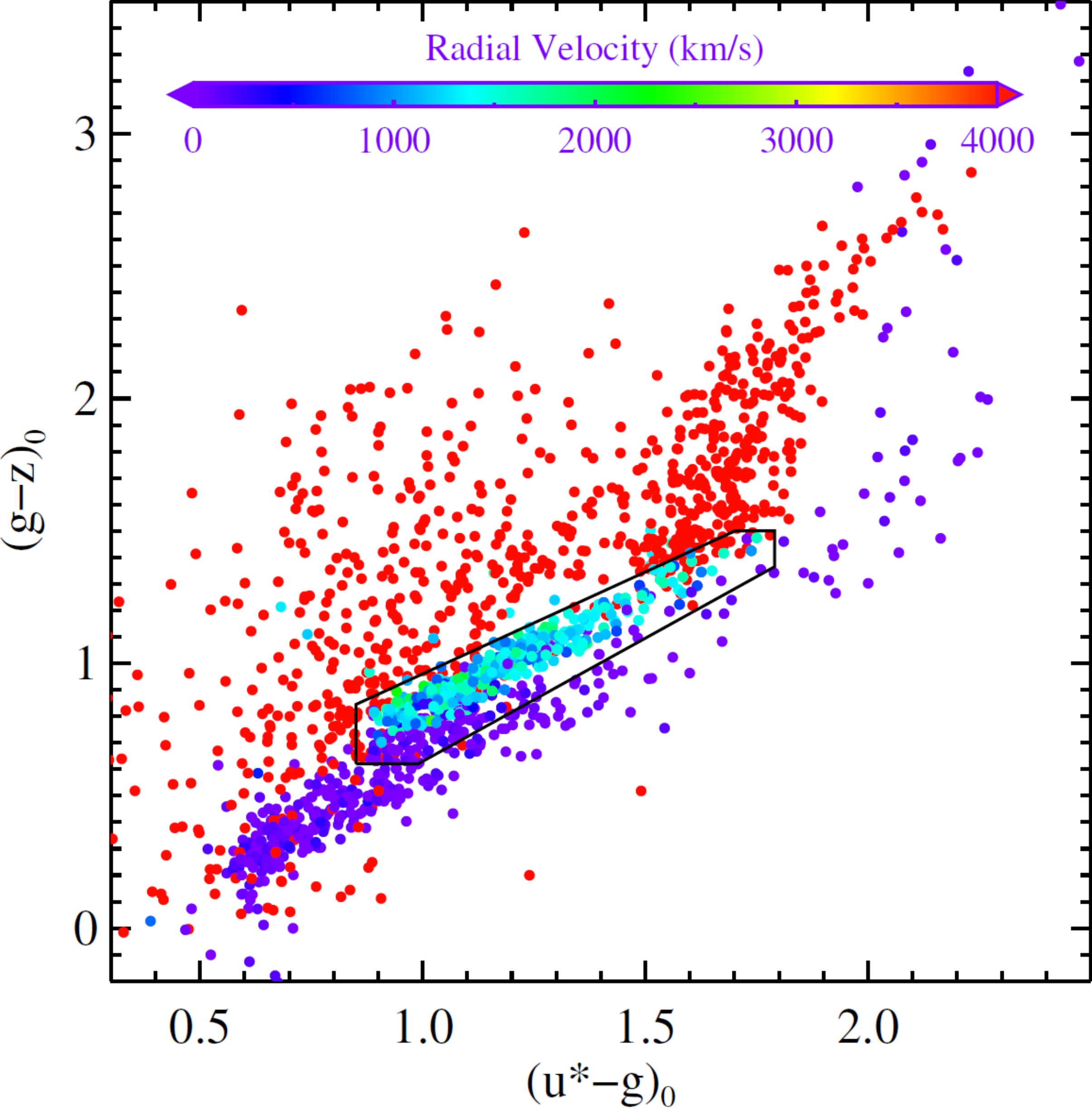}
\epsscale{0.57}
\plotone{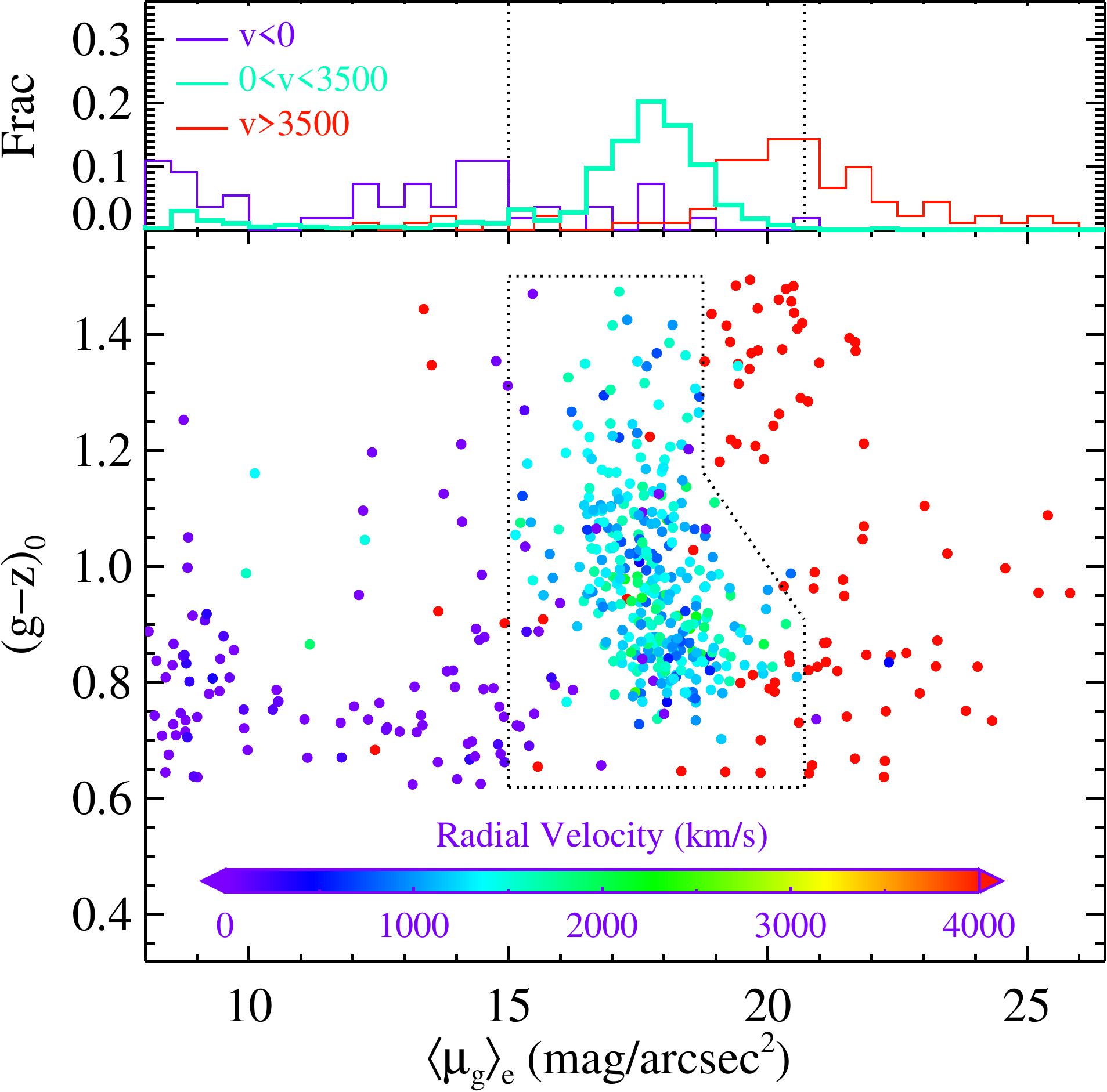}
\caption{As in Figure \ref{fig:uiK_based} but based on the $u^*gz$ color-color diagram instead.}
\label{fig:ugz_based}
\end{figure*}

\begin{figure}
\epsscale{1.17}
\plotone{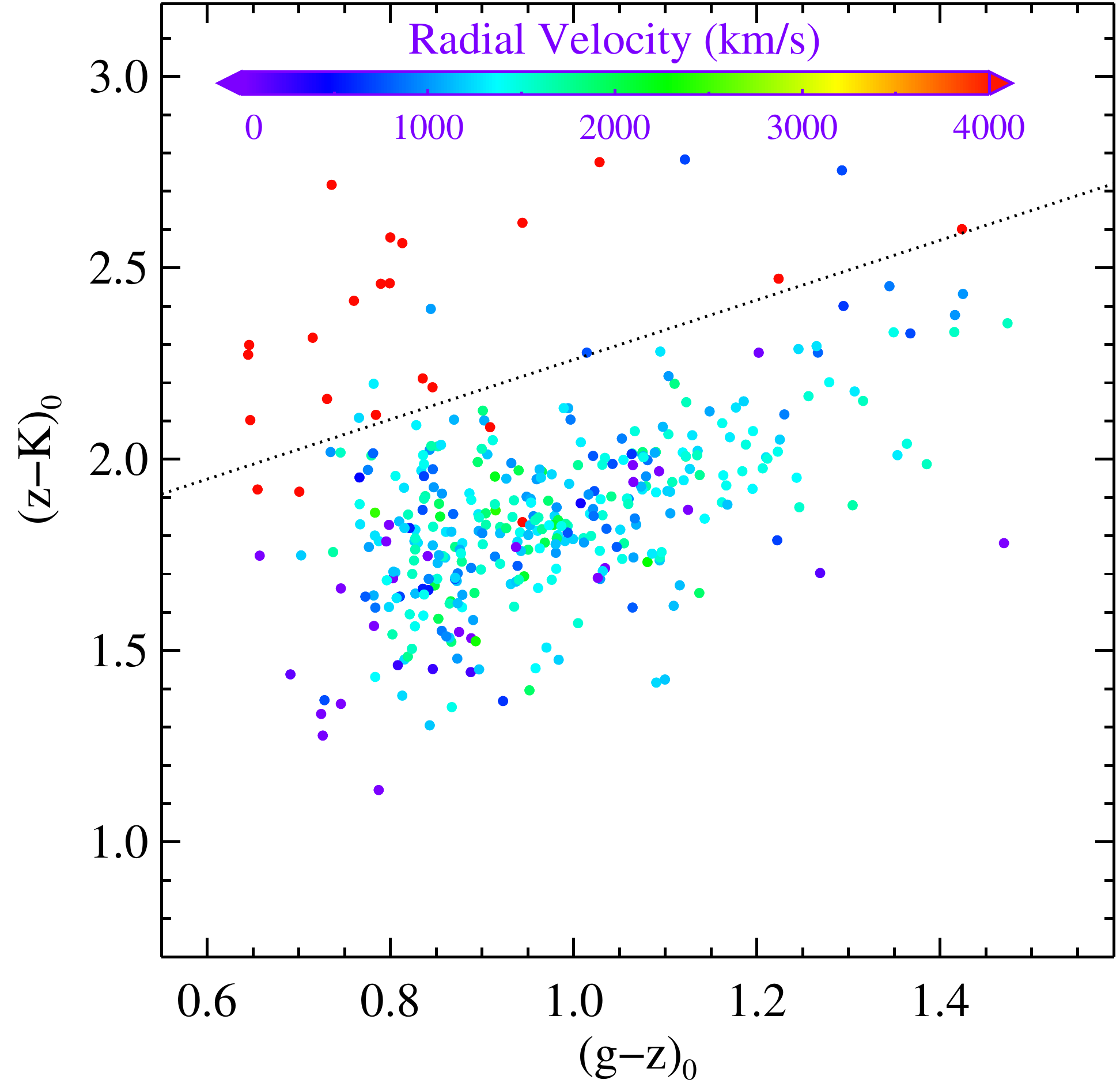}
\caption{$gzK$ color-color diagram for $u^*gz$-selected UCD candidates. Different colors indicate different radial velocities as given by color bar. The black dotted line is used to separate background galaxies and UCD candidates.}
\label{fig:gzk}
\end{figure}

\subsection{$u^*giz$- and $u^*gizK$-Based Selection}

Given the lack of deep $K_s$-band imaging over most of the NGVS footprint, we have developed an alternative strategy for selecting UCDs based on the $u^*$, $g$, $i$, $z$ and (where available) $K$ bands. As seen in Figure~\ref{fig:map_bands}, the NGVS imaging is not fully complete in the $u^*$ and $i$ bands for the short exposures. As argued in \citet{2014ApJS_210_4Munoz}, the $u^*$ band is essential when selecting UCDs due to its sensitivity to young/hot stellar populations, which allows for the removal of background star-forming galaixes. The left panel of Figure \ref{fig:ugz_based} shows the $u^*gz$ color-color diagram for the same spectroscopic sample (following the same color-coding) as in Figure \ref{fig:uiK_based}. In this case, we can see that the spaces occupied by background galaxies, Virgo members, and foreground stars more heavily overlap with each other. Nevertheless, the Virgo members still form a relatively tight sequence in this plane, such that we can select a sample of these objects with high completeness, albeit with more contamination. We adopt a color selection (represented by the polygon) within this diagram of the form:
  \begin{equation}
    \left\{
    \begin{aligned}
      & 0.850 \le (u^*-g)_0 \le 1.790; \\
      & 0.620 \le (g-z)_0 \le 1.500; \\
      & (g-z)_0 \le  0.190 + 0.770 \times (u^*-g)_0; \\
      & (g-z)_0 \ge -0.300 + 0.930 \times (u^*-g)_0.
   \end{aligned}
   \right.
   \label{eq:u^*gz}
  \end{equation}

The objects that satisfy these cuts are plotted in the plane of color versus surface brightness in the right-hand panel of Figure \ref{fig:ugz_based}. Although many contaminants pass our color-color cuts, it is possible to eliminate most of these following the same cuts on surface brightness that we applied to our $u^*giK_s$ sample (Equation \ref{eq:mug}; dotted polygon in this panel). These cuts are not as effective as before, however, and leave behind a number of background galaxies (red dots). Fortunately, we can remove most of these residual contaminants wherever we have $K$-band data. Figure \ref{fig:gzk} shows the $gzK$ color-color diagram for those objects from our spectroscopic catalog that satisfy our $u^*gz$ and surface brightness cuts. For a given $(g-z)_0$ color, background galaxies tend to be redder in $(z-K)_0$ than Virgo members, and we use the following relationship to isolate the latter:
  \begin{equation} 
      (z-K)_0 \le 1.480 + 0.780 \times (g-z)_0.
   \label{eq:gzk}
  \end{equation}
After this, we apply the same size cuts as before (Equation \ref{eq:rh}) to arrive at our sample of UCD candidates.

\subsection{Comparing our Selection Methods}
\label{subsec:compare_selections}

In this section, we compare the UCD catalogs derived from the two selection methods described above. For this comparison, the best region within the NGVS footprint is the center of sub-cluster A, where we have $u^*$, $g$, $i$, $z$, $K_{s}$ (NGVS-IR), and $K$ (UKIDSS) band fluxes and radial velocities for $\sim$2000 objects. Among this sample, 71 are spectroscopically-confirmed UCDs that were already known from previous work (mainly \citealt{2011ApJS_197_33Strader} and \citealt{2015ApJ_802_30Zhang}).

{\centering
\begin{deluxetable*}{l|rrrr|rrrr|rrrr}
 \tablewidth{0pt}
 \tabletypesize{\footnotesize}
 \tablecaption{Application of our Photometric UCD Selection Methods ($u^*giK_{s}$, $u^*giz$, $u^*gizK$) to a Spectroscopic Training Set.
 \label{tab:compare}}
 \tablehead{
 \colhead{ } &
 \multicolumn{4}{c}{$u^*giK_{s}$} &
 \multicolumn{4}{c}{$u^*giz$} &
 \multicolumn{4}{c}{$u^*gizK$}}
 \startdata
                      velocity &  $v < 0$ & \multicolumn{2}{c}{$0 < v < 3500$} & $v > 3500$ &  $v < 0$ & \multicolumn{2}{c}{$0 < v < 3500$} & $v > 3500$ &  $v < 0$ & \multicolumn{2}{c}{$0 < v < 3500$} & $v > 3500$   \\
                  Obj. Type    & Stars            & UCDs    & dE,Ns                    & BGs                & Stars            & UCDs     & dE,Ns                    & BGs                & Stars            & UCDs     & dE,Ns                    & BGs                \\
      $g_0<21.5$ \& $e<0.3$ &  183             &   71    &   17                     &  841               &  183             &   71     &   17                     &  841               &  183             &   71     &   17                     &  841               \\
           Color-Color Diagram &   21             &   71    &    9                     &   14               &   55             &   69     &   10                     &   91               &   55             &   69     &   10                     &   91               \\
     $\langle{\mu}_g\rangle_e$ &    8             &   69    &    6                     &    1               &   10             &   68     &    7                     &   17               &   10             &   68     &    7                     &   17               \\
                         $\langle r_h \rangle$ &    2             &   67    &    4                     &    0               &    2             &   66     &    5                     &   14               &    2             &   66     &    5                     &   14               \\
            $gzK$ & \nodata          & \nodata & \nodata                  & \nodata            & \nodata          & \nodata  & \nodata                  & \nodata            &    2$^{a}$       & 66$^{b}$ &    5$^{c}$               &    1               \\
\enddata
\tablenotetext{$a$}{~These two ``stars" lie in the NGVS-1+1 field, where many Virgo members have negative radial velocities. We consider these two objects as UCDs, in which case all three of our selection methods successfully cull the stars from our training set.}
\tablenotetext{$b$}{~Two of these objects are included in our final UCD catalog, while the remaining three have half-light radii of 0.0, 10.1 and 10.8 pc.}
\tablenotetext{$c$}{~These 5 objects are included in the Virgo Cluster Catalogue \citep[VCC;][]{1985AJ_90_1681Binggeli}.}
\end{deluxetable*}
}

As described in {\S}\ref{sec:ugik_selection}, we divide these objects into three groups according to their radial velocities: background galaxies ($v > 3500$ km/s), Virgo members ($0 < v < 3500$ km/s) and foreground stars ($v < 0$ km/s). Within the ``Virgo" group, we only consider two sub-classes of objects: UCDs and galaxies classified in previous or contemporaneous studies as nucleated dwarf elliptical galaxies, which have identifiable point sources at their centers \citep{1985AJ_90_1681Binggeli, 2019ApJ_878_18Sanchez-Janssen, 2020ApJ_890_128Ferrarese}. Diffuse, non-nucleated dwarfs are explicitly excluded in our selection pipeline.

As listed in Table \ref{tab:compare}, we have 183 stars, 71 UCDs, 17 dE,Ns, and 841 background galaxies (BGs) which satisfy our magnitude and ellipticity criteria (i.e., 14.0 $< g_0 <$ 21.5 mag and $e < 0.3$). Following this, we successively apply our color, surface brightness, and size cuts, with the choice of colors being the only variable. Table \ref{tab:compare} presents the number of objects from each group that survive each step of our selection functions.

For the $u^*giK_s$-based selection, the color and surface brightness cuts are quite effective at eliminating most of the contaminants. In the end, we find 2 stars ($\sim 1\%$), 67 UCDs ($\sim 94\%$), 4 dE,Ns ($\sim 24\%$) and zero BGs satisfying all of the criteria\footnote{These percentages represent the fraction of objects, within each group, with $g_0<21.5$ mag and $e<0.3$ that survive all of the selection criteria.}. For comparison, we find that 2 stars ($\sim 1\%$), 66 UCDs ($\sim 93\%$), 5 dE,N's ($\sim 29\%$) and 14 BGs ($\sim 2\%$) pass the cuts in our $u^*giz$-based selection. Thus, both methods achieve a completeness of $>90\%$ with respect to UCD selection. Conversely, the former method is much better than the latter in culling BGs from the sample. With the addition of $K$-band data though, we can reduce the number of BGs that pass our $u^*giK$-based selection from 14 to 1, which compares very favourably with the results of our $u^*giK_s$-based selection.

As for the other contaminants, we note that all three of our selection methods pass two objects from our training set that are labelled as foreground stars. In fact, these two objects are really UCDs that have negative radial velocities through their association with the M86 sub-group. This shows that the combination of optical and near-infrared photometry removes most foreground and background objects. Where this photometric combination falls short is in rejecting dE,Ns, but we can easily accomplish this through visual inspection since nuclei are surrounded by obvious envelopes.

To summarize, our tests in the central 4 deg$^2$ of sub-cluster A show that a $u^*gizK$-based selection of UCDs can be a reliable alternative to that based on $u^*giK_{s}$. Hereafter, we mainly rely on our $u^*gizK$ selection method to detect UCD candidates wherever $K$-band data is available, and $u^*giz$ otherwise.

\subsection{Data Inputs for the Full Catalog}

 {\centering
\begin{deluxetable*}{lrrrrrr}
 \tablewidth{0pt}
 \tabletypesize{\small}
 \tablecaption{Summary of Multi-Band Datasets Over the NGVS Footprint and the Corresponding UCD Selection Methods.
 \label{tab:bands_selection}}
 \tablehead{
  \colhead{Bands} &
  \colhead{$f_{\rm Objs}$} &
  \colhead{Selection Method} &
  \colhead{$N_{\rm UCDs}$} & 
  \colhead{$f_{\rm UCDs}$} & 
  \colhead{$N_{\rm UCDs, class=1}$} & 
  \colhead{$f_{\rm UCDs, class=1}$} \\ \vspace{-0.5cm}\\
  \colhead{(1)} &
  \colhead{(2)} &
  \colhead{(3)} &
  \colhead{(4)} &
  \colhead{(5)} &
  \colhead{(6)} &
  \colhead{(7)}} 
 \startdata
 $u^*gizK$  & $ 59.0\%$ & $u^*gzK+\langle{\mu_g}\rangle_e+\langle{r_h}\rangle$ & $ 286$ & $ 34.5\%$ & $ 235$ & $ 38.4\%$ \\
 $u^*giz$   & $ 29.3\%$ & $u^*gz+\langle{\mu_g}\rangle_e+\langle{r_h}\rangle$  & $ 465$ & $ 56.2\%$ & $ 363$ & $ 59.3\%$ \\
 $u^*gzK$   & $  2.4\%$ & $u^*gzK+\langle{\mu_g}\rangle_e+r_{h,g}$             & $   4$ & $  0.5\%$ & $   0$ & $  0.0\%$ \\
 $u^*gz$    & $  0.2\%$ & $u^*gz+\langle{\mu_g}\rangle_e+r_{h,g}$              & $   1$ & $  0.1\%$ & $   0$ & $  0.0\%$ \\
 $gizK$     & $  3.8\%$ & $gzK+\langle{\mu_g}\rangle_e+\langle{r_h}\rangle$    & $   1$ & $  0.1\%$ & $   0$ & $  0.0\%$ \\
 $giz$      & $  0.1\%$ & $gz+\langle{\mu_g}\rangle_e+\langle{r_h}\rangle$     & $   0$ & $  0.0\%$ & $   0$ & $  0.0\%$ \\
 $gzK$      & $  4.3\%$ & $gzK+\langle{\mu_g}\rangle_e+r_{h,g}$                & $   9$ & $  1.1\%$ & $   0$ & $  0.0\%$ \\
 $gz$       & $  0.9\%$ & $gz+\langle{\mu_g}\rangle_e+r_{h,g}$                 & $  13$ & $  1.6\%$ & $   0$ & $  0.0\%$ \\
 \vspace{-0.35cm} \\ 
 $gi$+spec  & \nodata                                 & $(v_r<3500~{\rm km/s})+\langle{r_h}\rangle$          & $  49$ & $  5.9\%$ & $  14$ & $  2.3\%$ \\
 \vspace{-0.35cm} \\  
 Total      & $100\%$ & \nodata & $ 828$ & $ 100\%$ & $ 612$ & $ 100\%$ \\
 \enddata
 \tablenotetext{}{NOTE-- (1) the available filter combinations; (2) the fraction of the parent sample covered by each filter combination; (3) the UCD selection method employed; (4) the number of UCD candidates selected; (5) the fraction of all UCD candidates selected; (6) the number of UCDs which survive visual inspection; (7) the fraction of all classified UCDs.}
\end{deluxetable*}
}

Based on the material presented to this point, the selection of UCDs would ideally rely on data in the $u^*gzK$ ~bands for color and surface brightness cuts, and in the $g$- and $i$-bands for size cuts. However, as shown in Figure \ref{fig:map_selection}, we do not possess imaging in the $u^*$, $i$ and $K$ ~bands over the full NGVS footprint. In Table \ref{tab:bands_selection}, we list the available combinations of photometric data and the corresponding selection method applied in each case. Approximately $90\%$ of bright objects ($g_0<21.5$ mag) detected in the NGVS have data in either the $u^*gizK$ or $u^*giz$ bands; all other objects are covered by some subsample of the five available bands.

Table~\ref{tab:bands_selection} also shows the number ($N_{\rm UCDs}$) and the fraction ($f_{\rm UCDs}$) of UCDs selected by each method. Most of our UCD candidates ($\sim$91\%) are selected through either the $u^*gizK$- or $u^*giz$-based methods, which were discussed in previous sections. It is noteworthy that only $\sim$37\% (286) of these candidates are selected from $\sim$60\% of our catalog that has $u^*gizK$ data. Proportionally, we expect to find $\sim$145 UCD candidates from $\sim$30\% of the catalog. However, 465 candidates ($\sim$60\%) are selected from the $\sim$30\% of the catalog that do not have $K$ band data. As described above, $K$ band data is efficient to eliminate BGs. This suggests that the portion of our UCD sample selected via the $u^*giz$-based method is inflated through contamination by BGs. Note that the other selection methods we have employed contribute just 28 objects to our UCD sample (or $\sim$3\%). In all, we have identified 779 UCD candidates based on photometry alone.

Finally we note that if only $g$- and $z$-band data are available, we select UCD candidates using the following color cut:
  \begin{equation}
      0.620 \le (g-z)_0 \le 1.500.
   \label{eq:gz}
  \end{equation}
Also, if $i$-band half-light radii are not available, we apply the following cuts to the corresponding $g$-band value:
   \begin{equation}
    \left\{
    \begin{aligned}
      & 11 < r_{h,g} < 100 \rm ~pc; \\
      & \frac{r_{h,g,\rm error}}{r_{h,g}} \le 15\%.
   \end{aligned}
   \right.
   \label{eq:rhg}
  \end{equation}

\subsection{Spectroscopically-Selected Virgo Members}
\label{subsec:spec_ucds}

\begin{figure}
\epsscale{1.15}
\plotone{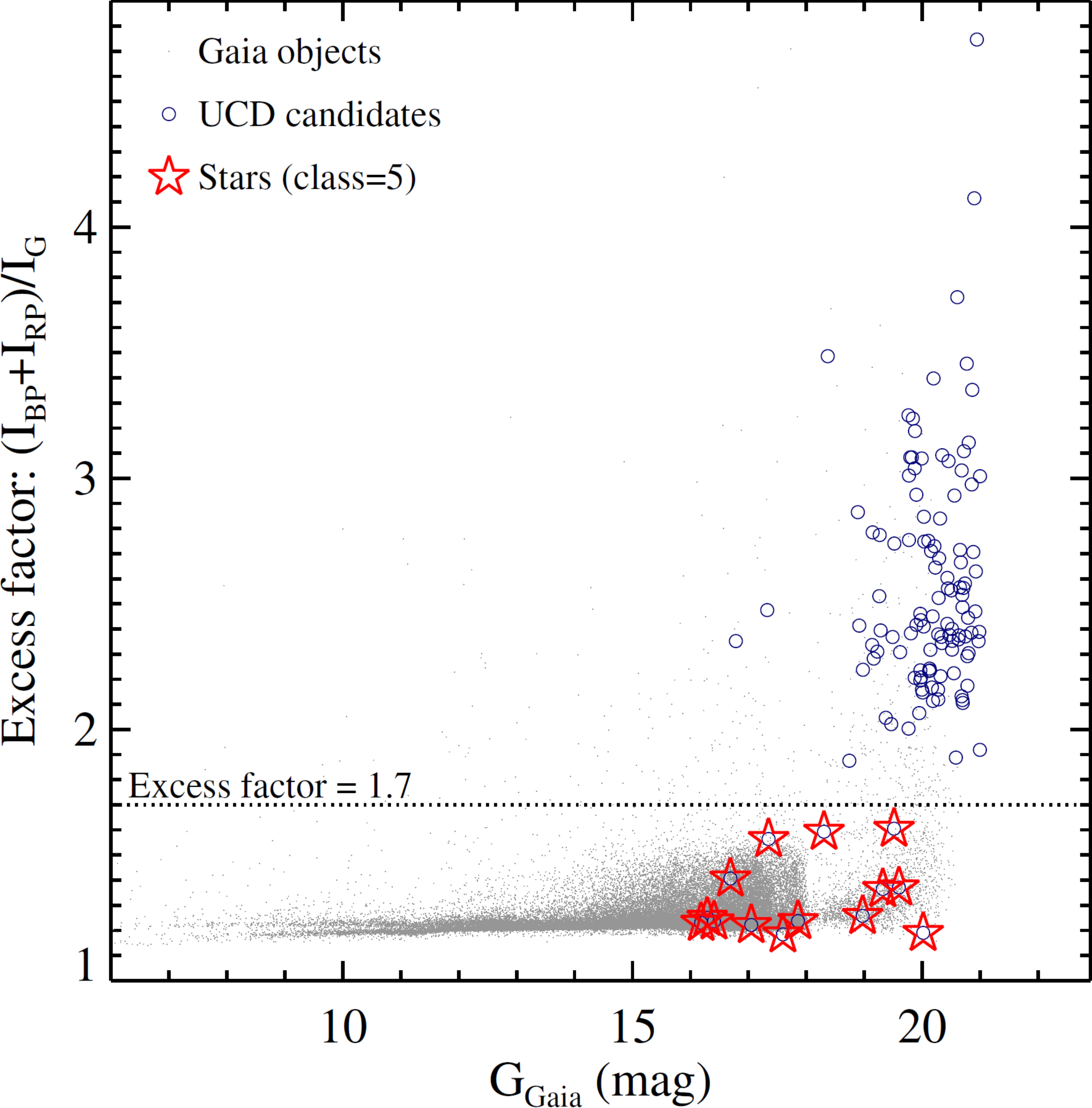}
\caption{Gaia BP/RP flux excess factor -- $(I_{BP}+I_{RP})/I_G$ -- as a function of $G$-band magnitude. Gray dots are Gaia sources in the NGVS footprint. Blue circles are UCD candidates which have a Gaia counterpart. Red stars are objects classified as stars ({\tt class = 5}) in this study. The black dotted line shows an excess factor of 1.7, a division that is found to separate stellar and extended sources.} 
\label{fig:gaia}
\end{figure}

\begin{figure*}
\epsscale{1.17}
\plotone{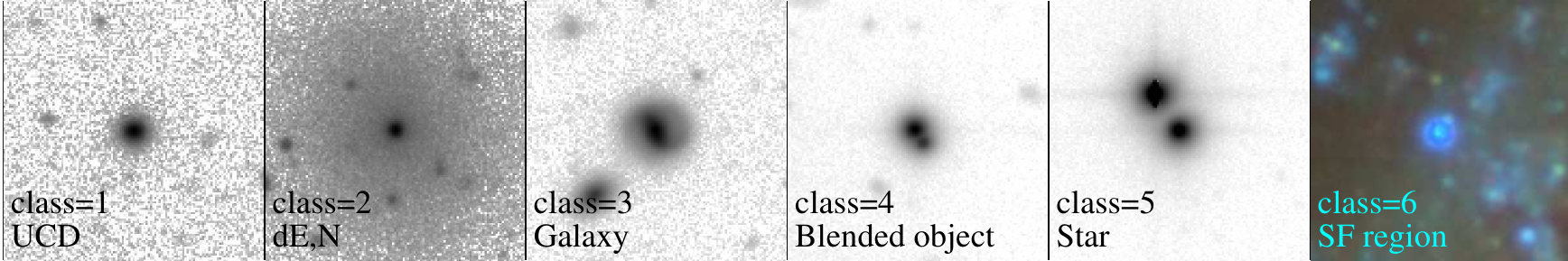}
\caption{UCD candidates are divided into six classes by visual inspection. From left to right: UCD -- ${\rm class}=1$; nucleated dwarf elliptical galaxy (dE,N) -- ${\rm class}=2$; background galaxy -- ${\rm class}=3$; blended object -- ${\rm class}=4$; star -- ${\rm class}=5$; star-forming region -- ${\rm class}=6$. The image size in each panel is $120\times120$ pixels, where $120$ pixels $\sim22.32$\arcsec~$\sim1.8$ kpc at Virgo distance.} 
\label{fig:classification}
\end{figure*}

In Figures \ref{fig:ugz_based} and \ref{fig:gzk}, we see that a few confirmed Virgo members fall outside our color selection windows. Some of these objects lie close to the window boundaries though and therefore could be UCD candidates. The purpose of our color selection criteria are, first and foremost, to reduce or eliminate contamination from background galaxies. For these few cases, if we know from their radial velocities that they are not background objects, then we do not require color criteria. We assume that the Virgo members are UCDs if they satisfy our size cuts (Equation \ref{eq:rh}, or \ref{eq:rhg} if $i$-band data are not available). This assumption is reasonable as most objects in the literature in this size range are UCDs. There are $\sim$8000 Virgo objects with $v<3500$ km/s and do not satisfy our color or/and surface brightness cuts. Among these objects, we find an additional 49 UCD candidates through this ``perturbation" method and henceforth refer to them as ``spectroscopically-selected UCD". 

In Table \ref{tab:compare}, we recover 66 of the 71 known UCDs using either our $u^*giz$- or $u^*gizK$-based methods (the remaining five do not meet all of our selection criteria). In reality, two of these five ``missing" UCDs are included in the spectroscopically-selected UCD sample. The three remaining UCDs have $\langle{r_h}\rangle$ = 0.0 (due to the bad image quality in this region), 10.1 and 10.8 pc individually: i.e., two of the three are slightly smaller than our size criterion.

\subsection{Visual Inspection}
\label{sec:visual_inspection}

As explained above, we have objectively selected 828 UCD candidates: 779 and 49 based on photometric and spectroscopic information, respectively. As a final hedge against contamination, we execute one last step --- visual inspection of the NGVS images, whereby we classify each candidate as either a: (1) = {\tt probable UCD}; (2) = {\tt dwarf nucleus}; (3) = {\tt background galaxy}; (4) = {\tt blended object}; (5) = {\tt star}; (6) = {\tt star-forming region}.

Some contaminants can be difficult to identify from visual inspection alone and, in such cases, we classify objects using additional information. For example, a nucleated dwarf galaxy ({\tt class = 2}) usually contains a nucleus at its photocenter (although some are slightly off-centered) and a stellar halo component surrounding it. Some UCDs are also embedded in low-surface brightness envelopes \citep[e.g.,][]{2005ApJ_627_203Hacsegan} making it difficult to distinguish dE,Ns from UCDs with envelopes. We therefore classify objects as dE,Ns only when they appear in either the VCC \citep{1985AJ_90_1681Binggeli} or NGVS galaxy catalogs \citep{2020ApJ_890_128Ferrarese}. Background galaxies are identified with the help of redshift measurements, and all objects with $v_r > 3500$ km/s are immediately classified as such ({\tt class = 3}). Blended objects ({\tt class = 4}) are relatively easy to identify, although they may not be separable in NGVS images if they are too close. We make use of Gaia DR2 data \citep{2018A+A_616_3Riello} to help further separate UCDs and stars ({\tt class = 5}), in that stars can have significant proper motions (i.e., $3\sigma$ significance) and UCDs can be resolved in Gaia imaging \citep[see][]{2001.02243}. Figure~\ref{fig:gaia} shows the $BP/RP$ flux excess factor, which is defined as flux ratio $(I_{BP}+I_{RP})/I_G$, as a function of Gaia $G$ band magnitude. Briefly, the fluxes in $BP$ and $RP$ bands ($I_{BP}$ and $I_{RP}$) are measured in a window of $3.5 \times 2.1$ arcsec$^2$ while the flux in $G$ band ($I_G$) is derived from PSF fitting. For point sources, the flux excess factor should be $\gtrsim$1 (the $BP$ and $RP$ filters overlap at around 6500\AA~and are broader than the $G$ band, especially at the red side, see \citealt{2018A+A_616_4Evans}). As shown in Figure~\ref{fig:gaia}, most of Gaia targets have small flux excess factors while the excess factor of extended objects can be much larger: e.g., most of our UCD candidates (blue circles) in Figure~\ref{fig:gaia} have excess factors of 2--4. We draw a line at an excess factor of 1.7 and classify sources below this level as stars. Finally, the star-forming regions ({\tt class = 6}) are also easy to identify due to their blue colors.

In summary, among our 828 UCD candidates, we find 612 probable UCDs (598 identified on the basis of photometry alone), 41 nucleated dwarf galaxies, 132 background galaxies, 12 blended objects, 14 stars and 17 star-forming regions. Representative images for these six types of objects are shown in Figure \ref{fig:classification}. As can be seen in the final two columns of Table~\ref{tab:bands_selection}, 235 of the photometrically identified UCDs are $u^*gizK$-selected while the remaining 363 are $u^*giz$-selected. The UCD candidates initially selected based on other methods (i.e., $u^*gzK$, $u^*gz$, $gizK$, $giz$, $gzK$, and $gz$) are classified as contaminants after visual inspection --- one of the reasons that we only test the $u^*giz$- and $u^*gizK$- selection methods in {\S}\ref{subsec:compare_selections}. 

\subsection{Catalog of UCD Candidates}

Tables~\ref{tab:ucd_photometry} and \ref{tab:ucd_redshift} present observed and derived parameters for all of our UCD candidates\footnote{See https://gax.sjtu.edu.cn/data/UCDs.html}. Since visual inspection is inevitably subjective, we list all 828 objectively-selected candidates in these tables. We plan to measure radial velocities in future spectroscopic campaigns and assess their Virgo membership directly. Indeed, some of these ``contaminants" (such as apparent star-forming regions with UCD-like sizes) are interesting in their own right and will be investigated in future papers.

\subsection{Summary of UCD Selection Function}

\begin{figure*}
\epsscale{1.03}
\plotone{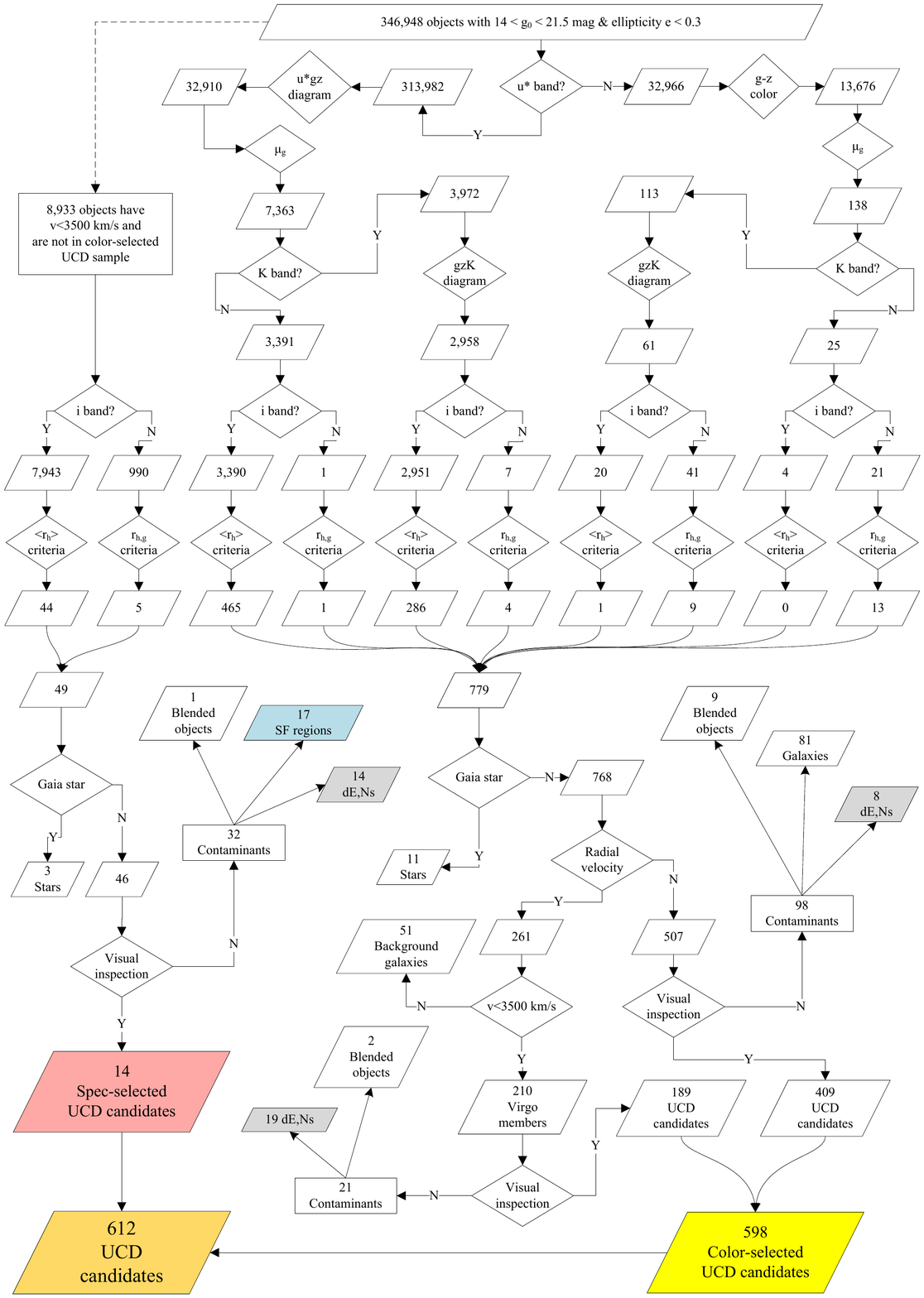}
\caption{The ``Pachinko Machine" plot illustrating the UCD selection methods used in this study. Our UCD selection is based on the combination of magnitude ($14.0<g_0<21.5$ mag), ellipticity ($e\equiv (1-b/a) < 0.3$), color-color diagram, surface brightness and half-light radius cuts and visual inspection. In total, a sample of 612 UCDs were identified in NGVS imaging.}
\label{fig:flowchart}
\end{figure*}

Given the complexity of our methodology, Figure~\ref{fig:flowchart} summarizes our UCD selection function in the form of a flowchart. An inspection of this figure shows that our selection process involves two main channels: one based on photometry (right) and the other on spectroscopy (left), which we now describe.

We begin with the photometric algorithm. There are 346,948 objects in the NGVS that satisfy our magnitude ($14.0 < g_0 < 21.5$ mag) and ellipticity ($e < 0.3$) cuts. All of these objects have $g$- and $z$-band data, while $u^*$-, $i$- and $K$-data are more limited. Note that only objects detected in our short exposures lack $u^*$- and $i$-band data (see Figure~\ref{fig:map_selection}), and so are drawn from the bright end of the UCD luminosity function. We then apply successive cuts based on each object's color (or colors), surface brightness (Equation~\ref{eq:mug}), position in the $gzK$ diagram (Equation~\ref{eq:gzk}; if $K$-band data available), and half-light radius (Equation~\ref{eq:rh} if $i$-band data available; otherwise Equation~\ref{eq:rhg}). Our color cuts are based on the $u^*gz$ diagram (Equation \ref{eq:u^*gz}) for those objects having $u^*$-band data; otherwise, we consider the $(g-z)$ color alone (Equation~\ref{eq:gz}).

These selection criteria dramatically reduce our original photometric sample from 346,948 objects to 779 candidate UCDs. With the assist of Gaia (see Figure~\ref{fig:gaia}), 11 of these candidates are re-classified as stars. To further reduce the number of contaminants, we place a cut on radial velocities (when available) and visually inspect the candidates (see Figure~\ref{fig:classification}). Another fifty-one candidates are removed for having a radial velocity in excess of $v = 3500$ km/s and 119 more are removed following visual inspection. The latter group is comprised of 27 dE,Ns, 81 background galaxies and 11 blended objects. Our final photometric UCD sample contains 598 UCD candidates.

We now describe the spectroscopic selection function depicted in Figure~\ref{fig:flowchart}. As stated earlier, this path is intended to select UCD candidates that fall outside our color and/or surface brightness selection windows. Using this approach, we find 49 more candidates which meet our radius cuts, while visual inspection shows that they are comprised of 14 probable UCDs, 17 star-forming regions, 14 dE,Ns, 3 stars and 1 blended object.

Overall, we have identified 612 UCD candidates, the majority of which (598) come from our photometric selection function. This constitutes the largest sample of UCD candidates identified to date. While we would prefer a simple and homogeneous selection, the lack of data in the $u^*$, $i$, and $K$ bands over the full NGVS footprint necessitates certain compromises. That said, the great majority (598/612 $\simeq$ 97.7\%) of these UCD candidates have been selected based on their $u^*giz$ or $u^*gizK$ colors.

We refer to the full group of candidates as the ``{\tt all UCD sample}". Within this group, 235 candidates have been selected on the basis of their $u^*gizK$ data, so we refer to this as the ``{\tt $u^*gizK$ UCD sample}". We also have a ``{\tt spec-UCD sample}", which contains the 203 candidates that have been spectroscopically confirmed as members of the Virgo cluster. The ``{\tt all UCD sample}" has the highest completeness but the largest number of contaminants. Conversely, the ``{\tt spec-UCD sample}" has the fewest contaminants but is inevitably biased by the choice of spectroscopic targets. Finally, the ``{\tt $u^*gizK$ UCD sample}" strikes the greatest balance between completeness, contamination, and homogeneity (see Table~\ref{tab:compare}).

\subsection{UCDs from Previous Studies}

\setcounter{table}{4}
{\centering
\begin{deluxetable*}{lrrrrrrll}
 \tablewidth{0pt}
 \tabletypesize{\tiny}
 \tablecaption{The properties of UCDs which are in previous studies but not included in this study.
 \label{tab:ucd_previous}}
 \tablehead{
  \colhead{Name} &
  \colhead{$\alpha_{\rm J2000}$} &
  \colhead{$\delta_{\rm J2000}$} &
  \colhead{$e$} &
  \colhead{$g_0$} &
  \colhead{$\langle r_h \rangle$} &
  \colhead{$v_r$} &
  \colhead{$v_{\rm source}$} &
  \colhead{Note} \\ \vspace{-0.6cm}\\
  \colhead{ } &
  \colhead{(deg)} &
  \colhead{(deg)} &
  \colhead{ } &
  \colhead{(mag))} &
  \colhead{(pc))} &
  \colhead{(km/s)} &
  \colhead{ } &
  \colhead{ }  \\ \vspace{-0.6cm}\\
  \colhead{(1)} &
  \colhead{(2)} &
  \colhead{(3)} &
  \colhead{(4)} &
  \colhead{(5)} &
  \colhead{(6)} &
  \colhead{(7)} &
  \colhead{(8)} &
  \colhead{(9)} }
 \startdata
 S1508                    & 187.6308727 &  12.4235661 &      0.07 &    22.018 &   \nodata &      2419 &S11                 &                                         $g_0>21.5~$mag \\
 S825                     & 187.7126346 &  12.3554004 &      0.16 &    21.613 &   \nodata &      1142 &S11                 &                                         $g_0>21.5~$mag \\
 S723                     & 187.7239811 &  12.3393933 &      0.07 &    21.737 &   \nodata &      1398 &NED,S11             &                                         $g_0>21.5~$mag \\
 H44905                   & 187.7378222 &  12.3944049 &      0.20 &    21.921 &   \nodata &      1564 &NED,S11             &                                         $g_0>21.5~$mag \\
 H30401                   & 187.8279366 &  12.2624564 &      0.18 &    21.593 &   \nodata &      1323 &NED,S11             &                                         $g_0>21.5~$mag \\
 T15886                   & 188.1520299 &  12.3491986 &      0.23 &    22.974 &   \nodata &      1349 &NED,S11             &                                         $g_0>21.5~$mag \\
 S991                     & 187.6937514 &  12.3382605 &      0.34 &    20.701 &     16.67 &      1004 &S11                 &                                                $e>0.3$ \\
 S672                     & 187.7280306 &  12.3606434 &      0.31 &    20.739 &     15.50 &       735 &S11                 &                                                $e>0.3$ \\
 NGVS-J124002.08+105517.2 & 190.0086686 &  10.9214418 &      0.50 &    21.338 &      3.48 &   \nodata &             \nodata &                                                $e>0.3$ \\
 NGVS-J122926.23+081658.8 & 187.3593098 &   8.2829991 &      0.05 &    21.373 &      9.06 &       465 &MMT14               &                             $\langle r_h\rangle<11~$pc \\
 NGVS-J123009.17+074127.5 & 187.5381980 &   7.6909633 &      0.08 &    19.449 &     10.76 &   \nodata &             \nodata &                            $\langle r_h\rangle<11~$pc \\
 NGVS-J123015.56+083445.0 & 187.5648407 &   8.5791729 &      0.07 &    20.822 &     10.33 &      -129 &MMT14               &                             $\langle r_h\rangle<11~$pc \\
 H20718                   & 187.5818069 &  12.1568298 &      0.03 &    20.989 &     10.76 &       876 &MMT09,S11           &                             $\langle r_h\rangle<11~$pc \\
 M87UCD-31                & 187.7305441 &  12.4110907 &      0.02 &    20.860 &     10.09 &      1301 &MMT09               &                             $\langle r_h\rangle<11~$pc \\
 M87UCD-14                & 187.7681413 &  13.1784860 &      0.12 &    20.200 &     10.83 &      1345 &MMT09,AAT12         &                             $\langle r_h\rangle<11~$pc \\
 NGVS-J122849.25+075919.4 & 187.2051905 &   7.9887156 &      0.03 &    21.179 &     13.37 &   \nodata &             \nodata &               $|r_{h,g}-r_{h,i}|>(r_{h,g}+r_{h,i})/2$ \\
 NGVS-J123004.35+073932.2 & 187.5181421 &   7.6589449 &      0.09 &    21.437 &     11.19 &      1043 &MMT14               &                $|r_{h,g}-r_{h,i}|>(r_{h,g}+r_{h,i})/2$ \\
 S8005                    & 187.6925322 &  12.4064233 &      0.02 &    20.444 &     27.91 &      1883 &S11                 &                $|r_{h,g}-r_{h,i}|>(r_{h,g}+r_{h,i})/2$ \\
 S9053                    & 187.7013369 &  12.4946708 &      0.16 &    21.313 &     30.25 &       829 &S11,IMACS16         &                $|r_{h,g}-r_{h,i}|>(r_{h,g}+r_{h,i})/2$ \\
 S686                     & 187.7242079 &  12.4718872 &      0.05 &    20.499 &     16.31 &       817 &S11                 &                $|r_{h,g}-r_{h,i}|>(r_{h,g}+r_{h,i})/2$ \\
 S6003                    & 187.7922587 &  12.2744282 &      0.08 &    21.277 &     14.19 &      1818 &S11                 &                $|r_{h,g}-r_{h,i}|>(r_{h,g}+r_{h,i})/2$ \\
 NGVS-J122806.14+065909.0 & 187.0255686 &   6.9858243 &      0.19 &    20.200 &     27.66 &   \nodata &             \nodata &                         $(g-z)_0~\rm{VS.}~\mu_g~$diagram \\
 NGVS-J123949.21+110135.3 & 189.9550398 &  11.0264758 &      0.03 &    21.080 &     12.73 &   \nodata &             \nodata &                         $(g-z)_0~\rm{VS.}~\mu_g~$diagram \\
 NGVS-J124216.65+114428.6 & 190.5693756 &  11.7412681 &      0.06 &    21.273 &     12.44 &   \nodata &             \nodata &                         $(g-z)_0~\rm{VS.}~\mu_g~$diagram \\
 NGVS-J124244.72+111240.5 & 190.6863239 &  11.2112384 &      0.12 &    20.491 &     32.90 &   \nodata &             \nodata &                         $(g-z)_0~\rm{VS.}~\mu_g~$diagram \\
 NGVS-J122708.94+074228.3 & 186.7872607 &   7.7078639 &      0.08 &    19.340 &     47.81 &   \nodata &             \nodata &           $(g-z)_0~\rm{VS.}~(z-K_{\rm{UKIDSS}})_0~$diagram \\
 NGVS-J122828.60+070228.2 & 187.1191854 &   7.0411563 &      0.11 &    20.646 &     28.26 &   \nodata &             \nodata &           $(g-z)_0~\rm{VS.}~(z-K_{\rm{UKIDSS}})_0~$diagram \\
 S887                     & 187.7038900 &  12.3654400 &   \nodata &    21.190 &   \nodata &      1811 &S11                 &                                         Fail~to~detect \\
 H30772                   & 187.7419100 &  12.2672800 &   \nodata &    20.750 &   \nodata &      1225 &NED,S11             &                                         Fail~to~detect \\
\enddata
\tablenotetext{}{NOTE-- (1) UCD name; (2) RA; (3) Dec; (4) Ellipticity; (5) $g$ band magnitude; (6) Half-light radius; (7) Radial velocity; (8)The source of velocity measurement; (9) The reason why this UCD is not included in this study.}
\end{deluxetable*}
}

As nearest rich cluster of galaxies, many previous investigations have studied UCDs in virgo \citep[e.g.,][]{2005ApJ_627_203Hacsegan,2006AJ_131_312Jones,2007AJ_133_1722Evstigneeva,2011ApJS_197_33Strader,2015ApJ_812_34Liu,2015ApJ_802_30Zhang,2017ApJ_835_212Ko}. Our sample contains 186 UCDs that have already been reported, while 29 systems from previous work are excluded. We list the properties of each member of the latter group in Table \ref{tab:ucd_previous}, including name, right ascension, declination, ellipticity, magnitude, half-light radius, radial velocity, the sources of velocity measurements, and the primary reason why it is not included in our sample. Among these 29 UCDs, six are fainter than $g_0=21.5$ mag, three have high ellipticity ($e>0.3$), 12 do not satisfy the radius criteria, four are located just outside the selection window in $(g-z)_0~\rm{vs.}~\mu_g$ diagram, two are classified as background galaxies by $gzK$ diagram, and two escaped detection in the NGVS images because they are projected close to saturated stars. We also note that 9 of these 29 UCDs lack radial velocity measurements.

\section{Results}
\label{sec:results}

To understand the nature and origin of UCDs --- and the extent to which they differ from other compact stellar systems --- it is important to first understand how their properties compare to those of GCs and nuclei, where evolutionary links to UCDs may exist. To this end, we now describe various samples of each class of compact stellar system in Virgo available to us.

\begin{description}
\item[Globular Clusters]
\end{description}
\begin{itemize}
\item {\tt ACSVCS GCs}: The sample of GCs from the HST ACS Virgo Cluster Survey. The objects with $p_{gc}>0.95$ are included in this sample, where $p_{gc}$ represents the probability that an object is a GC \citep{2006ApJ_639_95Peng,2009ApJS_180_54Jordan}.
\item {\tt Bright GCs}: A magnitude-limited ($g_0<21.5$ mag) sample of NGVS objects that satisfy our UCD selection criteria but have $\langle{r_h}\rangle <11$ pc. This sample includes GC candidates without velocity measurements and GCs with $v_r<3500$ km/s. Background galaxies with $v_r>3500$ km/s are removed from the sample.
\item {\tt Spec GCs}: The subset of {\tt bright GCs} with radial velocities $v_r<3500$ km/s. 
\end{itemize}

\begin{description}
\item[Ultra Compact Dwarfs]
\end{description}
\begin{itemize}
\item {\tt All UCDs}: Our full sample of 612 UCD candidates, which satisfy all of our selection criteria and pass our visual inspection. Amongst this sample, 203 ($\sim$1/3) have $v_r<3500$ km/s; all others lack radial velocity information. 
\item {\tt $u^*gizK$ UCDs}: The subset of 235 UCDs selected from $u^*gizK$ data, a fraction of which have velocity measurements.
\item {\tt Spec UCDs}: The subset of 203 UCDs with $v_r<3500$ km/s.
\end{itemize}

\begin{description}
\item[Stellar Nuclei]
\end{description}
\begin{itemize}
\item {\tt All nuclei}: The entire sample of 551 nuclei from the NGVS galaxy catalog \citep[][]{2019ApJ_878_18Sanchez-Janssen, 2020ApJ_890_128Ferrarese}.
\item {\tt Bright nuclei}: The subset of 339 bright ($g_0<21.5$ mag) nuclei from the NGVS galaxy catalog.
\item {\tt $u^*gizK$ nuclei}: The sample of 41 nuclei that satisfy our UCD selection criteria and are classified as dE,Ns by visual inspection (i.e., {\tt class=2}).
\end{itemize}

\begin{figure*}
\epsscale{0.55}
\plotone{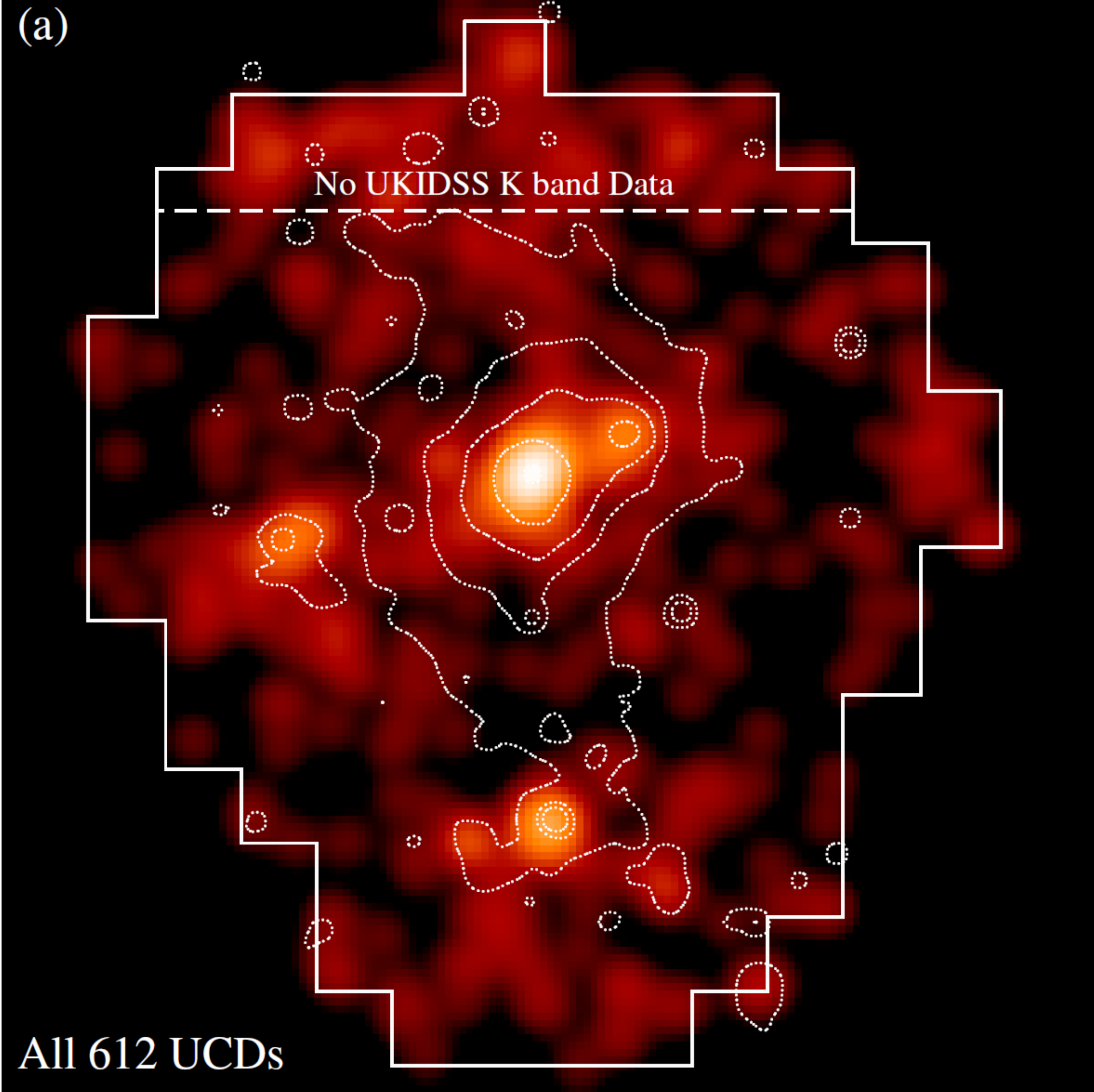}
\epsscale{0.55}
\plotone{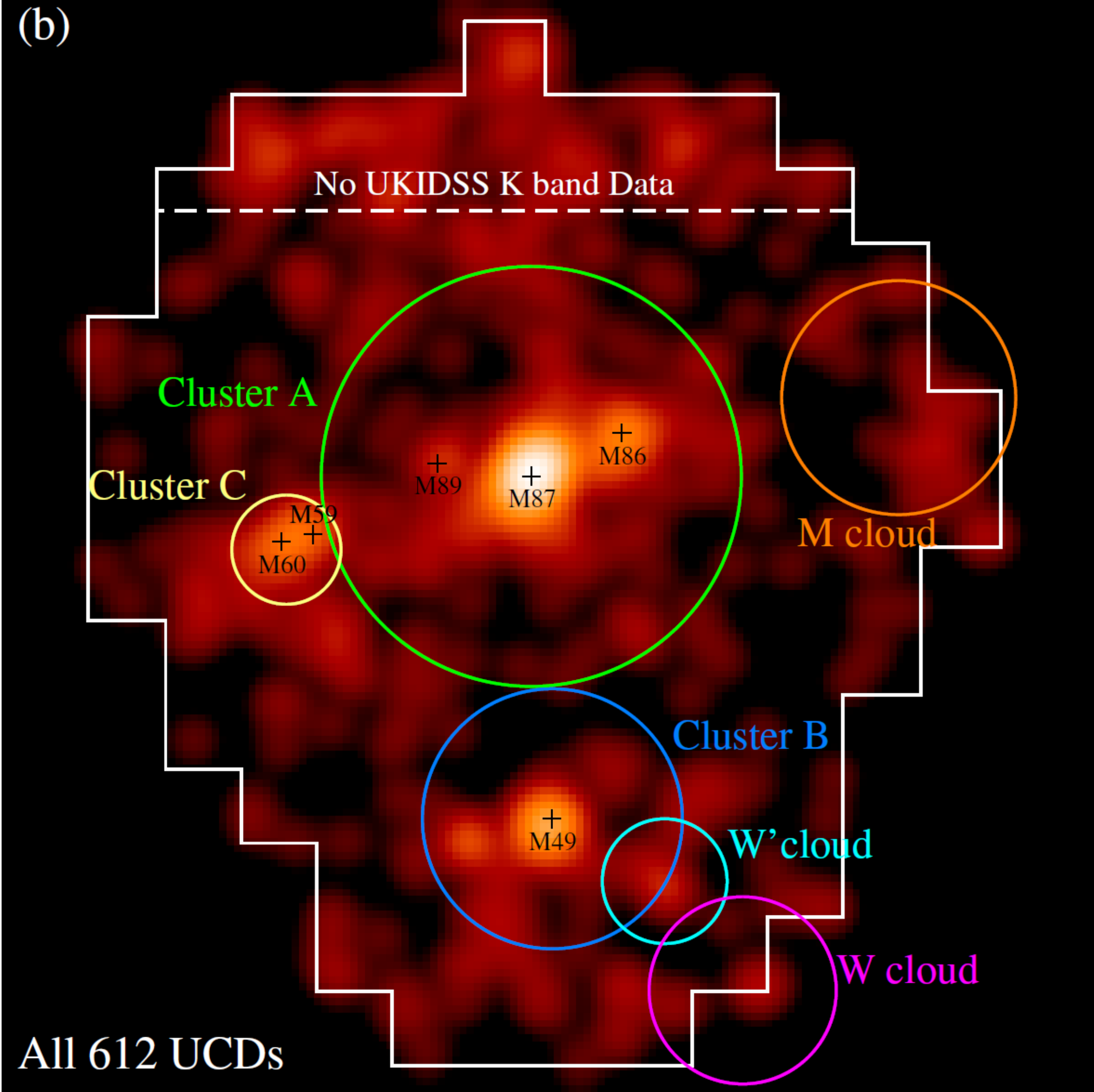}\\
\vspace{0.05cm}
\epsscale{0.55}
\plotone{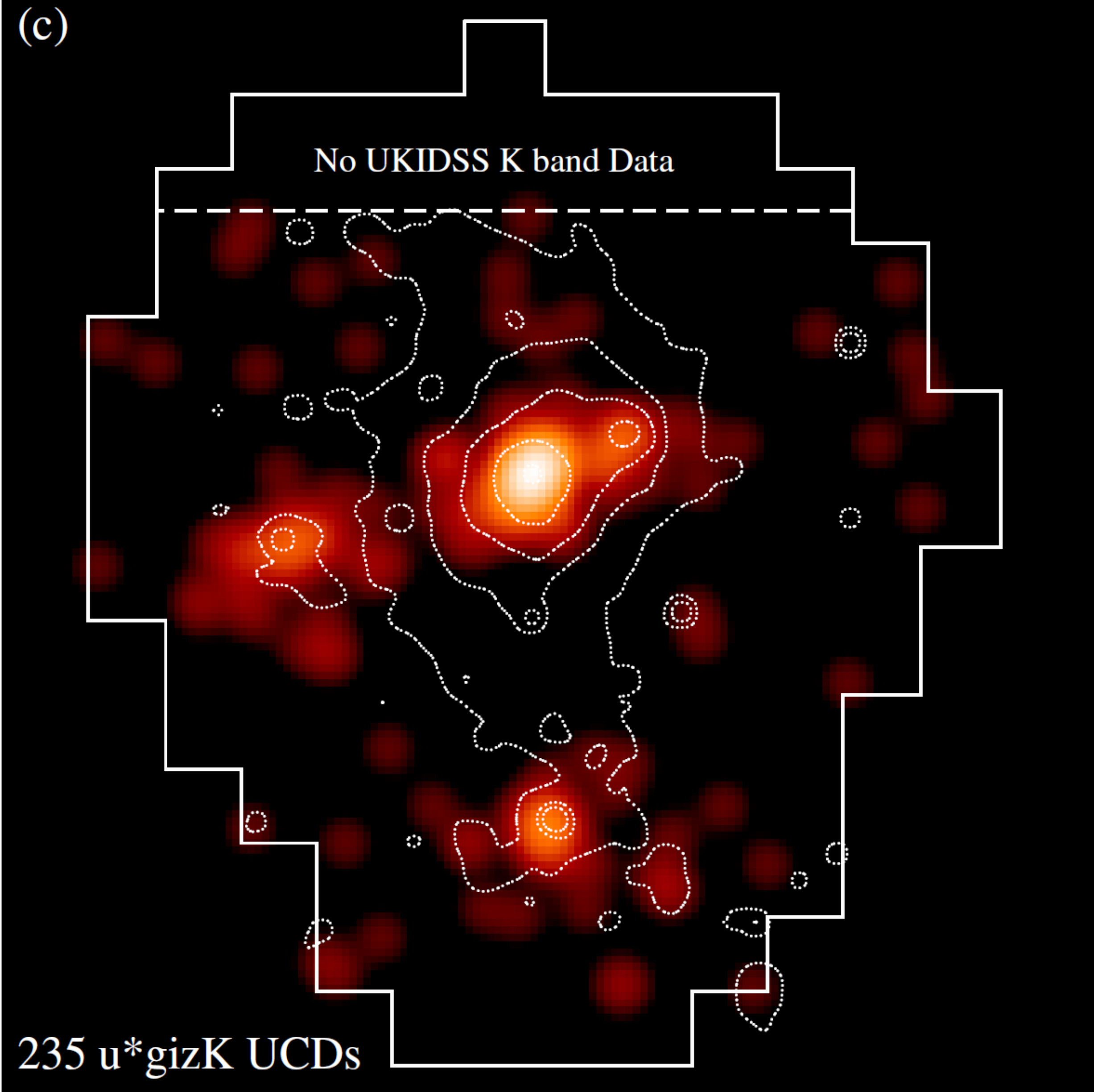}
\epsscale{0.55}
\plotone{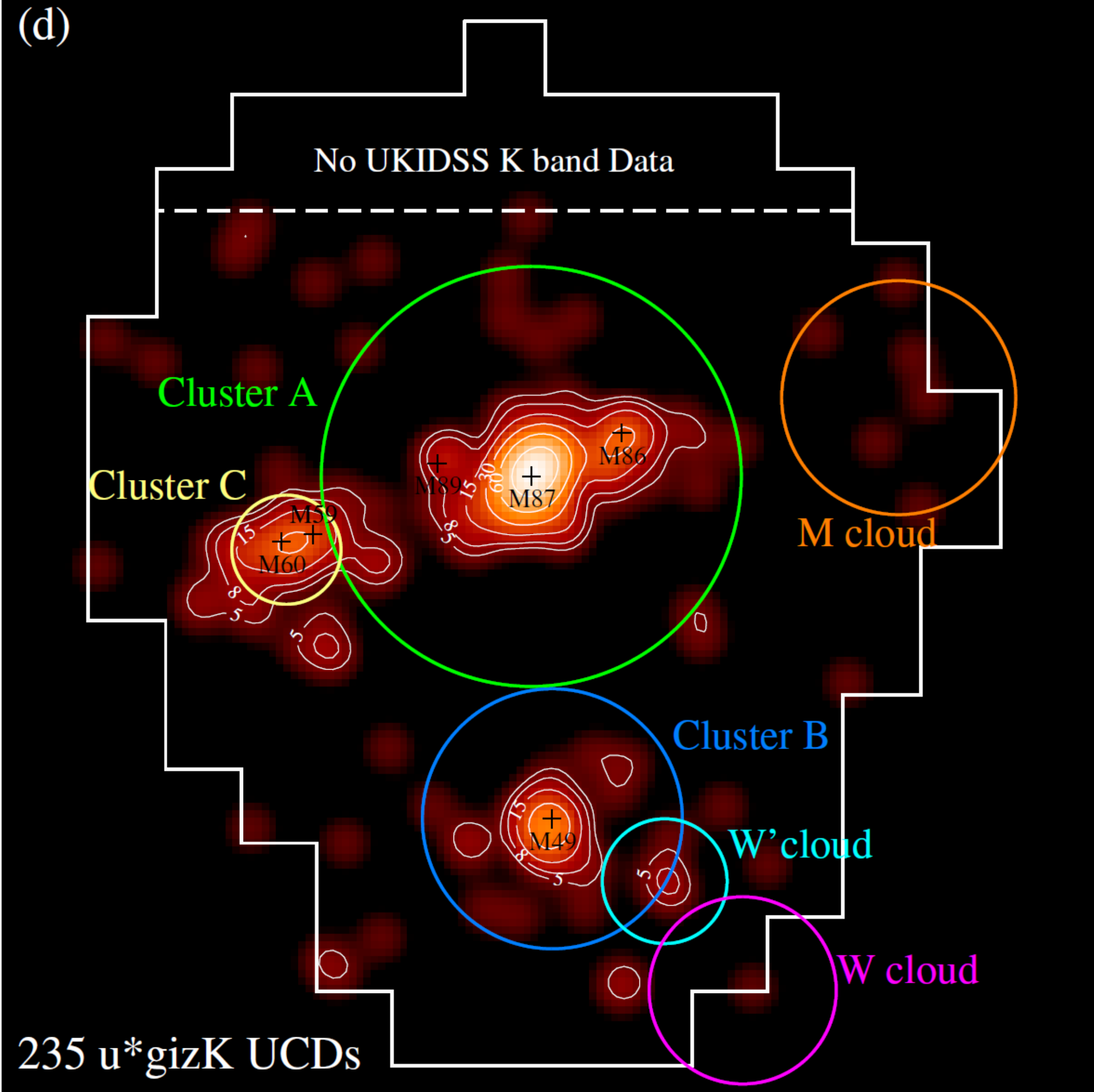}\\
\caption{The number density map of 612 {\tt all UCDs} (upper panels) and 235 {\tt $u^*gizK$ UCDs} (lower panels) identified in the NGVS. The maps have been smoothed with a Gaussian kernel with fixed $\rm{FWHM}\approx 12'$. The white solid polygon in each panel denotes the NGVS survey region. We have no UKIDSS $K$-band data above the white dashed line in each panel. The contours in left panels show the distribution of X-rays from the Rosat all sky survey in the 0.4-2.4 keV energy range \citep{1994Natur_368_828Bohringer}. In the right panels, we show the locations of a few bright galaxies (black crosses) and known galaxy sub-structures (color circles). The contours in panel (d) show the number density contours of $\Sigma_{\rm UCD}=5, 8, 15, 30$ and $60~{\rm deg}^{-2}$.}
\label{fig:map_ucds}
\end{figure*}

In the analysis that follows, we mainly focus on the {\tt bright GCs}, {\tt $u^*gizK$ UCDs} and {\tt bright nuclei} samples. This means that, for the first time, we are using homogeneous samples of GCs, UCDs and nuclei {\it selected from the same data set using consistent selection criteria}. We also use the {\tt ACSVCS GCs}, {\tt all UCDs} and {\tt all nuclei} samples when we require higher completeness, while the {\tt spec GCs} and {\tt spec UCDs} samples are drawn on when cleaner samples are required.

A complementary approach is to study individual UCDs with special properties that can shed light on their origins. For example, \citet{2014Natur_513_398Seth} found a supermassive black hole (SMBH) in M60-UCD1, which comprises $15\%$ of the system's total mass. This single piece of evidence strongly suggests that the progenitor of M60-UCD1 was a nucleated dwarf galaxy. 

In this section, we will summarize the basic statistical properties of our UCD samples, compare to those of GCs and nuclei, and examine interesting sub-samples of UCDs found in the NGVS \citep[e.g.,][]{2015ApJL_812_2Liu}.

\subsection{Spatial Distribution}

The upper two panels of Figure~\ref{fig:map_ucds} show number density maps for our {\tt all UCDs} sample in Virgo. \citet{2014ApJ_794_103Durrell} found that the spatial distribution of Virgo GCs is similar to that of its X-ray-emitting gas, so we compare the distribution of UCDs (color-coded number density map) and X-ray gas (contours) in panel (a) of Figure~\ref{fig:map_ucds}. Consistent with \citet{2014ApJ_794_103Durrell}, we find the densest concentrations of UCDs (i.e., the brightest regions in the map) are located in the regions that have the greatest amount of X-ray gas. This finding is also consistent with the observed correlation between number of UCDs and the X-ray gas mass of their host ($N_{\rm UCD}$ vs. $M_{\rm gas}$; see Figure 17 in \citealt{2015ApJ_812_34Liu}).

\begin{figure*}
\epsscale{0.58}
\plotone{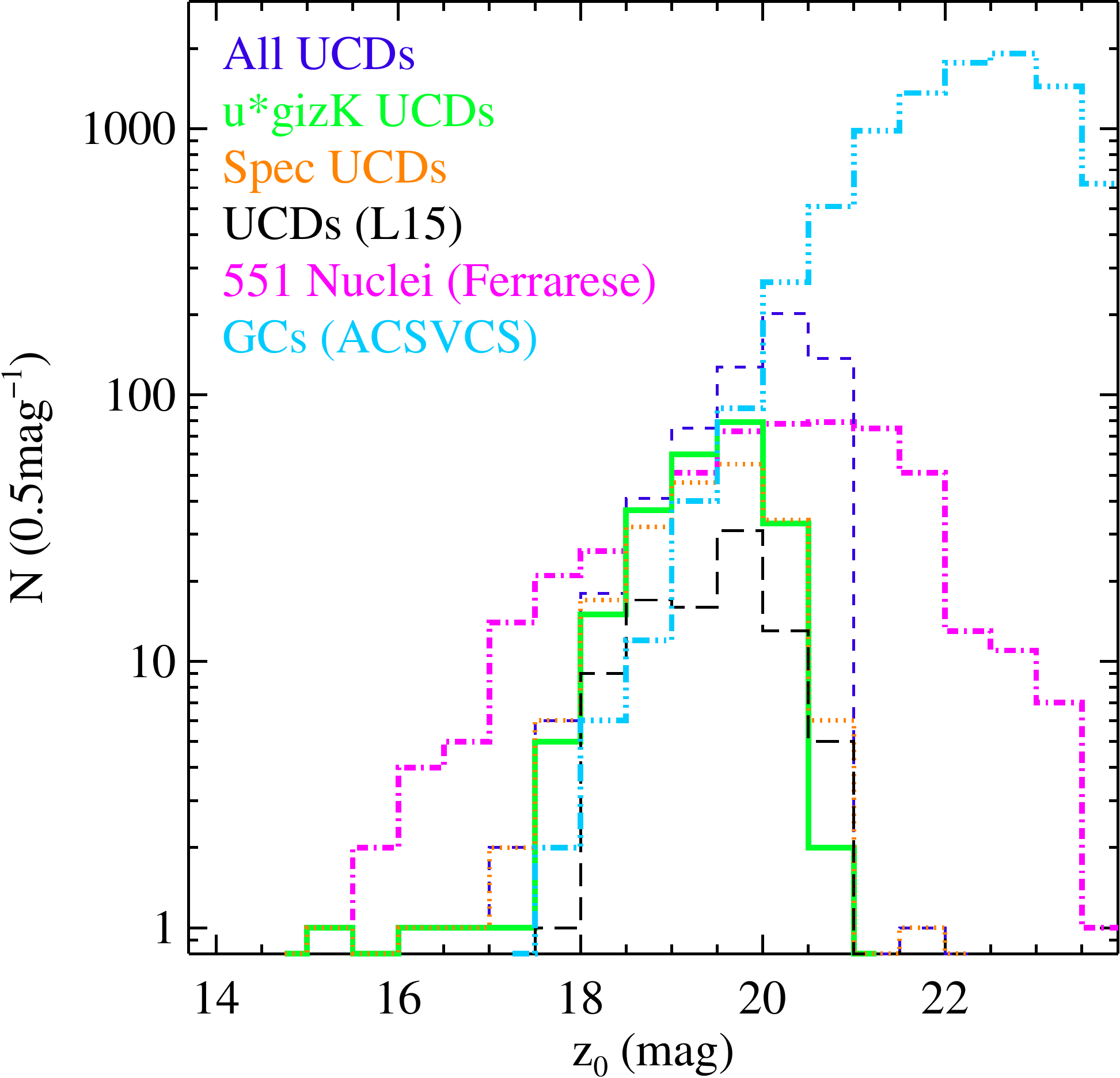}
\epsscale{0.563}
\plotone{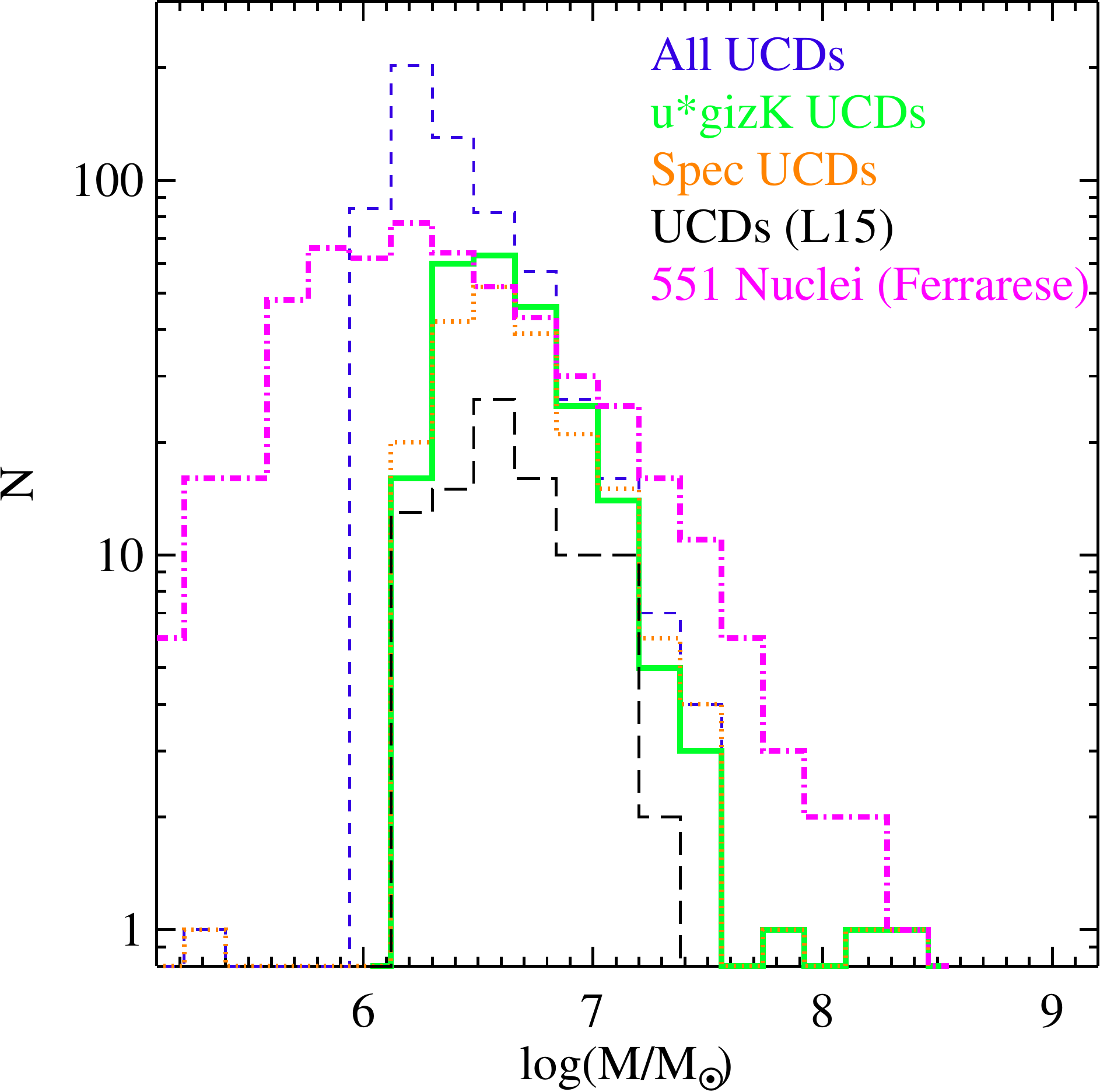}
\caption{The distribution of $z$-band magnitudes ({\it left panel}) and stellar masses ({\it right panel}) for compact stellar systems in Virgo. Blue short-dashed line: {\tt all UCDs} sample (612 candidates); green solid line: {\tt $u^*gizK$ UCDs} sample (235 candidates); orange dotted line: {\tt spec UCDs} sample (203 candidates); black long-dashed line: 92 UCDs from \citet{2015ApJ_812_34Liu}; cyan line: {\tt ACSVCS GCs} sample \citep{2009ApJS_180_54Jordan}; magenta line: {\tt all nuclei} sample \citep{2020ApJ_890_128Ferrarese}.}
\label{fig:distribution_mag}
\end{figure*}

The number density map shown in the panel (b) is the same as in the panel (a), but overlaid with known substructures in the cluster (i.e., the colored circles). The locations and radii of the substructures are taken from \citet{2014A+A_570_69Boselli}. Most of the UCDs are concentrated in sub-cluster A (green), sub-cluster B (blue) and sub-cluster C (yellow) (see also \citealt{2015ApJ_812_34Liu}). There are far fewer UCDs in the other three substructures: the W' (cyan), W (purple) and M clouds (red). This may be due, in part, to the fact that these three substructures are located farther away from us \citep{2007ApJ_655_144Mei, 2018ApJ_856_126Cantiello}, making it more difficult to identify UCDs (by size). At the same time, the reduced gas mass (see the contours in panel a) in these substructures are also likely to be a factor. This is especially true for the M cloud, which shows no significant amount of X-ray gas. Based on the $N_{\rm UCD}$-$M_{\rm gas}$ scaling relation \citep{2015ApJ_812_34Liu} then, the absence of UCDs in this region would not be surprising. In addition, the edges of the NGVS footprint cut through the M and W clouds. This may be another reason why we do not detect many UCDs in these regions.

The lower two panels of Figure~\ref{fig:map_ucds} show number density maps for our {\tt $u^*gizK$ UCDs} sample. As noted above, and as can be seen in the plots, this sample is noticeably cleaner. We also point out that the number densities from the {\tt all UCDs} sample are higher in the regions without UKIDSS K-band data (above the white dashed lines in panels (a) and (b)), indicating elevated contamination in this region. The {\tt $u^*gizK$ UCDs} are mainly concentrated around a handful of luminous galaxies: e.g., M87, M49, M60, M59, M86 and M89 (black crosses). The galaxies with low densities of UCDs cannot be seen in this map as we smooth the distribution using a large kernel ($\sim$12' $\sim 60$~kpc). 

Table 8 of \citet{2015ApJ_812_34Liu} presents estimates of the number density of contaminants in the NGVS based on four control fields. The mean density for our $u^*giz$-selection method is 2.25 deg$^{-2}$. As demonstrated in {\S}\ref{subsec:compare_selections}, our $u^*gizK$-selection method should be much cleaner by comparison. Indeed, \citet{2015ApJ_812_34Liu} estimated the contaminant number density using their $u^*giK_{\rm VISTA}$ selection method to be just 0.9 deg$^{-2}$ (the last row of their Table~8). The contours in panel (d) represent number densities of $\Sigma_{\rm UCD}=5,8,15,30$ and $60~{\rm deg}^{-2}$. We note that there are a few regions with $\Sigma_{\rm UCD}>5$~${\rm deg}^{-2}$ that fall outside of any known sub-structures. They may be intracluster UCDs or UCDs associated with low- or intermediate-mass galaxies \citep[e.g.,][]{2019A+A_625_50Fahrion}. We intend to focus on these regions in future papers. 

\subsection{Luminosity and Stellar Mass Distribution}

Our homogeneous sample provides us with an opportunity to carry out a study of the UCD luminosity function (UCDLF). The left panel of Figure~\ref{fig:distribution_mag} shows histograms of $z$-band magnitudes for a variety of UCD samples, which are identified in the upper-left corner. The distribution for the {\tt all UCDs} sample (blue, short-dashed histogram) has a higher peak value than all other UCD samples, which we suspect is due to contamination (especially at the faint end). Note that the truncation at $z_0\sim21.0$ mag is the result of our magnitude cut at $g_0 = 21.5$ mag and is therefore artificial. 

For the other, smaller samples of UCDS, we find a peak at roughly $z_0 \sim 19.75$ mag. Note that the {\tt spec UCDs} sample (orange dotted line) may be biased because brighter UCDs are more amenable to spectroscopic follow-up. The {\tt $u^*gizK$ UCDs} sample (green solid line) has higher completeness than the {\tt spec UCDs} sample but is limited by the shallow UKIDSS $K$-band imaging, which may cause us to miss some objects at the faint end. The UCD sample from \citet{2015ApJ_812_34Liu} (black long dashed line) is the best available based strictly on the $u^*giK_s$ selection method, where the $K_s$-band data reach to $\sim24$ mag. For comparison, we also show in Figure~\ref{fig:distribution_mag} the GC luminosity function (GCLF) from the ACSVCS (cyan line), which exhibits the well-known turnover for this population \citep[e.g.,][]{2007ApJS_171_101Jordan, 2010ApJ_717_603Villegas}. Despite the fact that three of our UCDLFs also exhibit turnovers, we cannot ensure that these features are authentic. It is quite probable that, depending on where we place our (subjective) size cut (set at $r_h$ = 11 pc), the prominence and location of this peak would change.

The right panel of Figure~\ref{fig:distribution_mag} shows distributions of stellar mass for UCDs and nuclei. Stellar masses have been determined using the relationships between stellar mass-to-light ratios and colors from \citet{2003ApJS_149_289Bell}. We calculate four sets of stellar masses, based on the relations for $(u-g)$, $(u-i)$, $(u-z)$ and $(g-z)$ colors (see their Table~7), and use the mean value for each object in the figure. Other than the {\tt all UCDs} sample, there appears to be a peak in the UCD stellar mass function at $M_* \sim 10^{6.6}$ M$_{\odot}$, roughly equivalent to that seen in the UCDLF. Again though, because of our selection criteria, we cannot be sure whether this is a genuine characteristic of the UCD population.

\begin{figure*}
\epsscale{1.17}
\plotone{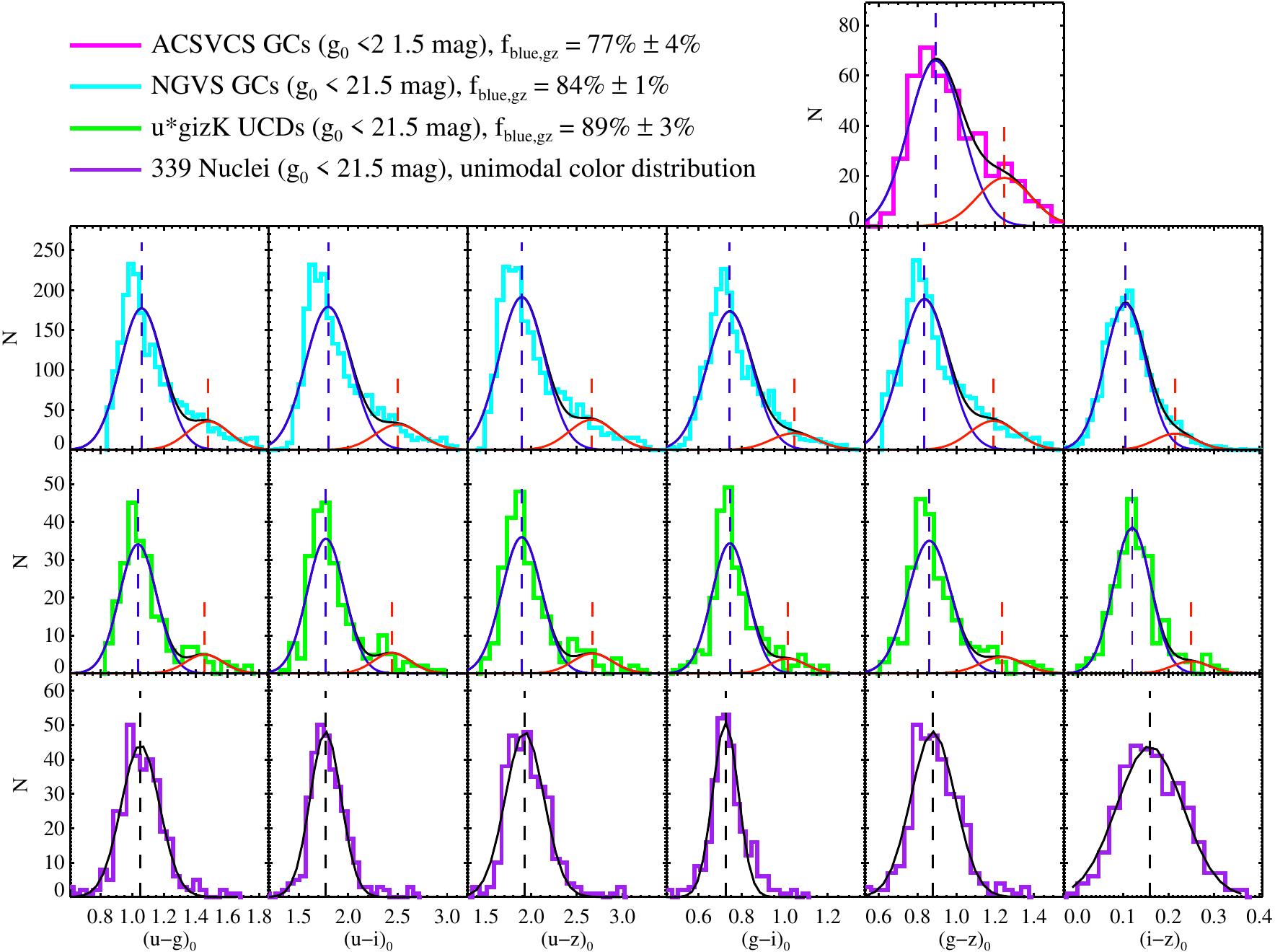}
\caption{The color histograms for {\tt ACSVCS GCs} (the first row, magenta histogram), {\tt bright GCs} (the second row, cyan histograms), {\tt $u^*gizK$ UCDs} (the third row, green histograms) and {\tt bright nuclei} (the forth row, purple histogram). All objects are brighter than $g_0=21.5$ mag. Colors plotted, from left to right, are: $(u-g)_0$, $(u-i)_0$, $(u-z)_0$, $(g-i)_0$, $(g-z)_0$ and $(i-z)_0$. For the bimodal color distributions, we show blue and red Gaussian components (blue and red curves) and their sums (black curves). For the unimodal color distributions, we plot the best fit single Gaussians using black curves. The vertical dashed line(s) shows the mean(s) of the associated color distribution(s).}
\label{fig:color_distribution}
\end{figure*}

The magenta lines in both panels of Figure~\ref{fig:distribution_mag} show the luminosity and mass functions for the {\tt all nuclei} sample that consists of 551 dE,N nuclei. The nuclei LF shows a clear turnover around $z_0 \sim 20.5$ mag and $M_* \sim 10^{6.2}$ M$_{\odot}$ and covers a larger stellar mass range ($10^{5.0} \lesssim M_* \lesssim 10^{8.5}$ M$_{\odot}$) than UCDs. \citet{2019ApJ_878_18Sanchez-Janssen} studied the mass function of nuclei in the Virgo core region and found a peak at $M_* \sim 10^{6.08}$ M$_{\odot}$, which is consistent with the result in this study.

\subsection{Color Distribution}
\label{sec:color_distribution}

It is well established that bimodal color distributions are a common feature of GC systems in massive elliptical galaxies \citep[e.g.,][]{1999AJ_118_1526Gebhardt, 2001AJ_121_2950Kundu, 2006ApJ_639_95Peng}. Interestingly, a bimodal color distribution has also been observed for the UCDs surrounding M87 \citep{2015ApJ_812_34Liu}, making it the only galaxy in the Virgo cluster that shows significant UCD color bimodality.\footnote{The bimodal UCD color distribution for M49 reported in \citet{2015ApJ_812_34Liu} has low statistical significance.} With our new UCD sample, we can examine for the first time the color distribution for UCDs distributed over the entire Virgo cluster.

Figure~\ref{fig:color_distribution} shows the color distribution for the {\tt ACSVCS GCs} (the first row, magenta histogram), {\tt bright GCs} (the second row, cyan histograms), {\tt $u^*gizK$ UCDs} (the third row, green histograms) and {\tt bright nuclei} (the forth row, purple histogram). All objects are brighter than $g_0=21.5$ mag. Because we have only $g$ and $z$ data for the {\tt ACSVCS GCs}, the first row shows only a $(g-z)_0$ distribution. By contrast, a total of six color indices -- $(u^*-g)_0$, $(u^*-i)_0$, $(u^*-z)_0$, $(g-i)_0$, $(g-z)_0$ and $(i-z)_0$ -- are shown (from left to right) in the second, third and forth rows. The distributions of GCs, UCDs and nuclei have red tails for all indices except $(i-z)_0$ for the nuclei. To quantify the bimodality of color distributions, we use Gaussian Mixture Modeling (GMM; \citealt{2010ApJ_718_1266Muratov}) and assume a homoscedastic distribution: i.e., two sub-populations having the same dispersions. As described by \citet{2010ApJ_718_1266Muratov}, a dimensionless peak separation ratio $D$ (see Equation~A3 in their paper) should be larger than 2 for a bimodal distribution. We measure $D$ for the color distribution in each panel of Figure~\ref{fig:color_distribution}. The $D$ parameters for GCs and UCDs are larger than 2 while those for nuclei are smaller than 2. In addition, the unimodal distribution is rejected at a level better than $1\%$ for {\tt bright GCs} and better than $0.1\%$ for {\tt ACSVCS GCs} and {\tt $u^*gizK$ UCDs}. 

The best-fit GMM models are shown in Figure~\ref{fig:color_distribution} as well. For those distributions found to be bimodal, blue and red curves are used to represent the individual components (with vertical dashed lines marking their respective means), while the black curves represent their sums. Conversely, where unimodal distributions are favored, we show the best-fit Gaussians (black curves) and the corresponding means (vertical dashed lines).

In the case of a bimodal color distribution, GMM also yields the probability that a given object belongs to the blue or red component. We therefore calculate the fraction of objects belonging to the blue sub-population, $f_{\rm blue}$. The $f_{\rm blue}$, based on the $(g-z)_0$ color index, is $77\%$ ($\pm 4\%$), $84\%$ ($\pm 1\%$) and $89\%$ ($\pm 3\%$) for {\tt ACSVCS GCs}, {\tt bright GCs} and UCDs. Based on ACSVCS data, \citet{2006ApJ_639_95Peng} concluded that the blue GC fraction for individual galaxies ranges from $85\%$ to $40\%$, and decreases with increasing host galaxy luminosity.

There are two main reasons why the blue fractions are different between {\tt ACSVCS} and {\tt bright GCs}. First, the blue fraction of {\tt bright GCs} is calculated over the entire cluster which contains many more dwarf galaxies (which have higher $f_{\rm blue}$) than the ACSVCS galaxy sample. Second, as described in \citet{2005ApJ_634_1002Jordan}, almost all the GCs in the Virgo cluster can be resolved in ACSVCS imaging, which enables a very clean GC selection \citep[see][]{2006ApJ_639_95Peng, 2009ApJS_180_54Jordan}, while our {\tt bright GC} sample is contaminated by many background galaxies when the $K$-band data is not available (see Table~\ref{tab:compare}). Therefore, our {\tt bright GC} sample contains more dwarf galaxies and is less pure than the {\tt ACSVCS GC} sample, causing it to have a higher blue fraction. For nuclei, we can see from the last row of Figure~\ref{fig:color_distribution} that the vast majority are blue with just a few having red colors. Since a unimodal distribution is preferred for this group, we are unable to calculate a blue fraction on the basis of GMM alone. However, considering that the blue and red components for GCs and UCDs cross at $(g-z)_0 \sim 1.1$, we can split the nuclei distribution at this color to calculate their blue fraction. In so doing, we find an $f_{\rm blue}~\sim 95\%$.

To summarize, for objects brighter than $g_0=21.5$, GMM fitting shows that GCs and UCDs exhibit bimodal color distributions while the nuclei follow a unimodal distribution with a small tail to red colors. Ordering these groups of compact stellar systems by $f_{\rm blue}$, from low to high, yields: {\tt ACSVCS GCs} $<$ {\tt bright GCs} $\lesssim$ UCDs $<$ {\tt bright nuclei}.

\subsection{Size Distribution}

\begin{figure}
\epsscale{1.16}
\plotone{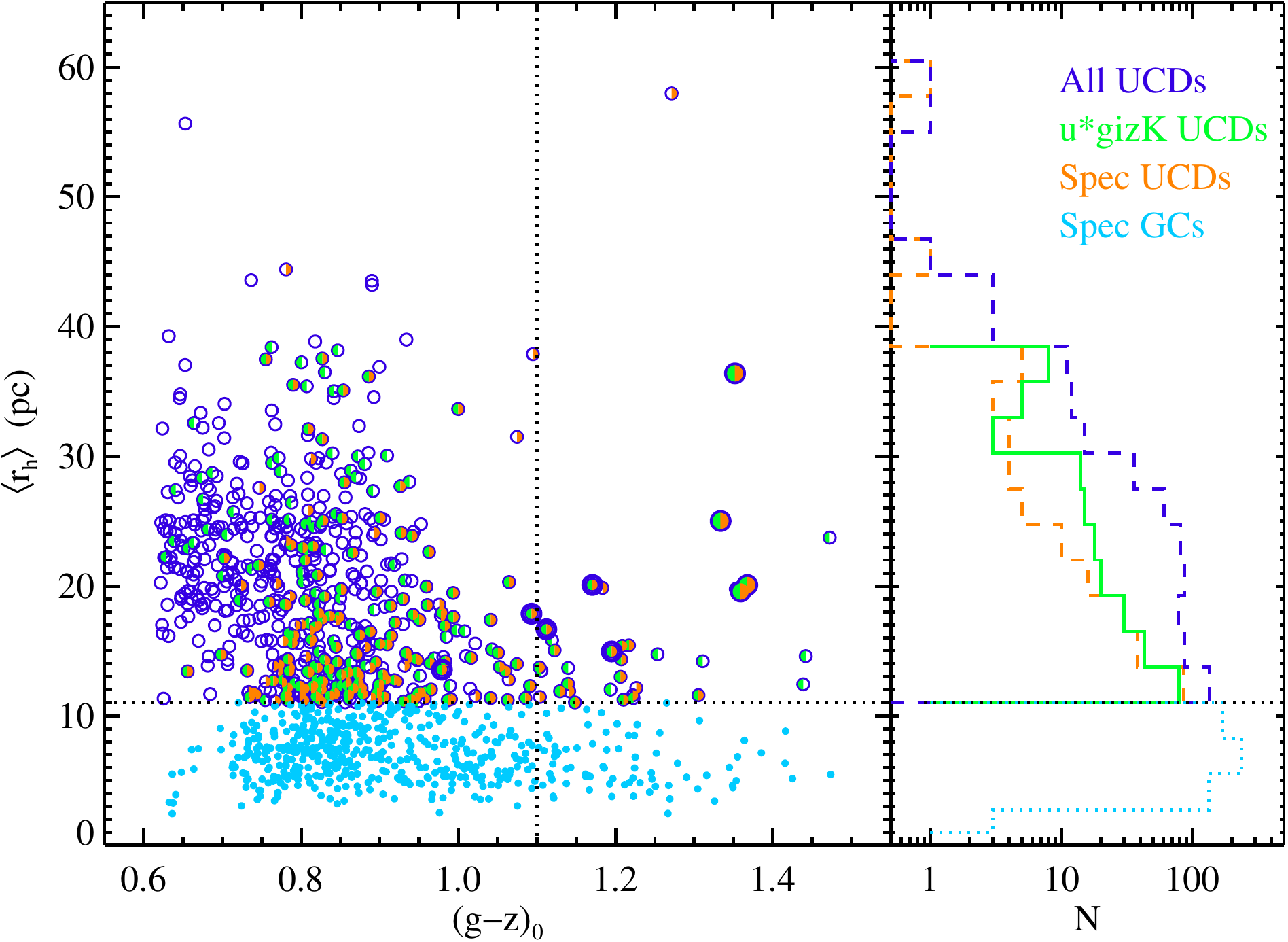}
\caption{{\it Left panel:} Mean half-light radius vs. $(g-z)_0$ color for {\tt spec GCs} (cyan filled circles) and UCDs ({\tt all UCDs}: blue open circles; {\tt $u^*gizK$ UCDs}: green filled semi-circles; {\tt spec UCDs}: orange filled semi-circles). The four larger circles are those UCDs that are known to host SMBHs: i.e., M60-UCD1 \citep{2014Natur_513_398Seth}, M59-UCD3 \citep{2018ApJ_858_102Ahn}, M59cO \citep{2017ApJ_839_72Ahn} and VUCD3 \citep{2017ApJ_839_72Ahn}. The five thicker blue circles denote the five bright UCDS in Figure~\ref{fig:ucd_bright} that have not yet been studied for SMBHs. The vertical black dotted line shows the $(g-z)_0$ color, which is used to divide the GCs and UCDs into red and blue sub-populations. {\it Right panel}: $r_h$ distribution for {\tt all UCDs} (blue long-dashed line), {\tt $u^*gizK$ UCDs} (green solid line), {\tt spec UCDs} (orange short-dashed line), and {\tt spec GCs} (cyan dotted line). The horizontal black dotted line in both panels shows a half-light radius of 11 pc, which is used to separate UCDs from GCs.}
\label{fig:distribution_rh}
\end{figure}

In order to select UCDs, we have measured half-light radii for all bright objects ($g_0<21.5$) in the NGVS footprint using KINGPHOT. The left panel of Figure~\ref{fig:distribution_rh} plots $\langle r_h \rangle$ versus $(g-z)_0$ for our {\tt all} (blue open circles), {\tt $u^*gizK$} (green filled semi-circles), and {\tt spec UCDs} samples (orange filled semi-circles), and {\tt spec GCs} sample (cyan dots). It is worth noting that the majority of UCDs with very blue colors, $(g-z)_0 \lesssim 0.7$, lack radial velocity measurements, a situation we aim to improve upon through dedicated spectroscopic follow-up. The right panel shows the distribution of half-light radii for both UCDs and GCs. The dotted horizontal line marks $\langle r_h \rangle = 11$ pc, the size used to separate GCs from UCDs, while the dotted vertical line indicates the color, $(g-z)_0=1.1$, used to divide the UCDs into blue and red sub-populations (see \S \ref{sec:color_distribution}). 

The mean $r_h$ of UCDs is $\overline{r_{h}}=19.8$ pc with a standard deviation of $\sigma_{r_h}=6.8$ pc. For sub-populations, the mean $r_h$ and standard deviations are $\overline{r_h}=20.0$ pc and $\sigma_{r_h}=6.8$ pc for the blue UCDs, and $\overline{r_h}=14.6$ pc and $\sigma_{r_h}=3.8$ pc for the red UCDs. The blue UCDs are larger and cover a wider range in half-light radius than red ones. 

As discussed in \citet{2015ApJ_812_34Liu}, when comparing to HST-based values, we find that we overestimate the half-light radii of smaller objects ($r_h \lesssim 11$ pc). Thus, we suspect that many diffuse GCs are included in our UCD sample. This may explain why we see such a high degree of similarity between the color distributions of our NGVS-based samples of GCs and UCDs.

At the time of writing, SMBHs have been detected in four Virgo UCDs. These objects (all of which were objectively identified by our selection function) are indicated by the large circles in the left panel of Figure~\ref{fig:distribution_rh}). These are M60-UCD1 \citep{2014Natur_513_398Seth}, M59cO \citep{2017ApJ_839_72Ahn}, VUCD3 \citep{2017ApJ_839_72Ahn} and M59-UCD3 \citep{2018ApJ_858_102Ahn}. Interestingly, all four objects are quite red, with colors of $(g-z)_0 \sim 1.35$. Based on the color-magnitude relation for UCDs, we know that the redder systems tend to be brighter. To date, no blue UCD is known to contain a SMBH, although we have many promising bright, blue UCD candidates in our sample. These are obvious targets for future spectroscopic searches for SMBHs. 

\subsection{Sub-samples of Unique UCDs}

As noted above, we find no significant color differences between our UCD and GC samples. However, some UCDs, by virtue of their extreme or unusual properties, have a special significance for understanding the origin of these systems and their relationship to ``normal" GCs. In this section, we pause to consider these unique UCDs. 

\subsubsection{The Brightest UCDs}

\begin{figure}
\epsscale{1.17}
\plotone{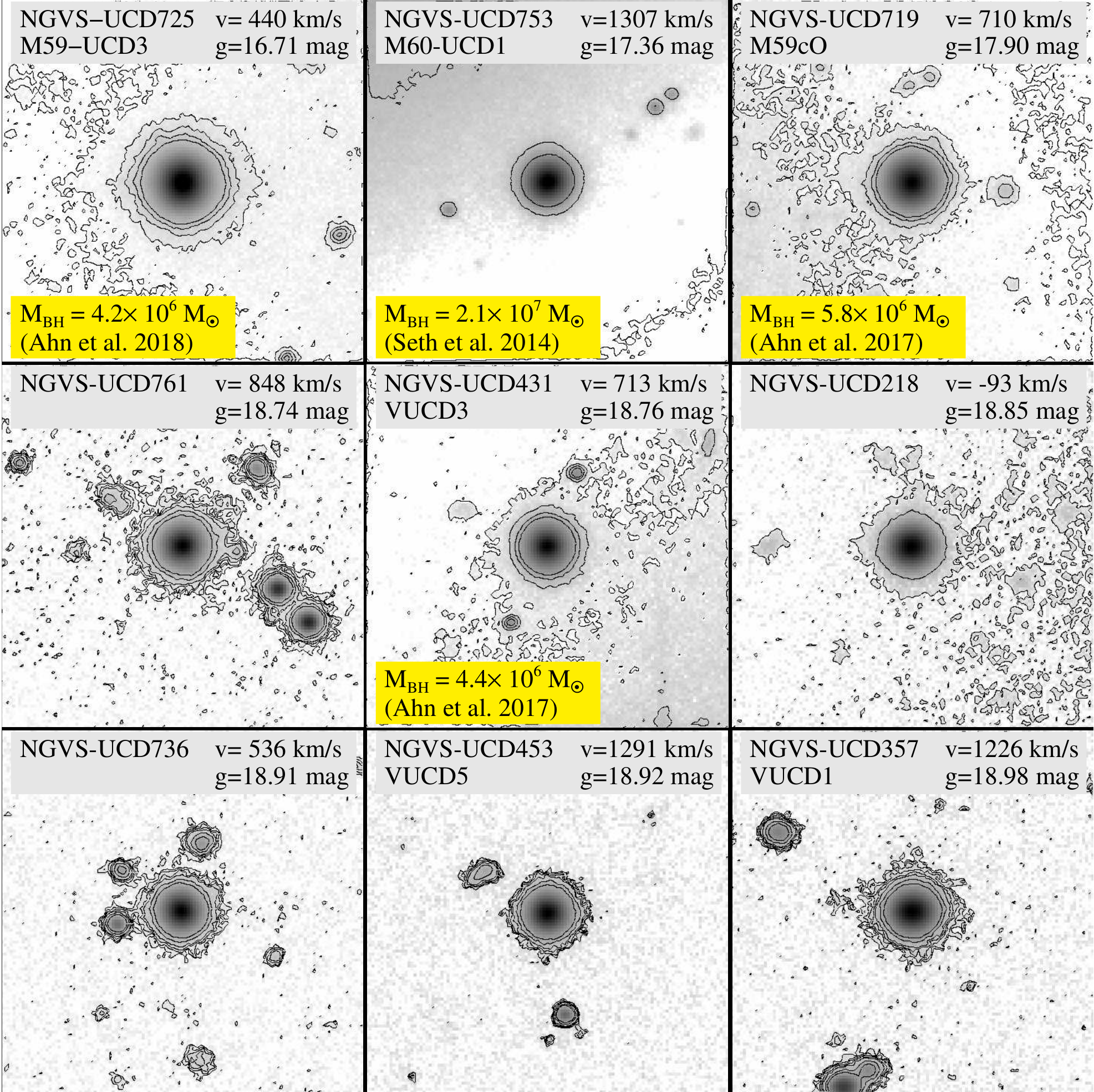}
\caption{NGVS $g$-band images for the brightest nine UCDs in Virgo cluster. Radial velocities used to confirm cluster membership are available for each of these objects. Four of these UCDs are known to host SMBHs, the masses of which are reported in the panels. The image size in each panel is $120\times120$ pixels, where $120$ pixels $\sim22.32$\arcsec~ $\sim1.8$ kpc at Virgo distance.}
\label{fig:ucd_bright}
\end{figure}

\begin{figure}
\epsscale{1.165}
\plotone{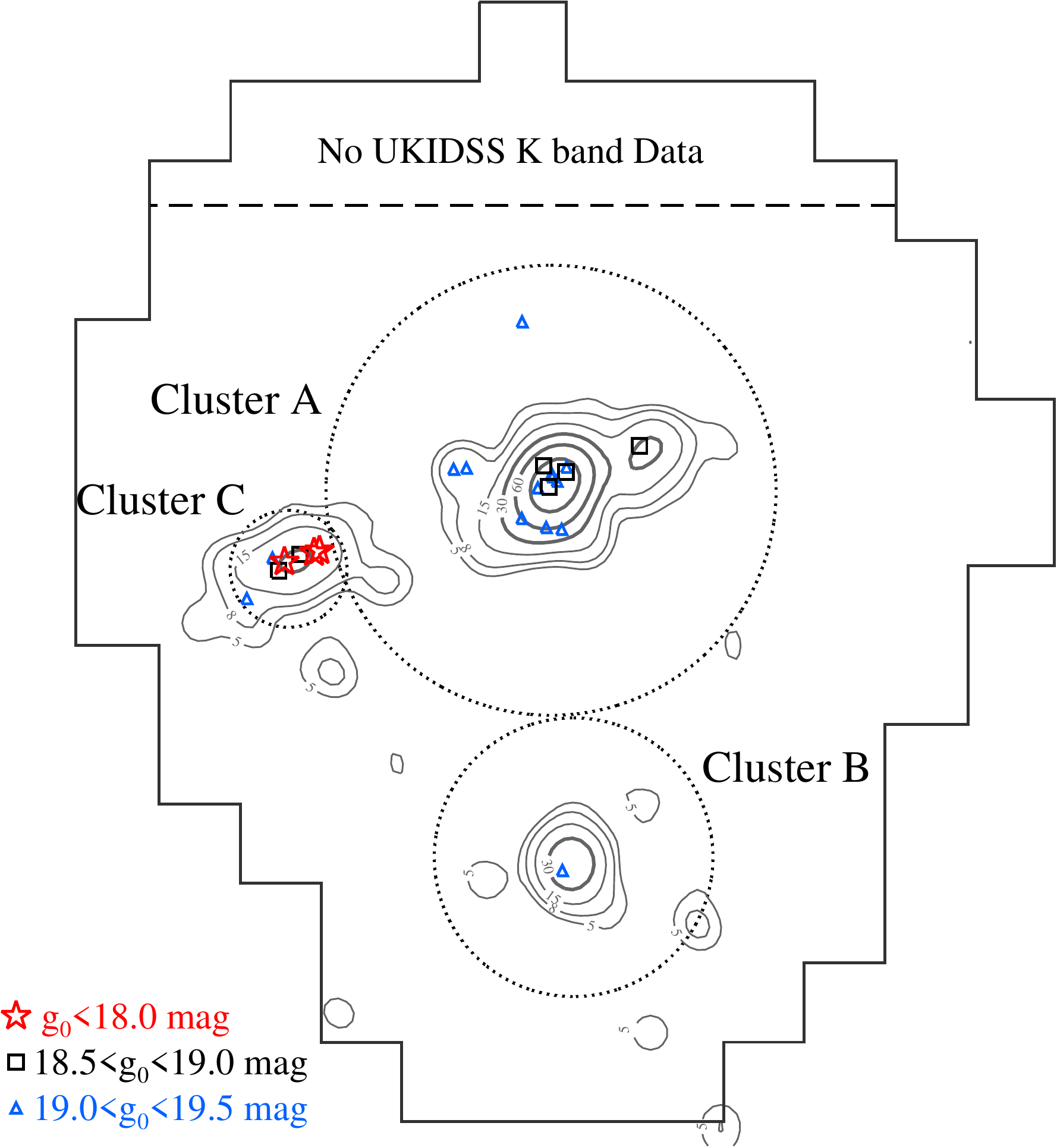}
\caption{The spatial distribution of bright {\tt $u^*gizK$ UCDs} with $g_0<19.5$ mag. Red stars show the three UCDs with $g_0<18.0$ mag. Black squares show the six UCDs with $18.5<g_0<19.0$ mag (note that there are no UCDs are in the magnitude range $18.0<g_0<18.5$ mag). Blue triangles show the fourteen UCDs with $19.0<g_0<19.5$ mag. Contours show the surface density distributions for the sample of 235 {\tt $u^*gizK$ UCDs} (see Figure~\ref{fig:map_ucds}). The large dotted circles show Virgo's three main sub-clusters: i.e., clusters A, B and C, which are centred on M87, M49 and M60, respectively.}
\label{fig:map_bright_ucd}
\end{figure}

We begin with the subset of UCDs that are known to contain a SMBH. The first such detection was made in M60-UCD1, where \citet{2014Natur_513_398Seth} found a SMBH of mass $\sim2.1\times10^7$~M$_{\odot}$. The inferred mass fraction ($15\%$) suggests that the progenitor of M60-UCD1 was a dwarf galaxy. Soon afterwards, an even more massive system in Virgo, M59-UCD3, was co-reported by \citet{2015ApJL_812_2Liu} and \citet{2015ApJL_808_32Sandoval}. Later, \citet{2018ApJ_858_102Ahn} showed that this UCD also contains a SMBH of mass $4.2\times10^6$ M$_\odot$. SMBHs have also been found in two more Virgo UCDs, M59cO and VUCD3 \citep{2017ApJ_839_72Ahn}. The discovery that UCDs can harbor SMBHs is crucial evidence of a dE,N origin for at least some UCDs.

One property that unifies the four UCDs with known SMBHs is their luminosity: they are all very bright. Based on this, we have carried out a search for bright UCDs using the NGVS short-exposures. Figure~\ref{fig:ucd_bright} presents $g$-band cutouts for the nine brightest UCD candidates in our sample, which shows that M59-UCD3, M60-UCD1 and M59cO are the brightest three UCDs in all of Virgo, and the only three brighter than $g_0=18$ mag. The next brightest UCD, NGVS-UCD761, with $g_0=18.74$ and $(g-z)_0=1.11$, represents a new detection and is a tantalizing target for future SMBH searches.

Figure \ref{fig:map_bright_ucd} shows the locations of the 23 UCDs in Virgo that are brighter than $g_0=19.5$ mag. The red stars represent the three brightest systems. We note that the third- (M59cO) and fourth-brightest UCDs (NGVS-UCD761) are separated by a significant magnitude gap ($\Delta g_0$ = 0.84 mag). The fourth ($g_0=18.74$ mag) to ninth-ranked ($g_0=18.98$ mag) UCDs are shown as black squares in Figure~\ref{fig:map_bright_ucd}, while blue triangles represent UCDs in the range of $19.0<g_0<19.5$ mag.

Interestingly, all 23 UCDs with $g_0<19.5$ mag are located in Virgo's three main sub-clusters (black dotted circles). About half of these objects are in Cluster A (centred on M87) --- 4 with $18.5<g_0<19.0$ and 11 $19.0<g_0<19.5$. Curiously, the brightest three UCDs are all located in Cluster C, a much smaller sub-cluster centred on M60. On the contrary, only one bright UCD (20th-ranked; $g_0=19.39$) is found in Cluster B (centred on M49). It is puzzling why Cluster C is so special in this regard. Perhaps the fact that this substructure contains two massive galaxies in close proximity to one another makes it a conducive environment to forming bright UCDs (e.g. through tidal stripping).

\subsubsection{The Largest UCDs}

\begin{figure}
\epsscale{1.17}
\plotone{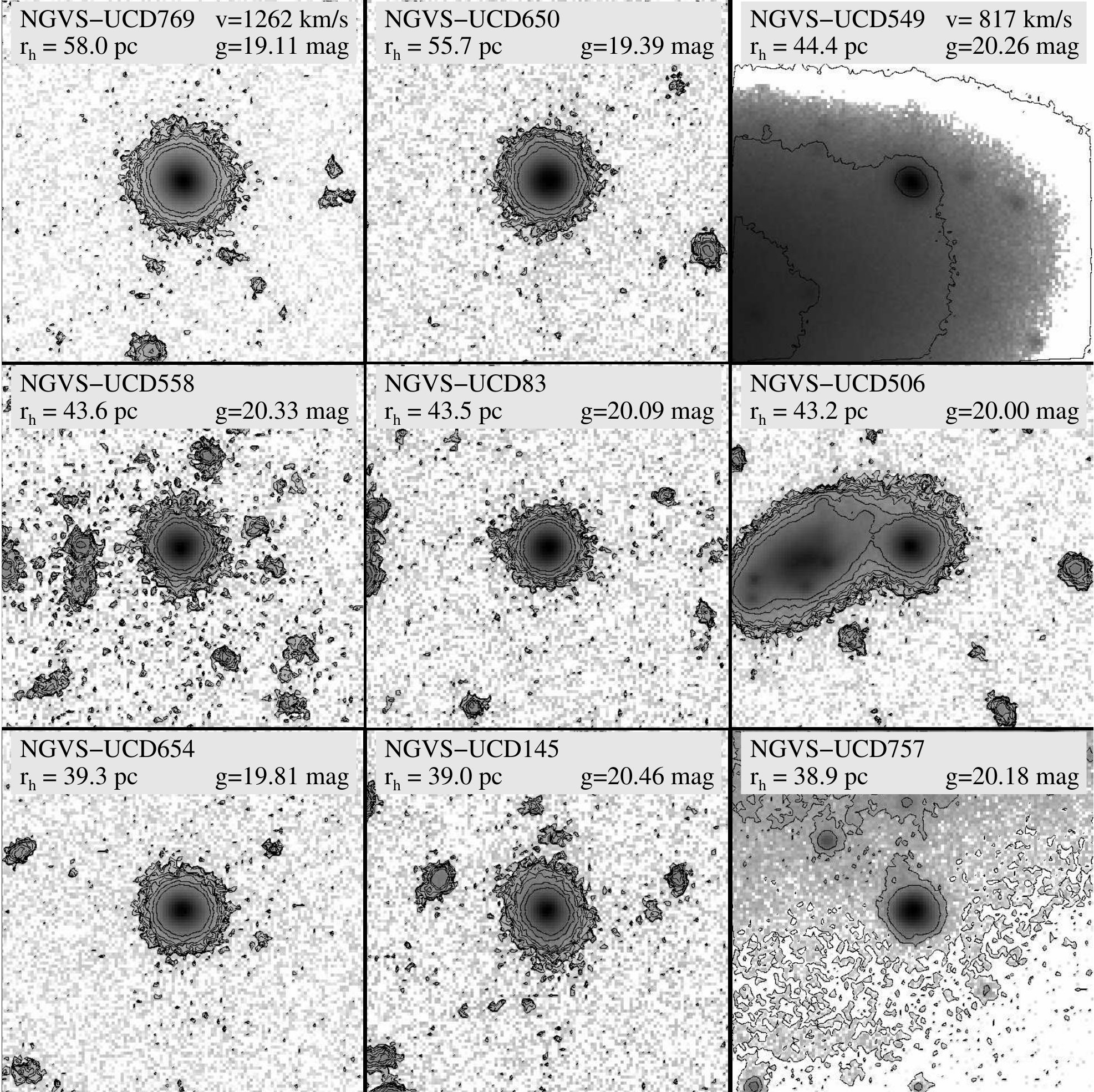}
\caption{NGVS $g$-band images for the largest nine UCDs in Virgo cluster. The image size in each panel is $120\times120$ pixels, where $120$ pixels $\sim22.32$\arcsec~ $\sim1.8$ kpc at the distance of Virgo.}
\label{fig:ucd_large}
\end{figure}

Figure~\ref{fig:ucd_large} shows $g$-band cutouts for the nine largest UCDs in our sample. The largest, NGVS-UCD769 ($\langle{r_h}\rangle=58$ pc\footnote{Note that the largest nucleus in Virgo has $r_h \sim 60$ pc \citep{2006ApJS_165_57Cote}}.), also happens to be the brightest ($g_0=19.11$ mag) and reddest one ($(g-z)_0 = 1.27$) of this sub-sample; the remaining eight all have $(g-z)_0 \lesssim 1.1$. There is, interestingly, no overlap between the nine brightest and nine largest UCDs in our sample. Only two of the largest UCDs currently have measured radial velocities. 

We note that three UCDs in Figure~\ref{fig:ucd_large} (NGVS-UCD549, 506 and 757) have bright neighbors. Recall that when we measure the half-light radius of an object, KINGPHOT fits the image using PSF-convolved King models within a fitting radius, $r_{\rm fit}$ (15 pixels here). In Figure~\ref{fig:rh_compare}, we compare $r_h$ measurements based on different fitting radii, $r_{\rm fit}=7$ or $15$ pixels. The measurements are fully consistent even for the UCDs surrounding the three most luminous galaxies in the Virgo cluster (M87, M49 and M60). We conclude then that light from neighboring galaxies does not seriously affect our $r_h$ measurements. In addition, we test our measurement procedure by injecting artificial UCDs across a range of environments covered by the NGVS footprint. We find that the KINGPHOT measurements are quite robust unless the UCD falls very close to a bright point source, like the blended objects shown in Figure~\ref{fig:classification}.

Using the ACSVCS data, \citet{2005ApJ_634_1002Jordan} found the vast majority of GCs in the Virgo cluster are smaller than $r_h$ = 10 pc, with their average size being $\langle r_h \rangle$ = 2.70 $\pm$ 0.35 pc. In addition, they found no correlation between half-light radius and luminosity for {\tt bright GCs} ($z \leq 22.9$ mag). Meanwhile, using ACSVCS data as well, \citet{2006ApJS_165_57Cote} found that nuclei follow a size-magnitude relation, with more luminous nuclei having larger half-light radii. A similar result has been found for UCDs, although not as tight (i.e., more luminous UCDs are usually larger; \citealt{2006ApJS_165_57Cote}; \citealt{2014MNRAS_439_3808Penny}). \citet{2012ApJ_747_72Dabringhausena} have also reported that UCDs follow a size-magnitude relation, while GCs do not.

It is worth pointing out that, several GCs in our MW are also larger than $r_h = 10$ pc \citep{2004MNRAS_354_713vandenBergh}. Such extended star clusters (ESCs) have also been found around other nearby galaxies, e.g., M31 \citep{2005MNRAS_360_1007Huxor, 2014MNRAS_442_2165Huxor}, Scl-dE1 \citep{2009AJ_137_4361DaCosta}, M51 \citep{2008AJ_135_1567Hwang} and NGC 6822 \citep{2011ApJ_738_58Hwang}. These ESCs mainly populate the faint end of the GCLF \citep{2006ApJ_639_838Peng, 2016ApJ_830_99Liu}, with those around the MW and M31 being fainter than $M_V = -7$ \citep{2004MNRAS_354_713vandenBergh, 2011ApJ_738_58Hwang}. Conversely, the typical UCD is brighter than ESCs by two magnitudes or more \cite[$-13 < M_V < -9$ mag;][]{2012AJ_144_76Willman}, and more still for the largest UCDs. Therefore, it is quite likely that the largest UCDs in Virgo originate under different circumstances than ``normal" GCs.

\subsubsection{UCDs with Asymmetric/Tidal Features}

If tidal stripping of dE,Ns produce UCDs, then we should be able to find some objects undergoing such a transformation (i.e., UCDs that exhibit asymmetries and/or tidal features). Of course, this exercise may be complicated by short transformation timescales and/or potentially low surface brightnesses of any stripped material \citep{2013MNRAS_433_1997Pfeffer}. Nonetheless, a few UCDs with asymmetric or tidal features have indeed been found in recent years through deep surveys \citep[e.g.,][]{2015ApJL_812_10Jennings, 2015ApJL_809_21Mihos, 2016A+A_586_102Voggel, 2018ApJ_853_54Schweizer}. Another potential complication surrounds the interpretation of such features -- for instance, several GCs around the MW are known to possess prominent tidal structures, such as NGC 6715 \citep{1994Natur_370_194Ibata, 2008AJ_136_1147Bellazzini}, Palomar 5 \citep{2001ApJL_548_165Odenkirchen}, and $\omega$ Cen \citep{2019NatAs_3_667Ibata}. However, most of these objects have been flagged by other studies as unusual members of the MW GC system \citep{2015AJ_150_63Johnson, 2017MNRAS_464_3636Milone, 2019A+ARv_27_8Gratton}, such that they are commonly held as remnants of nucleated satellites that were disrupted by the MW's tidal field \citep[e.g.,][]{2003MNRAS_346_11Bekki, 2003ApJ_599_1082Majewski, 2015ApJ_803_80Kuepper}.

\begin{figure}
\epsscale{1.17}
\plotone{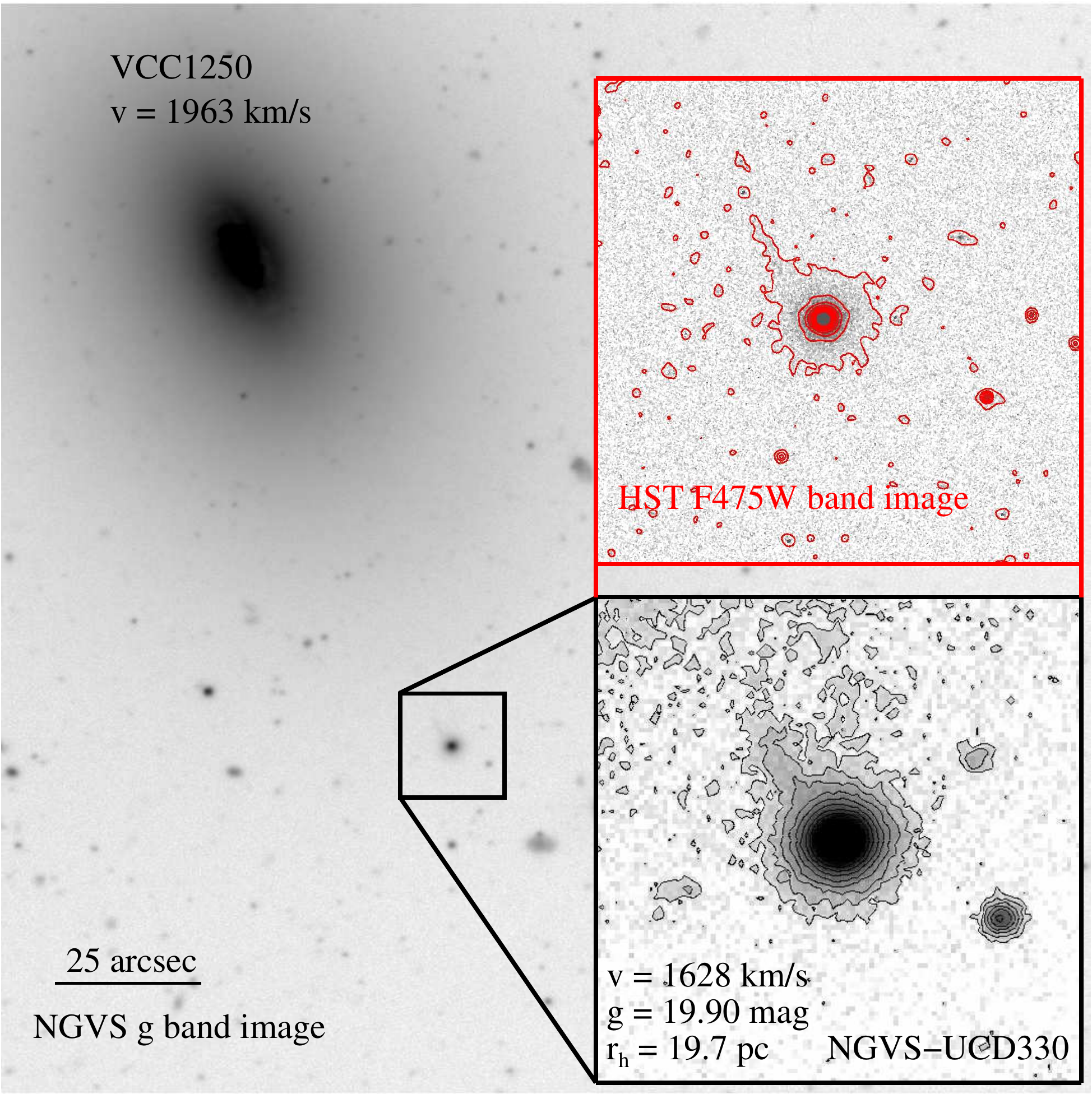}
\caption{NGVS $g$-band image for galaxy VCC 1250. The small panels shows the NGVS and HST images of NGVS-UCD330, a UCD that shows signs of an extra-tidal structure. The inset image with red contours is taken from the HST ACS Virgo Cluster Survey.}
\label{fig:asymmetry}
\end{figure}

The high sensitivity of the NGVS images ($\mu_g \lesssim 29$ mag arcsec$^{-2}$; \citealt{2012ApJS_200_4Ferrarese}) allows us to search for such features within our UCD sample. Our search indeed results in a handful of UCDs with apparent asymmetries. If confirmed as being tidal in origin, these features would offer direct evidence that at least some UCDs are the descendants of nucleated galaxies.

Figure~\ref{fig:asymmetry} shows one candidate tidal structure associated with NGVS-UCD330. This UCD ($v_r=1628$ km/s) is a satellite of VCC 1250 ($v_r=1963$ km/s). It was previously identified as VCC1250\_1 by \citet{2005ApJ_627_203Hacsegan}, who also noted that it appeared to be embedded in a diffuse envelope. Both the NGVS and HST images in Figure~\ref{fig:asymmetry} reveal an asymmetric structure emanating from this object and pointing towards VCC 1250. Thus, NGVS-UCD330 may be an example of a UCD caught in the act of losing what remains of its diffuse envelope. The putative tidal stream associated with this UCD is unusual in that we only detect one arm. However, we cannot rule out the second arm as being hidden by projection or surface brightness effects. Follow-up spectroscopy of this object would help us determine its origins.

\subsubsection{UCDs with Envelopes}
\label{sec:envelope}

\begin{figure}
\epsscale{1.17}
\plotone{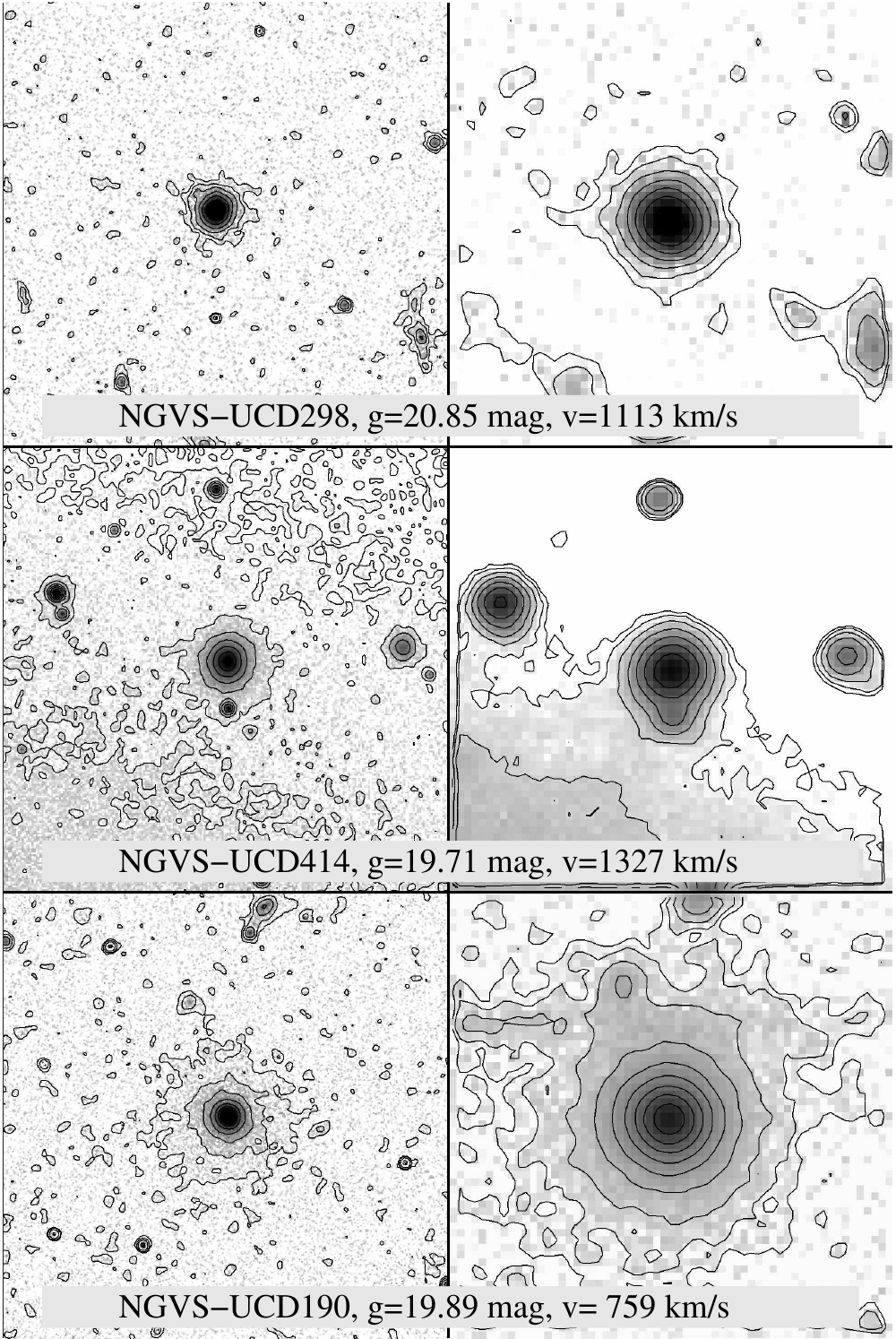}
\caption{HST F475W ($\sim$ SDSS $g$-band) images (left panels) and NGVS $g$-band images (right panels) for 3 UCD candidates, NGVS-UCD298, 414 and 190 (from top to bottom panel). Each of these objects are confirmed radial velocity members of the Virgo cluster (i.e., $v_r<3500$ km/s). The image size in each panel is $\sim 0.9$ kpc $\times 0.9$ kpc.}
\label{fig:ucd_large_env_acsvcs}
\end{figure}

Using HST imaging, \citet{2005ApJ_627_203Hacsegan} found three UCDs in Virgo that are embedded within shallow envelopes --- evidence that they may be related to the nuclei of dE,Ns with extremely low surface brightness and/or compact stellar halos. There are 22 UCD candidates in our sample that also have HST imaging from the ACSVCS \citep{2004ApJS_153_223Cote}. We have checked these HST images to look for envelopes. Some systems, like NGVS-UCD298 (top panel of Figure~\ref{fig:ucd_large_env_acsvcs}), show no evidence of envelopes in either the ACSVCS or NGVS images. Others, like NGVS-UCD414 (middle panel of Figure~\ref{fig:ucd_large_env_acsvcs}), have small envelopes that are visible only in the ACSVCS images thanks to its exceptional image quality. The remainder, like NGVS-UCD190 (bottom panel of Figure~\ref{fig:ucd_large_env_acsvcs}), have large (yet still diffuse) envelopes that can be seen in both ACSVCS and NGVS images. In all, we find about half of the 22 UCD candidates with HST imaging are embedded in diffuse envelopes that are visible in the space-based images. 

\begin{figure}
\epsscale{1.17}
\plotone{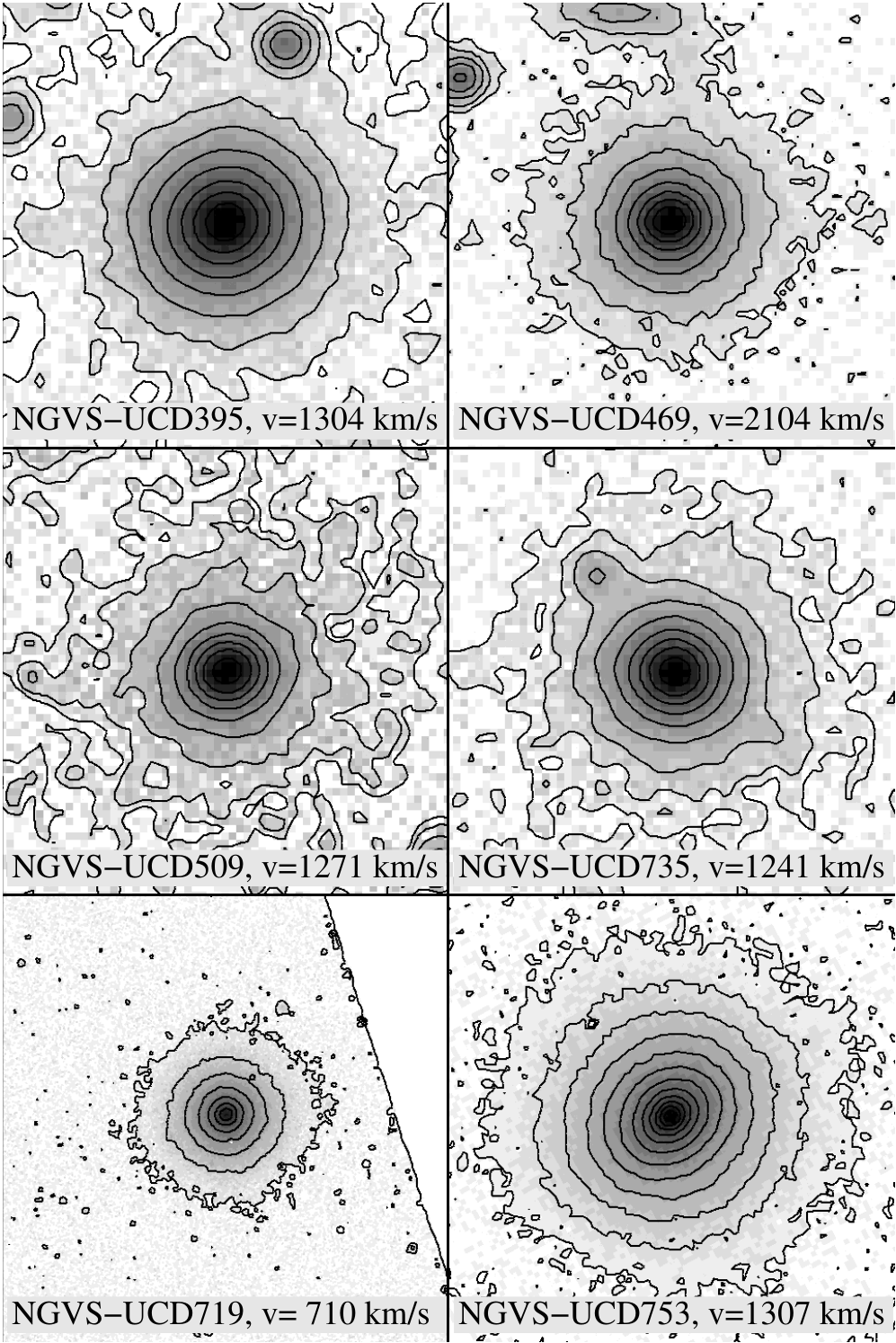}
\caption{NGVS $g$-band images for 4 UCDs (upper four panels) and HST F475W ($\sim$ SDSS $g$ band) images for 2 UCDs (lower two panels) having visible envelopes. Each of these objects are confirmed radial velocity members of the Virgo cluster (i.e., $v_r<3500$ km/s). The image size in each panel is the same with Figure~\ref{fig:ucd_large_env_acsvcs}.}
\label{fig:ucd_large_env}
\end{figure}

The above comparison demonstrates that we can detect envelopes around UCDs in the NGVS imaging, provided they are large enough. Our search reveals 41 instances of such features, and these cases have been accordingly flagged in Table~\ref{tab:ucd_redshift}. Most of these UCDs are found around M87 and M60/M59, with just a couple located sub-cluster B. Note that, in this section, we have focused on UCDs with envelopes that are obvious based on visual inspection --- which is admittedly subjective. Because this sample is not suitable for statistical analysis, we will postpone the detailed investigation of the properties of UCD envelopes to future work (Wang et al., in prep.). In the next section though we will introduce a parameter, $\Delta_{\rm env}$, to examine basic properties of the envelope. 

Figure~\ref{fig:ucd_large_env} presents NGVS $g$-band images for four UCDs (upper four panels) and HST F475W ($\sim$SDSS $g$-band) images for two UCDs (lower two panels) that are confirmed radial velocity members of the Virgo cluster and possess clear stellar envelopes. We see in both sets of imaging that these envelopes have near-zero ellipticity. From a morphological perspective, UCDs with diffuse envelopes look quite similar to dE,Ns, adding weight to the argument of an evolutionary connection between the two classes. Moreover, NGVS-UCD719 \citep[lower-left panel of Figure~\ref{fig:ucd_large_env}, also known as M59cO;][]{2008MNRAS_385_83Chilingarian} and NGVS-UCD753 \citep[lower-right panel of Figure~\ref{fig:ucd_large_env}, also known as M60-UCD1;][]{2013ApJL_775_6Strader} are two of the four UCDs known to possess central SMBHs \citep{2014Natur_513_398Seth, 2017ApJ_839_72Ahn}. We do not, however, observe any tidal features around them \citep[e.g.,][]{2010MNRAS_401_105Kuepper, 2015ApJL_812_10Jennings, 2018ApJ_853_54Schweizer}.

As with tidal features, the interpretation of envelopes around Virgo UCDs is also potentially complicated by the fact that similar structures have recently been detected around the MW GCs NGC 7089 \citep{2016MNRAS_461_3639Kuzma} and NGC 1851 \citep{2018MNRAS_473_2881Kuzma}. Once again though, these GCs are unusual in their chemistries (e.g., possessing broad dispersions in their abundances of iron and neutron-capture elements; \citealt{2015AJ_150_63Johnson}; \citealt{2017MNRAS_464_3636Milone}), suggesting that they are actually the remnant nuclei of disrupted nucleated galaxies.

\section{Discussion}
\label{sec:discussion}

\begin{figure}
\epsscale{1.15}
\plotone{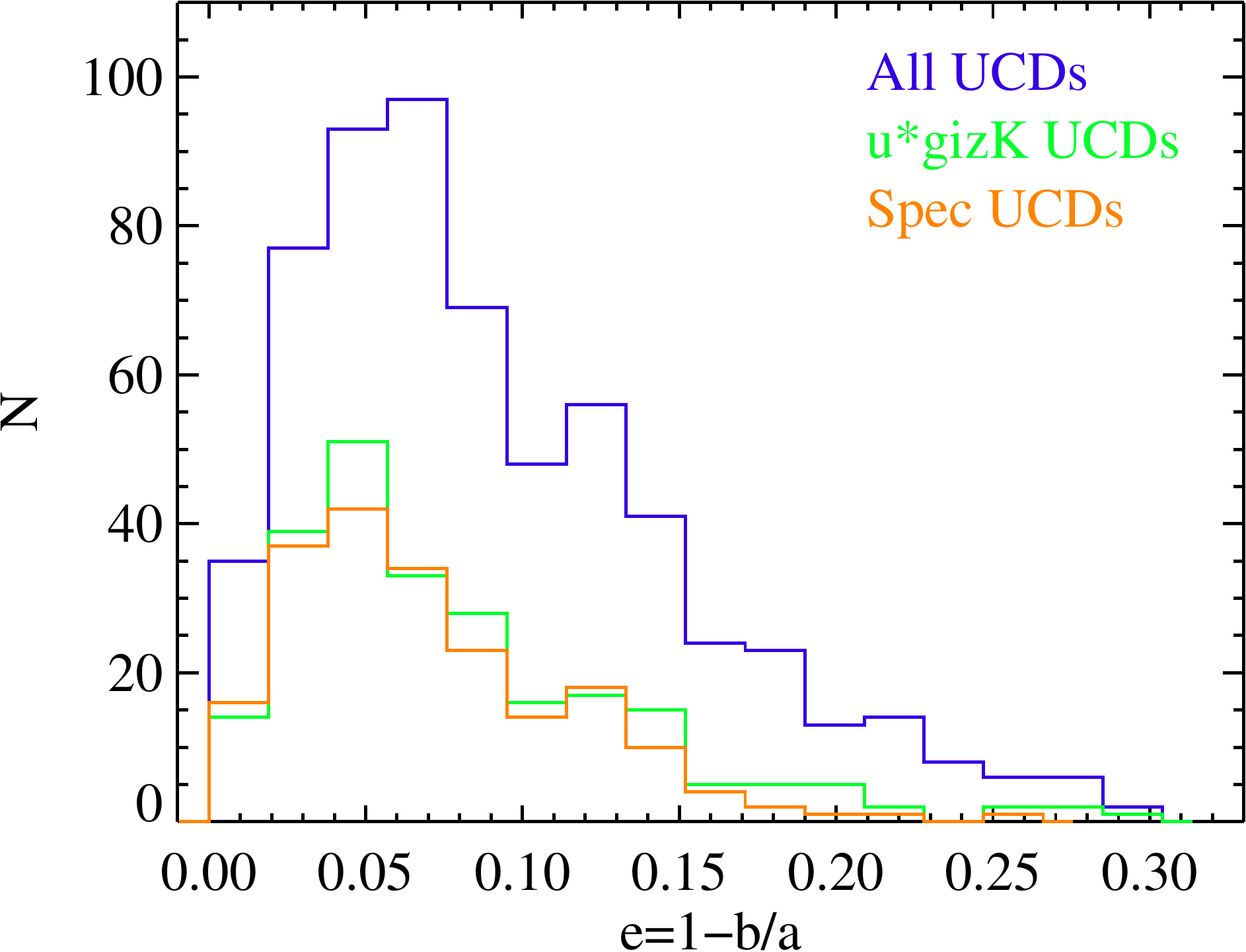}
\caption{The ellipticity distributions of {\tt all UCDs} (blue histogram), {\tt $u^*gizK$-UCDs} (green histogram) and {\tt spec-UCDs} sample (orange histogram).}
\label{fig:ellipticity}
\end{figure}

\begin{figure*}
\epsscale{1.17}
\plotone{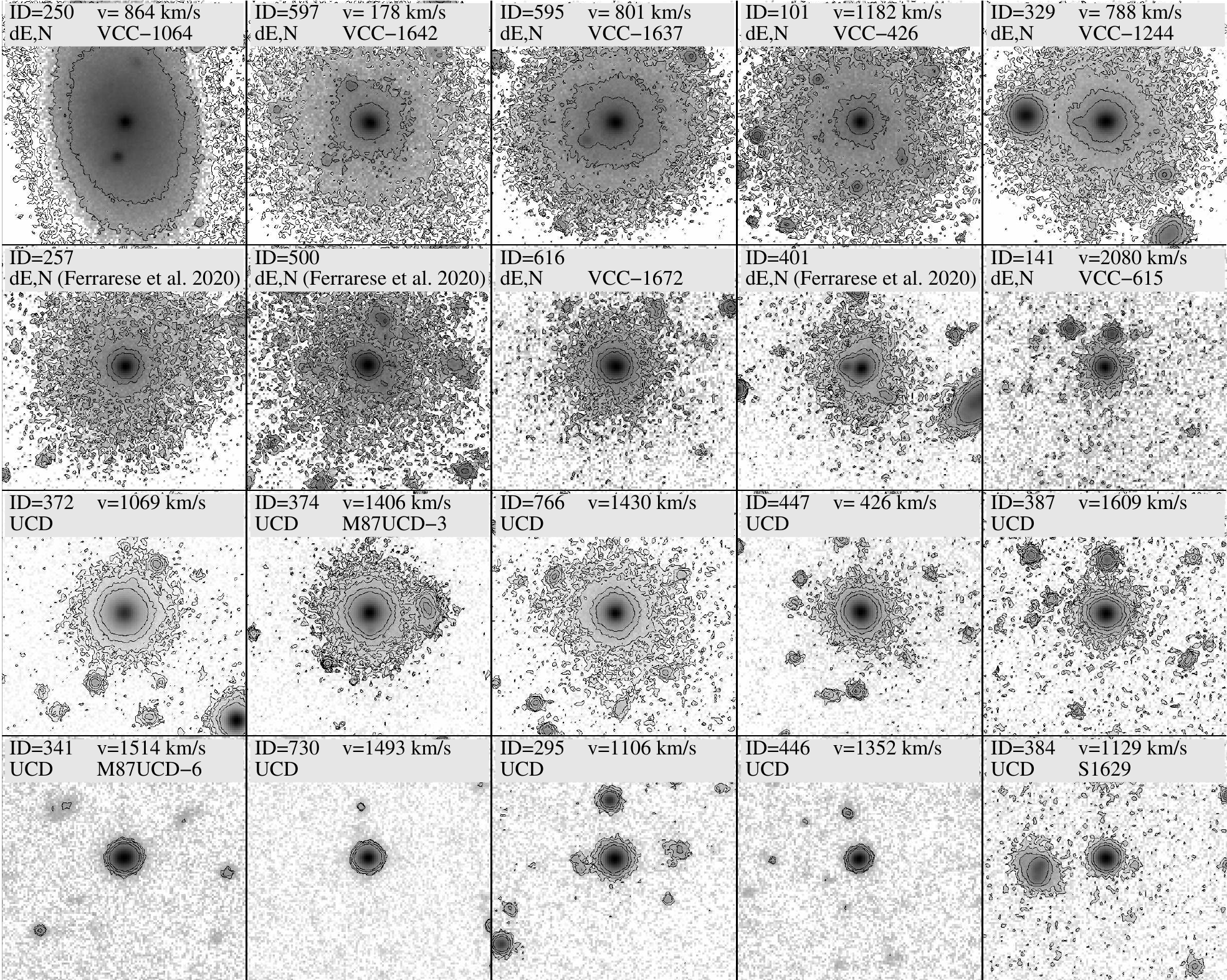}
\caption{Mosaic of $g$-band images for 20 objectively selected UCD candidates from the NGVS. The image size is $120\times120$ pixels (where $120$ pixels $=22.32$\arcsec~ $\sim1.8$ kpc in Virgo cluster). The upper row shows five nucleated dwarf galaxies (dE,N) from the VCC catalog \citep{1985AJ_90_1681Binggeli}. All of them are confirmed spectroscopic members of the Virgo cluster (i.e., $v_r <3500$ km/s). Another five dE,Ns shown in the second row include two VCC galaxies (ID=613 and 141) and three newly discovered galaxies from the NGVS (ID=257, 497 and 401; \citealt{2020ApJ_890_128Ferrarese}). Four of these five dE,Ns do not have radial velocity measurements, but they are likely cluster members given their extended, low surface brightness envelopes. The rightmost galaxy in this row (ID=141) is an ultra-diffuse galaxy (UDG) that has a large extent (larger than the figure size) and very diffuse structure \citep[e.g.,][]{2018ApJL_856_31Toloba}. The third row shows five UCDs with apparent envelopes while the bottom row shows another five UCDs that have no discernible envelope. All these ten UCDs are radial velocity members of the cluster ($v_r<3500$ km/s). The contours in each of the first 15 panels show the isophotes with constant surface brightness level. The innermost isophote is 25 mag/arcsec$^2$ and each interval between successive isophotes is 0.5 mag/arcsec$^2$. The 10 dE,Ns (the top two rows) have been sorted by the size of the stellar halo while the UCDs (the bottom two rows) by the envelope parameter, $\Delta_{\rm env}$.}
\label{fig:den2ucds}
\end{figure*}

We have selected candidate Virgo UCDs from a combination of ellipticity, magnitude, colors, surface brightness, half-light radii, visual inspection and, when available, radial velocity. At the outset, we imposed a requirement that ellipticity $e<0.3$ since most spectroscopically-confirmed UCDs are quite round \citep{2015ApJ_802_30Zhang}. We show the ellipticity distributions of our UCD samples in Figure~\ref{fig:ellipticity} to gauge the impact this has on our selection function. As can be seen, the ellipticity distributions peak at $\sim$0.05-0.07, and decrease to roughly zero by $e\approx0.3$. We conclude then that the criterion $e<0.3$ does not significantly bias our selection of UCDs.

\begin{figure*}
\epsscale{1.15}
\plotone{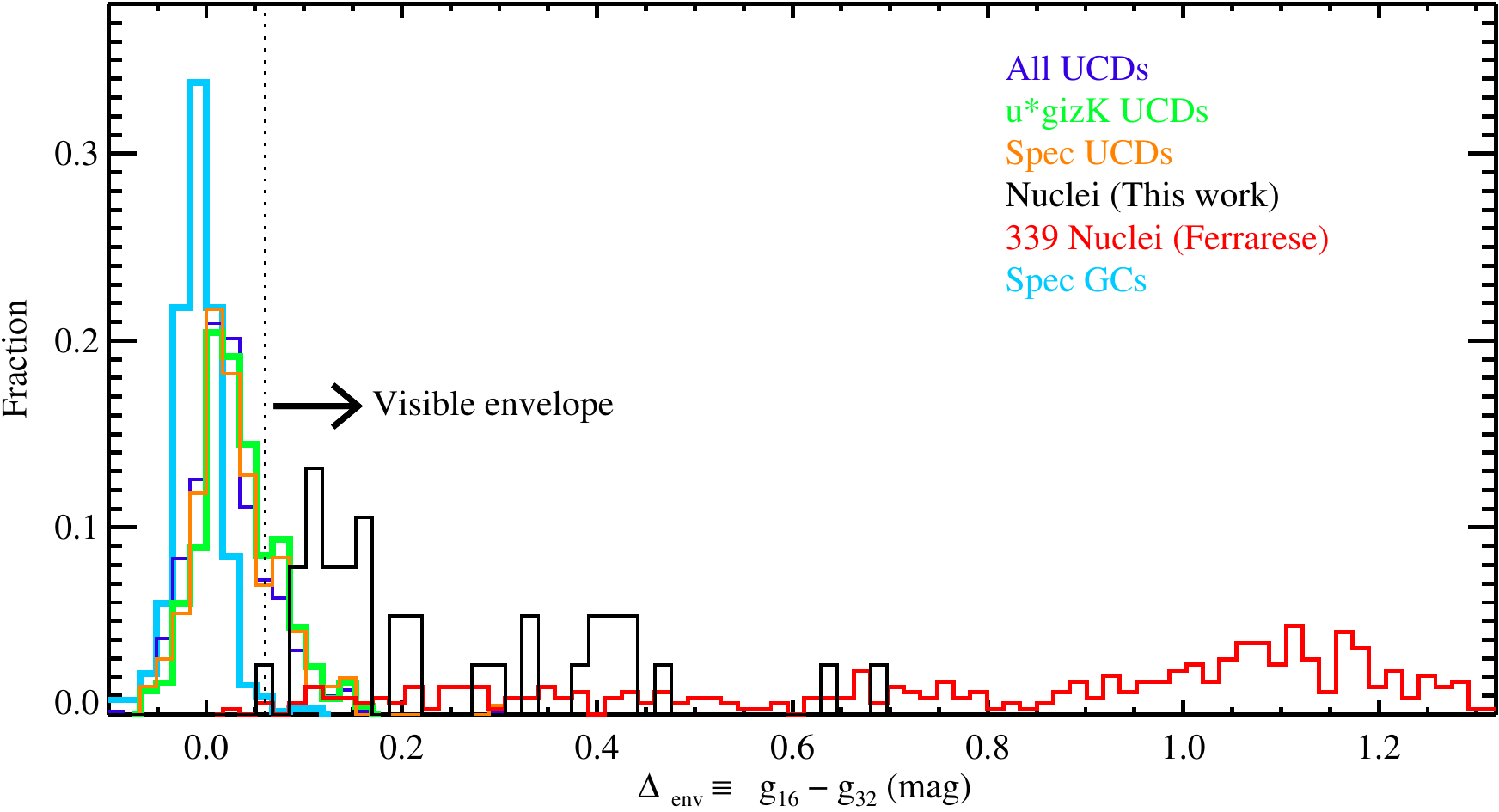}
\caption{The envelope parameter, $\Delta_{\rm env}$, distribution for the {\tt spec GCs} (cyan), {\tt all UCDs} (blue), {\tt $u^*gizK$ UCDs} (green), {\tt spec UCDs} (orange), {\tt $u^*gizK$ nuclei} (black) and {\tt bright nuclei} (red). We have divided the UCDs into two subgroups at $\Delta_{\rm env}=0.06$, shown by the dotted vertical line.}
\label{fig:distribution_env}
\end{figure*}

Our selection yields a catalog of more than 600 candidates within the $\sim$100 deg$^{2}$ footprint of the NGVS, making it the largest and most homogeneous UCD catalog for any environment to date. Moreover, our selection algorithm also produces samples of {\tt bright GCs} and galactic nuclei, such that we can compare directly the properties of these different populations.

Our large and complete sample also makes it possible to identify groups of UCDs with extreme or interesting properties, which may shed light on UCD origins in general. In this section, we examine our results using both approaches and discuss the implications for models of UCD formation. 

We begin by noting that the nuclei of dwarf galaxies are embedded in stellar envelopes that can vary widely in surface brightness \citep[see, e.g.,][]{2006ApJS_164_334Ferrarese, 2017ApJ_849_55Spengler, 2018ApJL_856_31Toloba}. In other words, when we visually classify our UCD candidates, it can be difficult to distinguish UCDs with faint envelopes from nuclei in low-mass galaxies. To solve this, we remove from our samples any objects that have been classified as galaxies in either the VCC \citep{1985AJ_90_1681Binggeli} or the NGVS galaxy catalogs \citep{2016ApJ_824_10Ferrarese, 2020ApJ_890_128Ferrarese}.

Figure~\ref{fig:den2ucds} shows a mosaic of NGVS $g$-band images for representative sub-samples of ten dE,Ns and ten UCDs. Most of these objects are confirmed radial velocity members of Virgo. Note that this mosaic is an updated version of Figure 32 in \citet{2015ApJ_812_34Liu}, which was based on just $\sim100$ UCD candidates near M87, and that the dE,Ns have been sorted by the prominence of their stellar envelopes. We can see that the envelopes of the first few dwarfs are significant, making it easy to identify these systems as galaxies. However, as the envelopes become progressively smaller and fainter, the distinction between dE,Ns and UCDs becomes blurred. For instance, it is difficult to distinguish the dE,Ns in the second row of Figure~\ref{fig:den2ucds} from the UCDs in the third row, based on morphology alone. Note that the systems in the second row are classified as galaxies in either the VCC \citep[ID = 613 and 141, ][]{1985AJ_90_1681Binggeli} or NGVS galaxy catalogs\footnote{Of course, in both of these catalogs, galaxy classifications were based on several properties, not just morphology.} \citep{2016ApJ_824_10Ferrarese, 2019ApJ_878_18Sanchez-Janssen, 2020ApJ_890_128Ferrarese}. Finally, as the bottom row of Figure~\ref{fig:den2ucds} illustrates, most of our UCD candidates do not have visible envelopes, or they are too diffuse to detect with NGVS imaging.

Figure~\ref{fig:den2ucds} prompts an obvious question: might there be an evolutionary sequence that links dE,N galaxies to UCDs, with the latter representing the end-state of severe and continuous tidal stripping? Such a link was suggested by \citet{2015ApJ_812_34Liu} who examined the distribution of dwarf galaxies and UCDs in the cores of the Virgo A and B sub-clusters. Armed with a cluster-wide sample of UCDs and galaxies, we can now revisit this claim. \citet{2012MNRAS_419_2063Bekki} found from their numerical simulations a morphological sequence similar to that shown in Figure~\ref{fig:den2ucds} (see the lower panel of their Figure~2). During the transformation from a dwarf galaxy to a bare nucleus, they showed that the stellar halo is stripped efficiently and decreases in size steadily over time. In other words, the physical extent of a given UCD envelope may indicate which stage the system falls along the evolutionary sequence from dE,N to UCD.

\begin{figure*}
\epsscale{1.165}
\plotone{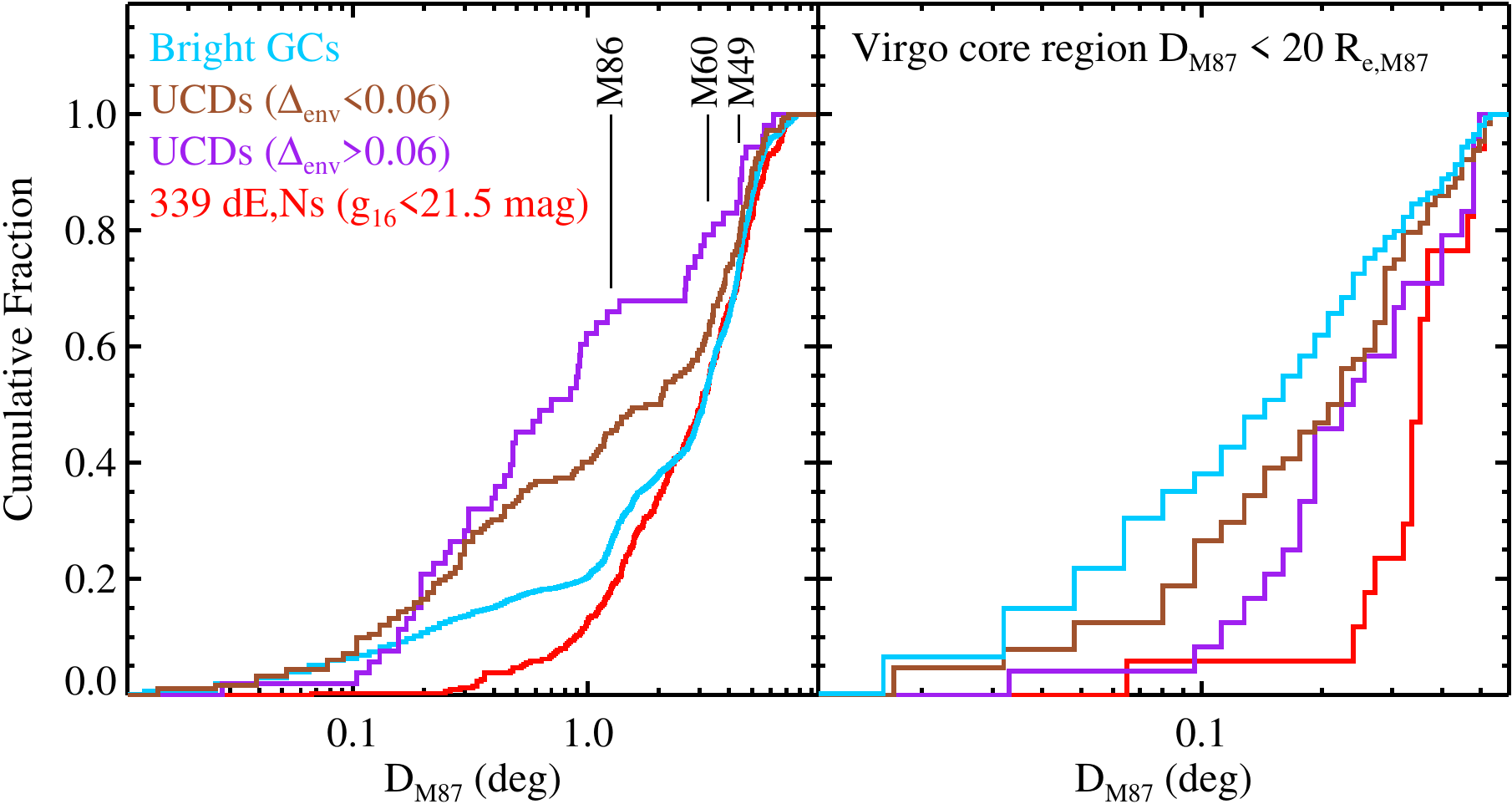}
\caption{Cumulative distributions for the projected distance from M87 ($D_{\rm M87}$) for the {\tt bright GCs} (cyan lines), UCDs with $\Delta_{\rm env}<0.06$ (brown lines), UCDs with $\Delta_{\rm env}>0.06$ (purple lines) and {\tt bright nuclei} (red lines). {\it Left panel:} the distribution for the entire NGVS survey area; {\it Right panel:} the distribution for the Virgo core region (within 20 $R_{e,\rm M87}$). The three black vertical lines in the left panel show the location of M86, M60 and M49.}
\label{fig:radial_distribution}
\end{figure*}

To further explore such a connection, \citet{2015ApJ_812_34Liu} introduced a parameter to describe the strength of a UCD's envelope. This parameter is defined as $\Delta_{\rm env} \equiv g_{16}-g_{32}$, where $g_{16}$ and $g_{32}$ are $g$-band magnitudes measured within apertures of 16 ($\sim$3.0\arcsec) and 32 pixels ($\sim$6.0\arcsec) diameter, respectively. Based on this definition, point-like sources should have $\Delta_{\rm env} \simeq 0$ while extended sources will tend to have $\Delta_{\rm env}>0$.

Figure \ref{fig:distribution_env} shows the distribution of envelope parameters for compact objects in the NGVS. The GCs are closely distributed around $\Delta_{\rm env}\sim0$, with a small tail to $\Delta_{\rm env} > 0.06$ (marked by the dotted vertical line). The low envelope strengths of GCs are to be expected given the vast majority of them are unresolved in NGVS imaging. Meanwhile, the dE,Ns show much larger envelope strengths ($\Delta_{\rm env} > 0.06$). For UCDs, the envelope strengths typically fall between those of GCs and dE,Ns. Following \citet{2015ApJ_812_34Liu}, we subdivide the UCDs into two groups at $\Delta_{\rm env}=0.06$. UCDs with $\Delta_{\rm env} > 0.06$, whose envelopes are clearly evident, more closely resemble nuclei and are thus distinct from GCs.

Previous studies have found dramatic differences in the cumulative radial distributions of UCDs and dE,Ns in galaxy clusters \citep[e.g.,][]{2002IAUS_207_287Drinkwater, 2004PASA_21_375Drinkwater, 2004AJ_128_1529Mieske, 2006AJ_131_312Jones, 2007A+A_472_111Mieske, 2015ApJ_812_34Liu, 2016MNRAS_458_2492Pfeffer}, but the situation between UCDs and GCs is less clear \citep[e.g.][]{2004A+A_418_445Mieske, 2012A+A_537_3Mieske}. Generally, GCs are highly concentrated towards cluster centers, while dE,Ns are less concentrated, and UCDs lie intermediate between the two. If the majority of UCDs are nothing more than the high luminosity tail of the GC population, then we might reasonably expect the radial distributions of GCs and UCDs to resemble each other. On the other hand, if most UCDs form through tidal stripping of dE,Ns, then we should expect the UCD radial profile to have a high central concentration: i.e., stripped galaxies would tend to lie deeper in the gravitational wells of their hosts. As noted in \S\ref{sec:results}, our UCD sample most certainly contains some number of GCs because bright, extended GCs (although rare) do exist and we systematically overestimate the sizes of GCs. We can, however, invoke $\Delta_{\rm env}$ to produce a UCD sample that is more heavily weighted to objects born of tidal stripping.

\begin{figure*}
\epsscale{1.17}
\plotone{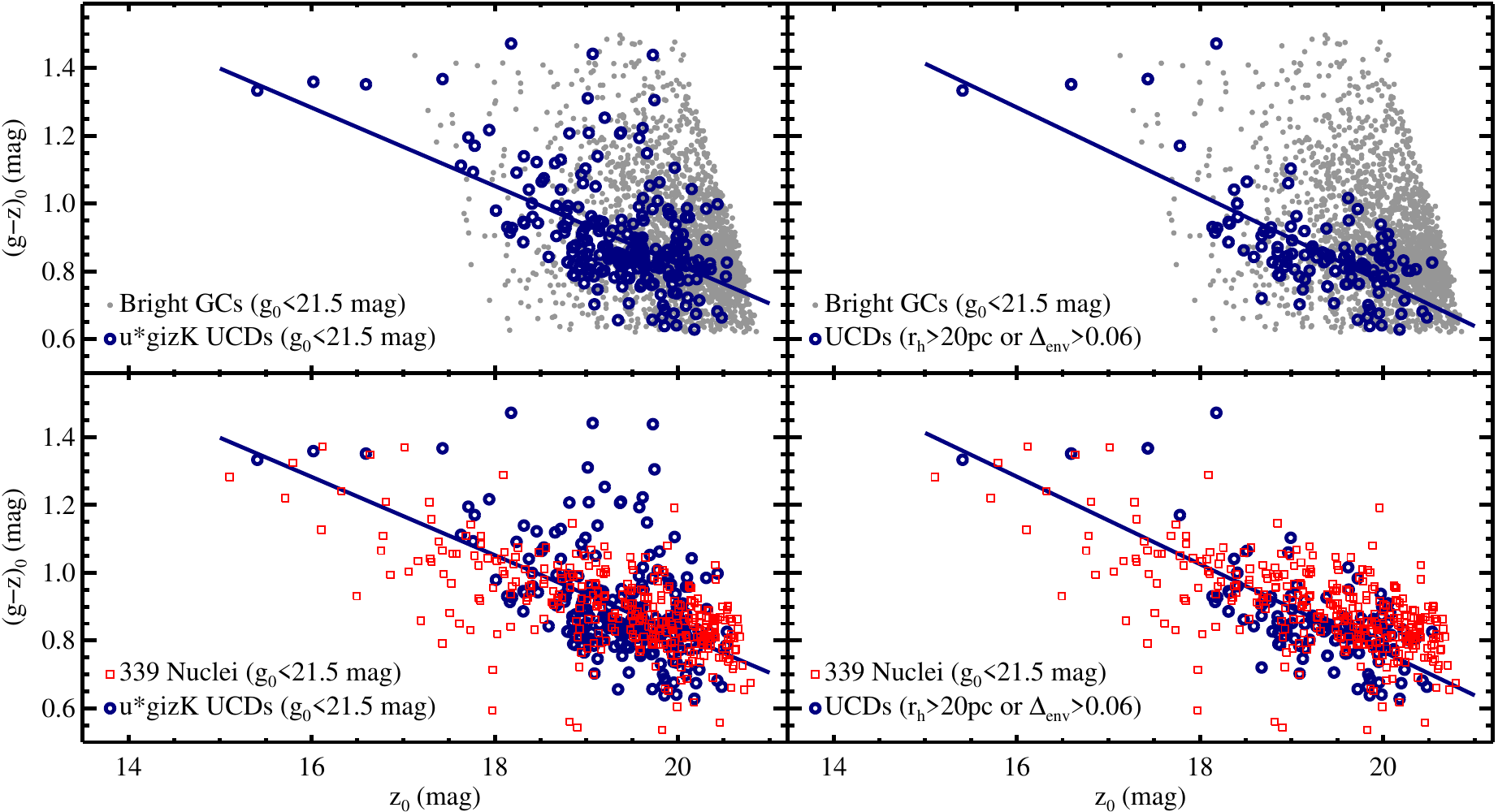}
\caption{Color-magnitude diagram (CMD) for {\tt bright GCs} (gray points), UCDs (blue open circles) and {\tt bright nuclei} (red open squares). \textit{Upper-left panel:} CMDs for {\tt bright GCs} and {\tt $u^*gizK$ UCDs}; \textit{Lower-left panel:} CMDs for {\tt bright nuclei} and {\tt $u^*gizK$ UCDs}; \textit{Upper-right panel:} CMDs for {\tt bright GCs} and UCDs with $r_h<20$ pc or $\Delta_{\rm env}>0.06$; \textit{Lower-right panel:} CMDs for {\tt bright nuclei} and UCDs with $r_h<20$ pc or $\Delta_{\rm env}>0.06$. The blue solid lines in the left two panels are the best linear fit for {\tt $u^*gizK$ UCDs}. The blue solid lines in the right two panels are the best linear fit for UCDs with $r_h<20$ pc or $\Delta_{\rm env}>0.06$.}
\label{fig:cmd_largeucd}
\end{figure*}

The right panel of Figure~\ref{fig:radial_distribution} shows the cumulative radial distribution of UCDs, {\tt bright GCs} (cyan lines), and {\tt bright nuclei} (red lines) in the core of sub-cluster A ($D_{\rm M87} < 20 R_{e,\rm M87}$). Here, we separate the UCDs into two groups: those without envelopes ($\Delta_{\rm env} \le 0.06$; brown line) and those with ($\Delta_{\rm env} > 0.06$; purple line). Consistent with previous studies, the ordered sequence of systems from high to low concentration in this region goes: GCs, UCDs without envelopes, UCDs with envelopes, and dE,Ns. As expected, the UCDs with envelopes are less centrally concentrated than those without. The difference in radial distributions between GCs and UCDs with $\Delta_{\rm env} > 0.06$ is clear.

In the left panel of Figure~\ref{fig:radial_distribution}, we show, for the first time, the radial distribution of UCDs, GCs, and nuclei over the entire Virgo cluster. In the outer regions, the distributions of {\tt bright GCs} and dE,Ns grow at similar rates. This makes sense since, at large cluster-centric distances, GCs will be found around the individual galaxy members. On the other hand, there is a clear difference in how bright GCs and UCDs with envelopes are distributed in the outskirts of the cluster. Most notably, the distribution of UCDs with envelopes exhibits clear bumps that coincide with the locations of the three luminous galaxies, M86, M60 and M49 (marked by the black vertical lines). This indicates that the UCDs with envelopes are mainly associated with giant galaxies, where the gravitational potential is deep enough to strip the diffuse components of dE,Ns. 

Another way we can attempt to isolate those UCDs formed via tidal stripping is by increasing the value of our size cut to $r_h > 20$ pc. In the right-hand panels of Figure~\ref{fig:cmd_largeucd}, we examine the color-magnitude relation of UCDs with $\Delta_{\rm env}>0.06$ or $r_h > 20$ pc and compare to those for the {\tt bright GCs} (upper-right panel) and {\tt bright nuclei} (lower-right panel). Meanwhile, the left-hand panels show a similar comparison, but using our {\tt $u^*gizK$ UCDs} sample instead. Placing more restrictive cuts on $r_h$ and $\Delta_{\rm env}$ has the effect of removing many red UCDs at faint luminosities ($z_0 \gtrsim 18.5$ mag). The color-magnitude relation for the {\tt $u^*gizK$ UCDs} sample is
\begin{equation}
(g-z)_0 = -0.12(\pm0.01)z_0+3.13(\pm0.05),
\end{equation}
while that for our more restricted UCD sample is,
\begin{equation}
(g-z)_0 = -0.13(\pm0.01)z_0+3.35(\pm0.04),
\end{equation}
which are shown as blue solid lines in Figure~\ref{fig:cmd_largeucd}.

At face value, it looks like the distributions of UCDs and nuclei overlap well in the color-magnitude plane. This compares well with previous work \citep[e.g.,][]{2006ApJS_165_57Cote,2011AJ_142_199Brodie} which found general agreement between the color-magnitude relations of nuclei and UCDs, in the sense that brighter UCDs and nuclei tend to be redder. However, as seen in lower-right panel of Figure~\ref{fig:cmd_largeucd}, there is a lack of blue UCDs at bright magnitudes ($z_0 \lesssim 18.0$) and that UCDs are bluer than nuclei at faint magnitudes ($z_0 \gtrsim 18.0$). To evaluate the significance of these differences, we have run a 2D Kolmogorov-Smirnov Test \citep{1983MNRAS_202_615Peacock, 1987MNRAS_225_155Fasano} on the distributions for UCDs and nuclei. The resulting $p$-values are quite small, implying that the distributions do not share a common parent. Also, in agreement with previous work \citep[e.g.,][]{2011AJ_142_199Brodie}, we find that UCDs occupy a narrower range of color than GCs, especially at faint magnitudes. The color-magnitude relations of UCDs, nuclei and GCs are thought to be the result of an underlying mass-metallicity relation for each population, with more massive systems tending to have higher metallicities, and GCs tending to have higher metallicities than UCDs and nuclei at a given mass (especially at low masses; \citealt{2018ApJ_858_37Zhang}).

As described in the introduction, recent studies are finding more support for the galactic nuclei origin for UCDs. For UCDs formed this way, one would expect to find some transition objects: e.g., UCDs with tidal streams. However, only a small number of UCDs have been found to show such tidal structures \citep[e.g., Figure~\ref{fig:asymmetry},][]{2015ApJL_812_10Jennings, 2015ApJL_809_21Mihos, 2016A+A_586_102Voggel, 2018ApJ_853_54Schweizer}. On the other hand, many more UCDs were found to show diffuse, but circular, envelopes as shown in Figure~\ref{fig:ucd_large_env}, Figure~\ref{fig:den2ucds} and also previous studies \citep[e.g.,][]{2003Natur_423_519Drinkwater, 2005ApJ_627_203Hacsegan}. \citet{2012MNRAS_419_2063Bekki} have established that diffuse and circular envelopes can indeed be found around nuclei when the stellar halos of dwarf galaxies have been mostly stripped by tidal forces. \citet{2013MNRAS_433_1997Pfeffer} have also demonstrated that while the stellar halos of dwarf galaxies are reduced during the tidal stripping, they can still be visible at later stages, especially for the more massive and extended UCDs. By contrast, tidal streams are often much fainter and only can be observable for those UCDs that are still experiencing significant stripping. 

Two decades have passed since the discovery of UCDs, and yet, we still lack an understanding of what fraction of them originate through the formation of GC systems versus tidal stripping. One of the chief reasons for this is the difficulty of identifying a given compact stellar system as a bare nuclear star cluster based on its integrated light. More locally though, following years of study using high-resolution spectroscopy and resolved photometry, we now know that the GC system of the MW includes several unusual objects. These anomalous GCs tend to have very high surface mass densities, large intrinsic metallicity dispersions, spreads in the abundances of s-process elements, complex sub-giant branches, kinematic subpopulations, and tidal streams (see the review of \citealt{2019A+ARv_27_8Gratton}, and references therein). Many of these (massive) GCs are thought to be remnants of nucleated dwarf galaxies (e.g., M54 \citep{1994Natur_370_194Ibata}, $\omega$ Cen \citep{2003MNRAS_346_11Bekki}), and thus can be considered local examples of UCDs.

Beyond the Local Group, however, there is no consensus definition for what is a UCD. Investigators typically base their selection on arbitrary size ($10 \lesssim r_h \lesssim 100$) and luminosity/mass cuts ($10^6 \lesssim M_* \lesssim 10^8$ M$_{\odot}$), or simply observational limits (e.g., $r_h>11$ pc in this study). As a result, it is very challenging to isolate within current UCD samples those that are GC-like from those that are nuclei-like, as discussed by \citet{2011EAS_48_219Hilker}. Nevertheless, recent work makes it clear that many UCDs (e.g., massive UCDs, UCDs with tidal structures or diffuse envelopes) are indeed the nuclei of stripped dwarfs. Moreover, based on the \citet{2011MNRAS_413_101Guo} semi-analytic model, \citet{2014MNRAS_444_3670Pfeffer} have demonstrated that both massive GCs ($M_* \gtrsim 10^5$ M$_{\odot}$) and UCDs can form via tidal disruption of dwarf galaxies. Also, Mayes et al. (in prep.) use the EAGLE simulation suite \citep{2015MNRAS_450_1937Crain} to show that there is an overlap between stripped nuclei and ``normal" GCs in the stellar mass range $M_* \lesssim 2 \times 10^6$ M$_{\odot}$.

To date, there are $\sim$10$^3$ known UCDs in the local universe, more than half of which were found on the basis of photometric data alone. Obtaining spectroscopic observations for larger samples of UCDs (to measure radial velocities, velocity dispersions, and stellar populations) will be essential for understanding their nature. Fortunately, the next generation of ground- and space-based observatories promise to ameliorate this current deficit.

\section{Summary and Conclusions}
\label{sec:summary}

Using deep, wide-field $u^*,g,i,z$ imaging from the Next Generation Virgo Cluster Survey (NGVS) and $K$ band data from the UKIDSS, we have carried out a systematic search for UCDs across the entire Virgo cluster ($\sim 104$ deg$^2$). We describe our search methodology --- which is based on a combination of photometric (magnitude and color) information, half-light radius and surface brightness measurements, and radial velocity measurements, when available --- and present a sample of 612 UCD candidates. Among this UCD sample are 235 candidates selected on the basis of deep $u^*gizK$ data (our highest purity subsample) and 203 UCDs that are confirmed radial velocity members of the cluster (i.e., $v_r<3500$ km/s). This is the largest and most homogeneous sample of UCDs presented to date for any cluster environment, and the first of its kind for the Virgo cluster. Our principal findings can be summarized as follows: 

\begin{enumerate}
\item We construct the first number density map for UCDs in the Virgo Cluster, and show that UCDs are highly concentrated towards the largest and brightest galaxies: e.g., M87, M49, M60-M59 and M86 \citep[see also][]{2015ApJ_812_34Liu}. 
\item The UCDs, as a population, have bimodal color distributions. The fraction of UCDs belonging to the blue population is $89\% \pm 3\%$. This is slightly higher than that of {\tt bright GCs} ($84\% \pm 1\%$ for NGVS GCs and $77\% \pm 4\%$ for {\tt ACSVCS GCs}) and slightly smaller than that of {\tt  bright nuclei} ($95\%$).
\item We measure the mean half-light radius for UCDs to be $19.8 \pm 6.8$ pc. The blue UCDs ($20.0 \pm 6.8$ pc) are systematically larger than their red counterparts ($14.6 \pm 3.8$ pc). The largest UCD candidate in our sample is NGVS-UCD769 with $r_h = 58.0$ pc.
\item Based on our analysis (i.e., number density maps and color distributions), we find no dramatic differences between UCDs and the brightest GCs (i.e., those objects with $g_0 < 21.5$ mag). However, when we rely on the cleanest possible UCD sample (with reduced contamination from GCs), some differences begin to appear (i.e., in their cumulative radial distributions and color magnitude relations).
\item We identify a number of UCDs having properties that point to a connection with the nuclei of dwarf galaxies. This includes the most luminous and largest UCDs, UCDs with obvious stellar envelopes, and UCDs embedded in diffuse asymmetric structures.
\item There are tight color-magnitude relations for UCDs and dwarf nuclei, with brighter objects being redder. At the faint end, UCDs and nuclei are bluer and have a narrower color range than GCs.
\end{enumerate}

Some obvious extensions of this work present themselves, most of which involve spectroscopic observations, e.g., the property of envelopes of UCDs, searching for SMBHs in massive UCDs, the specific frequencies for the UCDs around massive galaxies. Our radial velocity survey for UCD candidates brighter than $g_0 \sim 19.5$ mag is complete, allowing membership to be established for candidates brighter than $M_g \sim -12$, which corresponds to a stellar mass of $\sim10^{6.9}$~M$_{\odot}$ \citep{2003ApJS_149_289Bell}. It will be valuable to extend this work to the limit of our photometric catalog ($g\sim21.5$ mag) and thus obtain a complete sample of UCDs down to a stellar mass of $\sim10^{6.1}$~M$_{\odot}$. AO-assisted IFU spectroscopy for select UCDs (i.e., the brightest and largest objects, or those objects embedded in stellar envelopes) will allow the detection of SMBHs in these objects and provide a first glimpse into the SMBH occupation fraction in a magnitude-limited UCD sample. Finally, while the NGVS has made it possible to identify UCDs larger than $r_h \sim 11$~pc throughout the Virgo cluster, space-quality imaging will be needed to extend this work to the smaller radii and fainter magnitudes typical of GCs; in the future, high-resolution imaging from the Euclid or Roman space telescopes will allow the GC/UCD size-magnitude relation(s) to be mapped roughly to the level of the GCLF turnover using spatially complete samples and unbiased structural measurements.

\acknowledgments

The NGVS team owes a debt of gratitude to the director and the staff of the Canada France Hawaii Telescope who helped make the survey a reality. This work is based on observations obtained with MegaPrime/MegaCam, a joint project of CFHT and CEA/DAPNIA, at the Canada France Hawaii Telescope (CFHT) which is operated by the National Research Council (NRC) of Canada, the Institut National des Sciences de Univers of the Centre National de la Recherche Scientifique (CNRS) of France, and the University of Hawaii. This work is based in part on data products produced at Terapix available at the Canadian Astronomy Data Centre as part of the Canada-France-Hawaii Telescope Legacy Survey, a collaborative project of NRC and CNRS.

C.L. acknowledges support from the National Natural Science Foundation of China (NSFC, Grant No. 11673017, 11833005, 11933003, 11621303, 11973033 and 11203017). C.L. is supported by Key Laboratory for Particle Physics,Astrophysics and Cosmology, Ministry of Education, and Shanghai Key Laboratory for Particle Physics and Cosmology(SKLPPC). This work is supported by 111 project No. B20019. EWP acknowledges support from the National Natural Science Foundation of China through Grant No.\ 11573002. HXZ acknowledges a support from the CAS Pioneer Hundred Talents Program and the National Natural Science Foundation of China (NSFC, Grant No. 11421303, 11973039). AL acknowledges support from the French Centre National d'Etudes Spatiales (CNES). EWP thanks Karina Voggel for discussions in which she suggested the use of Gaia data in UCD selection.

This work was supported in part by the Sino-French LIA-Origins joint exchange program, by the French Agence Nationale de la Recherche (ANR) Grant Programme Blanc VIRAGE (ANR10-BLANC-0506-01), and by the Canadian Advanced Network for Astronomical Research (CANFAR) which has been made possible by funding from CANARIE under the Network-Enabled Platforms program. This research used the facilities of the Canadian Astronomy Data Centre operated by the National Research Council of Canada with the support of the Canadian Space Agency. This research has made use of the NASA/IPAC Extragalactic Database (NED), which is funded by the National Aeronautics and Space Administration and operated by the California Institute of Technology. This research has made use of the SIMBAD database, operated at CDS, Strasbourg, France. This publication has made use of data products from the Sloan Digital Sky Survey (SDSS). Funding for SDSS and SDSS-II has been provided by the Alfred P. Sloan Foundation, the Participating Institutions, the National Science Foundation, the U.S. Department of Energy, the National Aeronautics and Space Administration, the Japanese Monbukagakusho, the Max Planck Society, and the Higher Education Funding Council for England. This work is based in part on data obtained as part of the UKIRT Infrared Deep Sky Survey.

This research uses data obtained through the Telescope Access Program (TAP), which has been funded by the National Astronomical Observatories, Chinese Academy of Sciences, and the Special Fund for Astronomy from the Ministry of Finance. Observations reported here were obtained at the MMT Observatory, a joint facility of the University of Arizona and the Smithsonian Institution.

Facility: CFHT - Canada-France-Hawaii Telescope.

\appendix{

As described in \S\ref{sec:ucd_selection}, we use strict half-light radius criteria to select UCD candidates, including: (1) a radius cut ($11 < \langle{r_h}\rangle < 100 \rm ~pc$); (2) a requirement that the half-light radii measured in the $g$ and $i$ bands are in rough agreement ($|r_{h,g}-r_{h,i}|/\langle{r_h}\rangle \le 0.5$); and (3) a condition that the fraction radius errors are smaller than 15\% in both bands ($r_{h,g,\rm error}/r_{h,g} \le 15\%$ and $r_{h,i,\rm error}/r_{h,i} \le 15\%$). In this section, we take a closer look at objects that do not satisfy our radius criteria (but satisfy all other selection criteria). We divide such objects into three groups: objects with larger errors; objects with larger differences between two bands and; objects with smaller radius measurements. 

{\centering
\begin{deluxetable}{l|rrrrrrrcc}
 \tablewidth{0pt}
 \tabletypesize{\small}
 \tablecaption{The probable UCD candidates not included in the main sample.
 \label{tab:ucd_9to11pc}}
 \tablehead{
  \colhead{} &
  \colhead{ID} &
  \colhead{$\alpha_{\rm J2000}$} &
  \colhead{$\delta_{\rm J2000}$} &
  \colhead{$g_0$} &
  \colhead{$r_{h,g}$} &
  \colhead{$r_{h,i}$} &
  \colhead{$v_r$} &
  \colhead{$v_{\rm source}$}  \\  \vspace{-0.6cm}\\
  \colhead{ } &
  \colhead{ } &
  \colhead{(deg)} &
  \colhead{(deg)} &
  \colhead{(mag)} &
  \colhead{(pc)} &
  \colhead{(pc)} &
  \colhead{(km/s)} &
  \colhead{ }  \\ \vspace{-0.6cm}\\
  \colhead{} &
  \colhead{(1)} &
  \colhead{(2)} &
  \colhead{(3)} &
  \colhead{(4)} &
  \colhead{(5)} &
  \colhead{(6)} &
  \colhead{(7)} &
  \colhead{(8)} }
 \startdata
    \vspace{-0.35cm}  \\              
                & 1 & 186.5656471 & 16.0492075 &    19.857$\pm$0.001 &      17.75$\pm$1.37 &      35.32$\pm$1.70 &   \nodata &                  \nodata \\
 $r_h>11$ pc    & 2 & 185.1411400 & 15.8494361 &    21.548$\pm$0.005 &      15.95$\pm$2.05 &      28.76$\pm$1.21 &   \nodata &                  \nodata \\
 in both $g$    & 3 & 188.7565072 & 17.0687625 &    21.094$\pm$0.003 &      15.13$\pm$1.37 &      28.27$\pm$0.89 &   \nodata &                  \nodata \\
 and $i$ bands  & 4 & 188.2201250 & 16.0768933 &    21.431$\pm$0.006 &      13.32$\pm$1.06 &      28.21$\pm$1.73 &   \nodata &                  \nodata \\
 but large      & 5 & 190.6792998 & 16.5254659 &    21.468$\pm$0.005 &      11.78$\pm$0.70 &      25.20$\pm$0.76 &   \nodata &                  \nodata \\
 difference     & 6 & 186.1282677 & 16.7516812 &    21.528$\pm$0.004 &      16.96$\pm$0.55 &      28.31$\pm$1.00 &   \nodata &                  \nodata \\
 between the    & 7 & 188.3270896 & 16.7757418 &    21.418$\pm$0.004 &      11.26$\pm$0.70 &      22.63$\pm$0.60 &   \nodata &                  \nodata \\
 two            & 8 & 187.8761745 & 17.1490799 &    20.986$\pm$0.003 &      20.89$\pm$0.58 &      12.22$\pm$0.83 &   \nodata &                  \nodata \\
                & 9 & 188.7443865 & 17.1551750 &    20.584$\pm$0.002 &      11.69$\pm$0.54 &      24.27$\pm$1.09 &   \nodata &                  \nodata \\   
     \vspace{-0.35cm}  \\
     \hline 
     \vspace{-0.35cm}  \\
                & 1 & 187.2897173 & 12.6860435 &    19.596$\pm$0.001 &      33.40$\pm$0.95 &       0.61$\pm$0.00 &      1159 &MMT09                     \\
                & 2 & 188.8007729 &  9.3776914 &    19.531$\pm$0.001 &       2.84$\pm$1.59 &      13.63$\pm$1.10 &      1406 &AAT12                     \\
 Candidates     & 3 & 187.7426763 & 11.9746280 &    20.694$\pm$0.002 &      10.76$\pm$0.40 &      10.09$\pm$0.49 &      1230 &MMT09                     \\
 with visible   & 4 & 187.1924549 & 13.7198340 &    19.743$\pm$0.001 &      11.69$\pm$1.04 &       9.75$\pm$0.65 &      1022 &AAT12                     \\
 envelopes      & 5 & 188.3663498 & 10.6465580 &    20.490$\pm$0.002 &       9.95$\pm$0.24 &      11.13$\pm$0.45 &   \nodata &                  \nodata \\
                & 6 & 187.6404154 & 10.3641505 &    19.421$\pm$0.001 &      10.45$\pm$0.31 &       7.80$\pm$0.43 &       143 &MMT09,AAT12               \\
                & 7 & 187.1115755 & 13.0907153 &    19.996$\pm$0.001 &      10.06$\pm$0.31 &       8.54$\pm$0.33 &      1040 &MMT09,AAT12               \\
                & 8 & 187.4479438 &  9.8971556 &    20.915$\pm$0.002 &      10.16$\pm$0.55 &       8.44$\pm$0.40 &   \nodata &                  \nodata \\
\enddata
\tablenotetext{}{NOTE-- (1) Object ID number; (2) R.A.; (3) Decl.; (4) Aperture-corrected g magnitude within a 3-arcsec diameter aperture; (5) Half-light radius in $g$ band; (6) Half-light radius in $i$ band; (7) Radial velocity; (8) The source of velocity measurement: AAT12: Anglo-Australian Telescope (AAT) 2012 program; MMT09: Multiple Mirror Telescope (MMT) 2009 program \citep{2012ApJS_200_4Ferrarese}.}
\end{deluxetable}
}

{\bf Objects with larger errors} (i.e. $r_{h,g,\rm error}/r_{h,g} > 15\%$ or/and $r_{h,i,\rm error}/r_{h,i} > 15\%$): We visually inspected the imaging for objects with larger errors for their radius measurements. Most of these objects are blends or have poor image quality (i.e., sources located close to chips gaps or near saturated objects). We can not classify these objects as bonafide UCDs using the NGVS images alone.

{\bf Objects with larger differences} (i.e. $|r_{h,g}-r_{h,i}|/\langle{r_h}\rangle > 0.5$): The differences in image quality (PSF) is the primary reason for large differences in radius measurements. If an object is larger than 11 pc in both the $g$ and $i$ bands, then we believe it be may indeed be a UCD candidate although the radius difference between two bands is large. We list nine such objects in Table~\ref{tab:ucd_9to11pc}. 

{\bf Objects with smaller radius measurements} (i.e. $\langle{r_h}\rangle < 11 \rm ~pc$): Most objects with smaller measured radii are likely to be GCs. However, if they have visible envelopes, then they are viable UCD candidates. We find eight objects that have half-light radii slightly below our $r_h$ = 11 pc limit but appear to show diffuse envelopes. These are listed in Table \ref{tab:ucd_9to11pc}.

To ensure our samples are as homogeneous as possible, we have not used these 17 UCD candidates in our analysis but they are included here for completeness.

}

\bibliographystyle{aasjournal}

\clearpage
 
\setcounter{table}{2}
\begin{longrotatetable}
{\centering
\begin{deluxetable*}{lllcrrcrcrrrrc}
 \tablewidth{0pt}
 \tabletypesize{\tiny}
 \tablecaption{Photometric Properties of UCD Candidates
 \label{tab:ucd_photometry}}
 \tablehead{
  \colhead{ID} &
  \colhead{Name} &
  \colhead{NGVSID} &
  \colhead{$t_{\rm obs}$} &
  \colhead{$\alpha_{\rm J2000}$} &
  \colhead{$\delta_{\rm J2000}$} &
  \colhead{$E(B-V)$} &
  \colhead{$g_0$} &
  \colhead{$K$} &
  \colhead{$(u^*-g)_0$} &
  \colhead{$(g-i)_0$} &
  \colhead{$(g-z)_0$} &
  \colhead{$\Delta_{\rm env}$} &
  \colhead{UCD}  \\ \vspace{-0.5cm}\\
  \colhead{ } &
  \colhead{ } &
  \colhead{ } &
  \colhead{ } &
  \colhead{(deg)} &
  \colhead{(deg)} &
  \colhead{(mag)} &
  \colhead{(mag)} &
  \colhead{(mag)} &
  \colhead{(mag)} &
  \colhead{(mag)} &
  \colhead{(mag)} &
  \colhead{(mag)} &
  \colhead{ }  \\ \vspace{-0.5cm}\\
  \colhead{(1)} &
  \colhead{(2)} &
  \colhead{(3)} &
  \colhead{(4)} &
  \colhead{(5)} &
  \colhead{(6)} &
  \colhead{(7)} &
  \colhead{(8)} &
  \colhead{(9)} &
  \colhead{(10)} &
  \colhead{(11)} &
  \colhead{(12)} &
  \colhead{(13)} &
  \colhead{(14)}}
 \startdata
  1 &    NGVS-UCD1   & NGVS-J120652.65+113246.5 &    l & 181.7193703 & 11.5462618 & 0.028 & 19.212$\pm$0.002 &             \nodata &     0.554$\pm$0.004 &    -0.034$\pm$0.004 & -0.034$\pm$0.006 &  0.018 & 0 \\
  2 &    NGVS-UCD2   & NGVS-J120717.93+113846.7 &    l & 181.8247260 & 11.6463089 & 0.032 & 21.074$\pm$0.003 &             \nodata &     0.984$\pm$0.011 &     0.650$\pm$0.007 &  0.822$\pm$0.010 &  0.105 & 1 \\
  3 &    NGVS-UCD3   & NGVS-J120734.18+113626.0 &    l & 181.8924309 & 11.6072260 & 0.028 & 21.492$\pm$0.004 &             \nodata &     1.062$\pm$0.011 &     0.600$\pm$0.007 &  0.709$\pm$0.012 &  0.032 & 1 \\
  4 &    NGVS-UCD4   & NGVS-J120755.71+113921.2 &    l & 181.9821228 & 11.6558897 & 0.030 & 21.057$\pm$0.003 &             \nodata &     0.892$\pm$0.008 &     0.633$\pm$0.006 &  0.725$\pm$0.010 &  0.079 & 1 \\
  5 &    NGVS-UCD5   & NGVS-J120757.12+121954.2 &    l & 181.9879964 & 12.3317198 & 0.027 & 21.444$\pm$0.004 &             \nodata &     1.089$\pm$0.011 &     0.648$\pm$0.007 &  0.723$\pm$0.012 &  0.005 & 0 \\
  6 &    NGVS-UCD6   & NGVS-J120811.52+131327.6 &    l & 182.0480094 & 13.2243239 & 0.030 & 21.408$\pm$0.004 &             \nodata &     0.910$\pm$0.012 &     0.525$\pm$0.009 &  0.673$\pm$0.019 &  0.015 & 1 \\
  7 &    NGVS-UCD7   & NGVS-J120827.25+132407.4 &    l & 182.1135389 & 13.4020547 & 0.032 & 21.373$\pm$0.004 &             \nodata &     0.920$\pm$0.010 &     0.656$\pm$0.007 &  0.724$\pm$0.013 &  0.009 & 1 \\
  8 &    NGVS-UCD8   & NGVS-J120846.23+121935.8 &    l & 182.1926310 & 12.3266129 & 0.026 & 21.338$\pm$0.003 &             \nodata &     0.970$\pm$0.010 &     0.770$\pm$0.007 &  0.852$\pm$0.012 &  0.032 & 1 \\
\enddata
\tablenotetext{}{NOTE-- (1) Identification number; (2) UCD name; (3) Object name in NGVS catalog; (4) Exposure time: l $=$ selected using long exposure data; s $=$ selected using short exposure data; (5) Right ascension in decimal degrees; (6) Declination in decimal degrees; (7) Galactic extinction according to \citet{1998ApJ_500_525Schlegel}; (8) Galactic extinction-corrected, aperture-corrected $g$-band magnitude found within a 16 pixels ($\sim3$-arcsec) diameter aperture. For details, see \citet{2015ApJ_812_34Liu}; (9) UKIDSS \citep{2007MNRAS_379_1599Lawrence} $K$-band magnitude within a 3-arcsec diameter aperture; (10-12) Color index measured in an 8-pixel ($\sim1.5$-arcsec) diameter aperture, Galactic extinction corrected; (13) Envelop parameter: $\Delta_{\rm env} \equiv g_{16}-g_{32}$, where $g_{16}$ and $g_{32}$ are magnitudes in 16- and 32-pixel diameter apertures; (14) Object flag: 0$=${\tt contaminant}, 1$=${\tt confirmed or possible UCD}. }
\end{deluxetable*}
}

\end{longrotatetable}
\clearpage


\begin{longrotatetable}
{\centering
\begin{deluxetable*}{lrrrcrlccrc}
 \tablewidth{0pt}
 \tabletypesize{\tiny}
 \tablecaption{Structural Properties and Other Information for UCD Candidates
 \label{tab:ucd_redshift}}
 \tablehead{
  \colhead{Name} &
  \colhead{$M_B$} &
  \colhead{$M_V$} &
  \colhead{$\langle r_h \rangle$} &
  \colhead{$\log(M_{*}/M_{odot})$} &
  \colhead{$v_r$} &
  \colhead{$v_{\rm source}$} &
  \colhead{Class} &
  \colhead{Envelope} &
  \colhead{Method} &
  \colhead{OtherName}  \\ \vspace{-0.5cm} \\
  \colhead{ } &
  \colhead{(mag)} &
  \colhead{(mag)} &
  \colhead{(pc)} &
  \colhead{ } &
  \colhead{(km/s)} &
  \colhead{ } &
  \colhead{ } &
  \colhead{ } &
  \colhead{ } &
  \colhead{ }  \\ \vspace{-0.5cm} \\
  \colhead{(1)} &
  \colhead{(2)} &
  \colhead{(3)} &
  \colhead{(4)} &
  \colhead{(5)} &
  \colhead{(6)} &
  \colhead{(7)} &
  \colhead{(8)} &
  \colhead{(9)} &
  \colhead{(10)} &
  \colhead{(11)} }
 \startdata
    NGVS-UCD1   &              -11.62 &              -11.38 &     14.03$\pm$ 0.39 &       6.5 &       -38 &                                                   SDSS & 5 & 0 &          $(v_r < 3500 {\rm km/s})+\langle r_h \rangle$ &              \nodata \\
    NGVS-UCD2   &               -9.68 &               -9.21 &     29.86$\pm$ 0.76 &       6.3 &   \nodata &                                                \nodata & 1 & 0 &   $u^*gz +\langle{\mu_g}\rangle_e+\langle{r_h}\rangle$ &              \nodata \\
    NGVS-UCD3   &               -9.27 &               -8.76 &     15.34$\pm$ 0.38 &       6.1 &   \nodata &                                                \nodata & 1 & 0 &   $u^*gz +\langle{\mu_g}\rangle_e+\langle{r_h}\rangle$ &              \nodata \\
    NGVS-UCD4   &               -9.70 &               -9.27 &     29.48$\pm$ 0.54 &       6.2 &   \nodata &                                                \nodata & 1 & 0 &   $u^*gz +\langle{\mu_g}\rangle_e+\langle{r_h}\rangle$ &              \nodata \\
    NGVS-UCD5   &               -9.33 &               -8.80 &     22.44$\pm$ 0.28 &       6.1 &   \nodata &                                                \nodata & 3 & 0 &   $u^*gz +\langle{\mu_g}\rangle_e+\langle{r_h}\rangle$ &              \nodata \\
    NGVS-UCD6   &               -9.36 &               -8.93 &     24.29$\pm$ 1.07 &       6.0 &   \nodata &                                                \nodata & 1 & 0 &   $u^*gz +\langle{\mu_g}\rangle_e+\langle{r_h}\rangle$ &              \nodata \\
    NGVS-UCD7   &               -9.40 &               -8.97 &     24.51$\pm$ 0.37 &       6.1 &   \nodata &                                                \nodata & 1 & 0 &   $u^*gz +\langle{\mu_g}\rangle_e+\langle{r_h}\rangle$ &              \nodata \\
\enddata
\tablenotetext{}{NOTE-- (1) UCD name; (2-3) Absolute $B$ and $V$ band magnitude, calculated using the transformation equation $B=u^*-0.8116(u^*-g)+0.1313$, $V=g-0.2906(u^*-g)+0.0885$ (Lupton 2005) and a distance modulus of $31.1$ mag \citep{2007ApJ_655_144Mei, 2009ApJ_694_556Blakeslee}; (4) Weighted mean half-light radius in the $g$ and $i$ bands, $\langle r_h \rangle=r_{h,g}$ if an $i$-band measurement is not available; (5) Stellar mass; (6) Heliocentric radial velocity (or weighted mean value if multiple measurements are available); (7) The source of our adopted $v_r$ measurement, SDSS: \citet{2018ApJS_235_42Abolfathi}; NED: NASA/IPAC Extragalactic Database; SIMBAD: SIMBAD Astronomical Database \citep{2000A+AS_143_9Wenger}; VCC: \citet{1985AJ_90_1681Binggeli}; NTT17: NTT 2017 program; AAT12: AAT 2012 program; MMT09: MMT 2009 program; Toloba2018: \citet{2018ApJL_856_31Toloba}; Ko2017: \citet{2017ApJ_835_212Ko}; MMT14: MMT 2014 program; Keck: Keck program; C03: \citet{2003ApJ_591_850Cote}; IMACS16: Megallan/IMACS 2016 program; S11: \citet{2011ApJS_197_33Strader}; (8) Class parameter: 1$=${\tt probable UCD}, 2$=${\tt dwarf nucleus}, 3$=${\tt background galaxy}, 4$=${\tt blended object}, 5$=${\tt star}, 6$=${\tt star-forming region}; (9) Envelope: 0$=${\tt lacks obvious envelope}, 1$=${\tt has obvious envelope}; (10) Selection method; (11) Altenative names, if available. }
\end{deluxetable*}
}
\end{longrotatetable}
\clearpage

\end{CJK*}
\end{document}